\documentclass[aps,pra,reprint,superscriptaddress,amsmath,amssymb,floatfix,longbibliography]{revtex4-2}
\usepackage{amssymb}
\usepackage{amsmath}
\usepackage[utf8]{inputenc}
\usepackage{array}
\usepackage{multirow}
\usepackage{color}
\usepackage{esint}
\usepackage{bm}
\usepackage{color}
\usepackage{bbm}
\usepackage{babel}
\usepackage{titlesec}
\usepackage{graphicx,amssymb,color}
\usepackage{epsfig}
\usepackage{epstopdf}
\usepackage{mathtools}
\usepackage{soul}
\usepackage[dvipsnames]{xcolor}
\usepackage{tikz}
\usepackage{tcolorbox}
\usepackage{graphics,amssymb,amsmath,epsfig,color}
\usepackage{physics}
\usepackage[breaklinks,colorlinks,
linkcolor=blue,citecolor=blue,urlcolor=blue]{hyperref}
\usepackage{orcidlink}

\usepackage{graphicx}
\usepackage{xcolor}
\usetikzlibrary{calc,arrows.meta}

\begin{document}

\title{
Microscopic Origin of Emergent Elliptic Flow and Molecule Formation in Strongly Interacting Quasi-Two-Dimensional Few-Body Systems
}
\author{Xin-Yuan Gao\orcidlink{0000-0003-0608-9029}}
\thanks{These authors contributed equally.}
\affiliation{%
Department of Physics, The Chinese University of Hong Kong, Shatin, New Territories, Hong Kong, China
}%
\author{Kai Yuen Lee\orcidlink{0009-0005-9491-0340}}
\thanks{These authors contributed equally.}
\affiliation{%
Department of Physics, The Chinese University of Hong Kong, Shatin, New Territories, Hong Kong, China
}%
\author{Qingze Guan\orcidlink{0000-0003-4813-0114}}%
\email{qingze.guan@wsu.edu}
\affiliation{%
Department of Physics and Astronomy, Washington State University, Pullman, Washington 99164-2814, USA
}
\author{Yangqian Yan\orcidlink{0000-0002-3237-5945}}%
\email{yqyan@cuhk.edu.hk}
\affiliation{%
Department of Physics, The Chinese University of Hong Kong, Shatin, New Territories, Hong Kong, China
}
\affiliation{State Key Laboratory of Quantum Information Technologies and Materials, The Chinese University of Hong Kong, Hong Kong SAR, China}
\affiliation{
The Chinese University of Hong Kong Shenzhen Research Institute, 518057 Shenzhen, China
}%
\begin{abstract}
    Recent experiments simulating two-dimensional few-fermion systems have observed emergent hydrodynamic behavior, i.e., interaction-driven elliptic flow by adding fermions two at a time~\href{https://doi.org/10.1038/s41567-024-02705-8}{[S.~Brandstetter \textit{et al.}, Nat.\ Phys.\ (2025)]}. Due to the curse of dimensionality and strong correlations, capturing such phenomena beyond two particles remains challenging.
    Here, we use the \textit{ab initio} time-dependent explicitly correlated Gaussian (TDECG) method to quantitatively reproduce these experimental observations.
    With only a moderate number of correlated Gaussian basis functions, our approach obtains converged dynamical observables for systems up to six particles.
    Furthermore, real-time access to the many-body wavefunction and two-point correlation functions enables us to visualize the transformation from a strongly interacting gas to a stream of paired molecules, i.e., a dynamical BCS-BEC crossover.

\end{abstract}
\maketitle
\phantomsection\label{arxiv:letter:start}

\textit{Introduction.---}
The real-time dynamics of strongly interacting quantum few-body systems lies at the frontier where microscopic quantum mechanics meets emergent collective phenomena.
Recent ultracold-atom experiments using single-atom-resolved optical tweezers allow deterministic preparation and real-time observation of few-fermion dynamics~\cite{serwane2011deterministic,liu2018building,reynolds2020direct,andersen2022optical,florshaim2024spatial,brandstetter2025magnifying}.
A striking demonstration is the observation of interaction-induced elliptic flow in a two-dimensional few-fermion system~\cite{brandstetter2025emergent}.
There, strong attractive interactions invert the cloud's spatial anisotropy during time-of-flight expansion: it expands faster along the initial tight direction so that the aspect ratio crosses unity~\cite{ohara2002observation,cao2011universal}, a hydrodynamic hallmark of much larger systems~\cite{floerchinger2022qualifying,heyen2025quantum,giacalone2025anisotropic}.
The expansion also reveals the emergence of tightly bound molecules, reminiscent of chemical freeze-out in quark-gluon plasma~\cite{braun-munzinger2004chemical,andronic2018decoding} and hinting at deep few-body/many-body connections.

Reproducing such phenomena theoretically remains challenging: the exponential scaling of Hilbert space makes direct approaches intractable beyond the smallest systems, calling for variational methods that capture strong correlations efficiently.
Among various approaches, the explicitly correlated Gaussian (ECG) method stands out for its exceptional ability to represent inter-particle correlations through flexible Gaussian basis functions.
For time-independent problems, ECG methods augmented with stochastic optimization have achieved remarkable precision in calculating bound states, resonances, and scattering properties across diverse physical systems~\cite{suzuki1998stochastic,mitroy2013theory}.
The extension to time-dependent phenomena, however, is substantially more challenging.
Pioneering time-dependent ECG works based on McLachlan variational-principle~\cite{varga2019optimization,rowan2020simulation}, interaction-picture~\cite{varga2012solution,sekine2017timedependent}, and Rothe-method~\cite{schrader2024time,schrader2025multidimensional} formulations have been confined to one-to-three-particle systems or to linear response, leaving open whether ECG can tackle nonequilibrium dynamics in larger, strongly correlated systems.

In this Letter, we develop the Time-Dependent Explicitly Correlated Gaussian (TDECG) method to capture these dynamics quantitatively.
By combining imaginary-time ground-state preparation with real-time variational evolution, we simulate the time-of-flight expansion of two-dimensional fermion systems up to six particles.
Our calculations reproduce the observed elliptic flow and reveal its microscopic origin through time-resolved correlation functions and Monte Carlo sampling of the many-body wavefunction.
We identify a dynamical BCS-BEC crossover during expansion, where decreasing density drives the system from Cooper-pair correlations to molecular binding, providing a microscopic picture of chemical freeze-out in few-body systems.
These results establish TDECG as a powerful tool for exploring emergent collective behavior in quantum few-body systems.

\textit{TDECG Framework.---}
We consider spin-balanced $N$ fermions of mass $M$ in an anisotropic two-dimensional (2D) harmonic trap with frequencies $\omega_x=0.615\omega$ and $\omega_y=1.63\omega$, where the geometric mean $\omega=\sqrt{\omega_x\omega_y}$ defines the harmonic-oscillator length $a_{\mathrm{ho}}=\sqrt{\hbar/(M\omega)}$.
Particles with opposite spins interact via a short-range potential $V$, while same-spin particles do not interact directly.
Since the center-of-mass decouples as a Gaussian wavepacket, we focus on the relative wavefunction, expanded by a generalized ECG ansatz:
\begin{equation}
    \Psi_\text{rel}=\sum_{j=1}^{N_b}\hat{\mathcal{A}}[\bar{u}_j\exp(-{\mathbf{x}}^\mathrm{T}\bar{A}_j\mathbf{x}/2)\exp(-{\mathbf{y}}^\mathrm{T}\bar{B}_j\mathbf{y}/2)],\label{ECG_ansatz}
\end{equation}
where $\hat{\mathcal{A}}$ enforces fermionic antisymmetry, and $\mathbf{x}$ and $\mathbf{y}$ are $(N-1)$-dimensional H-tree Jacobi coordinates~\cite{yin2015thesis}.
The prefactors $\bar{u}_j$ and the symmetric matrices $\bar{A}_j$, $\bar{B}_j$, encoding each basis function's shape and inter-particle correlations, are all taken to be complex to improve basis efficiency~\cite{hiyama2003gaussian,bubin2007relativistic}.
The $N_p=N_b[1+N(N-1)]$ independent parameters $\{u_j, A_j, B_j\}$ are collected into a complex parameter vector $\mathbf{z}$.

Applying the Lagrangian variational principle to $\Psi_\text{rel}$ recasts the time-dependent Schr\"odinger equation as a set of classical canonical equations for $\mathbf{z}$ and $\bar{\mathbf{z}}$~\cite{kramer1981geometry,kramer2008review,companion_pra}:
\begin{equation}
    i\hbar \mathcal{C} \frac{d \bar{\mathbf{z}}}{dt}=\frac{\partial \mathcal{H}}{\partial \mathbf{z}},\quad
    -i\hbar\bar{\mathcal{C}} \frac{d {\mathbf{z}}}{dt}=\frac{\partial \mathcal{H}}{\partial \bar{\mathbf{z}}},
    \label{real_time_eqn}
\end{equation}
where $\bar{\cdot}$ denotes complex conjugation, the classical Hamiltonian is $\mathcal{H}={\langle\Psi(\mathbf{z}) |\hat{H}|\Psi(\bar{\mathbf{z}}) \rangle}/{\langle\Psi(\mathbf{z}) |\Psi(\bar{\mathbf{z}}) \rangle}$, and the Fubini--Study metric $\mathcal{C}$ characterizing the symplectic manifold reads
\begin{equation}
    \mathcal{C}_{\alpha \beta}=\frac{\partial^2}{\partial z_\alpha \partial \bar{z}_\beta}\ln\langle\Psi(\mathbf{z})|\Psi(\bar{\mathbf{z}})\rangle,\quad \alpha,\beta \in \{1,2,...,N_p\}.
\end{equation}
The Wick rotation $t\rightarrow -i\tau$ in Eq.~(\ref{real_time_eqn}) gives the imaginary-time equation of motion used for ground-state preparation.
In the continuous-time limit, our formulation is equivalent to the McLachlan-variational-principle based ECG~\cite{varga2012solution,sekine2017timedependent}; however, with finite time steps and regularization of ill-conditioned matrices, retaining the explicit Fubini--Study symplectic metric $\mathcal{C}$ provides a Hamiltonian structure that better constrains the numerical drift of conserved quantities and thereby improves long-time stability~\cite{companion_pra}.

\begin{figure*}[t]
    \centering
    \includegraphics[width=\textwidth]{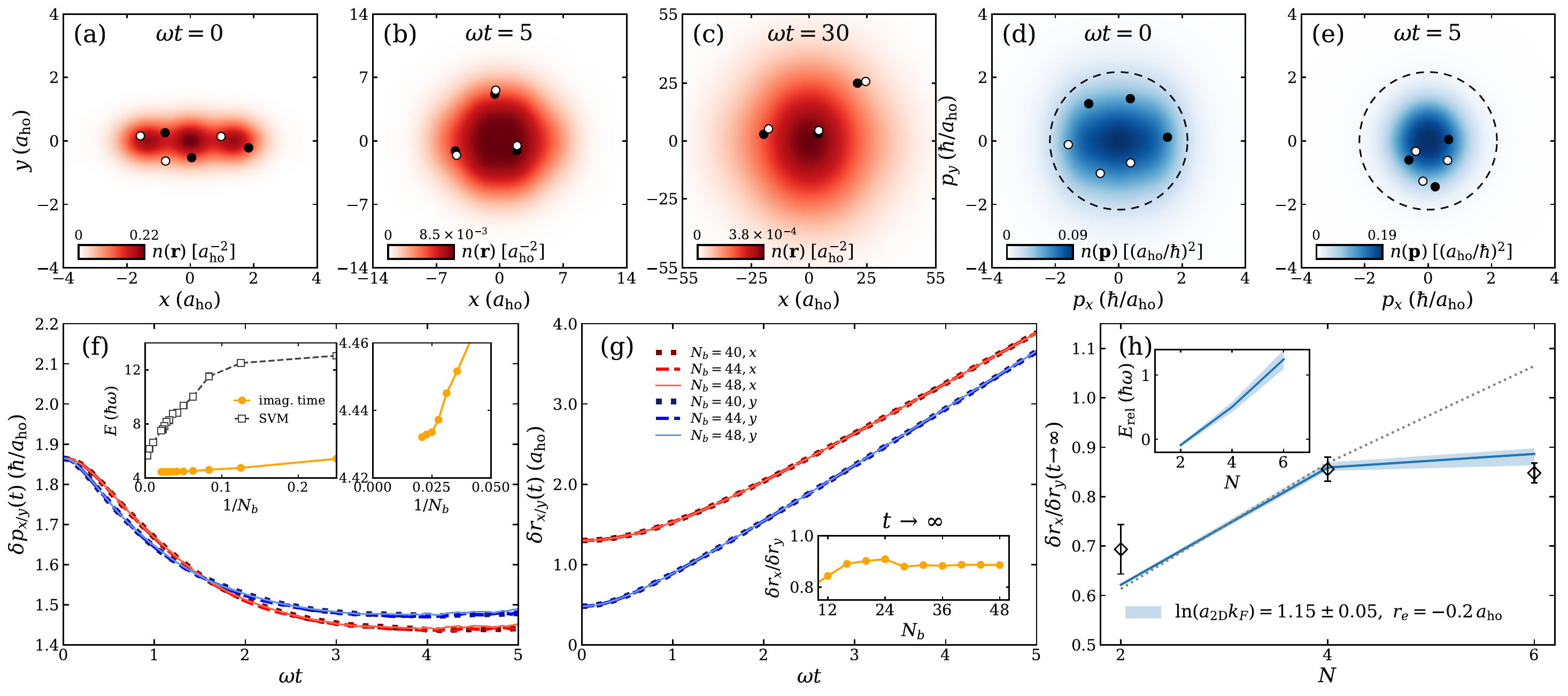}
    \caption{\textbf{Few-fermion 2D expansion: density snapshots, TDECG convergence, and emergence of elliptic flow.}
(a)-(e)~One-body densities for $N=3+3$ at $\omega t=0,5,30$ in real space and $\omega t=0,5$ in momentum space, with representative single-shot particle positions overlaid; black/white markers indicate sampled spin-up/spin-down particles. Dashed circles in (d),(e) mark the non-interacting Fermi momentum $\hbar k_F\simeq 2.17\,\hbar/a_{\mathrm{ho}}$ of the $3+3$ ground state in the anisotropic trap, corresponding to a Fermi energy $E_F=\hbar(5\omega_x+\omega_y)/2$.
(f)~Momentum-space widths $\delta p_{x/y}(t)$ after trap release, showing interaction-induced anisotropy, for basis sizes $N_b=40$ (dark, thick dots), $44$ (normal, dashed lines), and $48$ (bright, solid lines). Left inset: orange circles (gray squares) show the relative ground-state energy $E$ as a function of $1/N_b$ from imaginary-time propagation (from the stochastic variational method); imaginary-time propagation converges substantially faster. Right inset zooms in on the converged regime ($1/N_b\le 0.05$, $E\in[4.42,4.46]\,\hbar\omega$), demonstrating 2--3 significant digits of convergence by $N_b=48$.
(g)~Real-space widths $\delta r_{x/y}(t)$, showing linear ballistic expansion. Inset shows the long-time aspect ratio $\delta r_x/\delta r_y$ as a function of $N_b$.
(h)~Long-time aspect ratios $\delta r_x/\delta r_y(t\to\infty)$ for $N=2,4,6$. Black diamonds: experimental data~\cite{brandstetter2025emergent}; gray dashed line: non-interacting reference; blue band: TDECG results for $\ln(a_{\mathrm{2D}}k_F)=1.15\pm0.05$ at fixed $r_e=-0.2a_{\mathrm{ho}}^2$ (central line: $\ln(a_{\mathrm{2D}}k_F)=1.15$). Inset shows the initial relative energy $E_\mathrm{rel}$ as a function of $N$.
}
\label{fig:overview}
\end{figure*}

A key advantage of the Gaussian basis is that, paired with a Gaussian interaction, all matrix elements in $\mathcal{H}$ and $\mathcal{C}$ are analytical, enabling efficient evaluation of Eq.~(\ref{real_time_eqn}) for many basis functions.
The low-energy two-body scattering between opposite spins is characterized by the 2D scattering length $a_{\text{2D}}$ and effective range $r_e$~\cite{adhikari1986quantum,galea2017fermions,schonenberg2017effectiverange}, $\cot[\delta(k)]=\tfrac{2}{\pi}\ln\left(k a_{\text{2D}}\right)+\tfrac{k^2r_e}{2}+\cdots$.
To independently tune both $a_{\text{2D}}$ and $r_e$, we adopt a double-Gaussian pseudopotential for particles occupying different spin states:
\begin{equation}
    V(r)=-U_0\exp\left(-\frac{r^2}{2\sigma^2}\right)+U_1\exp\left[-\frac{r^2}{2(2\sigma)^2}\right]
    \label{pseudopotential}
\end{equation}
and vanishing interactions for particles in the same spin state. 
The second, repulsive Gaussian is essential for reproducing the experimentally relevant \textit{negative} effective range; a single-Gaussian potential can only yield $r_e>0$ in 2D (see the companion paper~\cite{companion_pra} and Ref.~\cite{yin2020fewbody} for a similar treatment).

Following the relevant experimental parameters~\cite{petrov2001interatomic,hu2019reduced,yang2024two,brandstetter2025emergent}, we target $\ln(a_{\text{2D}}k_F)\simeq 1.15$ and $r_e\simeq-0.2a_{\mathrm{ho}}^2$, with $k_F=\sqrt{2ME_F/\hbar^2}$, where $E_F$ is the highest single-particle energy occupied in the non-interacting ground state of the anisotropic trap; this places the system on the strongly interacting BCS side~\cite{ries2015observation,galea2017fermions,murthy2018hightemperature,sobirey2021observation}.
We fix the short-range scale at $\sigma=0.1a_{\mathrm{ho}}$ and tune $(U_0,U_1)$ for each particle number to match the target $a_{\text{2D}}$ and $r_e$.
Since $E_F$ depends on $N$, the well depth and barrier vary slightly: $(U_0,U_1)/\hbar\omega=(85.9,19.2),(95.9,22.7),(104,25.8)$ for $N=1{+}1,2{+}2,3{+}3$ systems.

\begin{figure*}[t]
    \centering
    \includegraphics[width=0.99\textwidth]{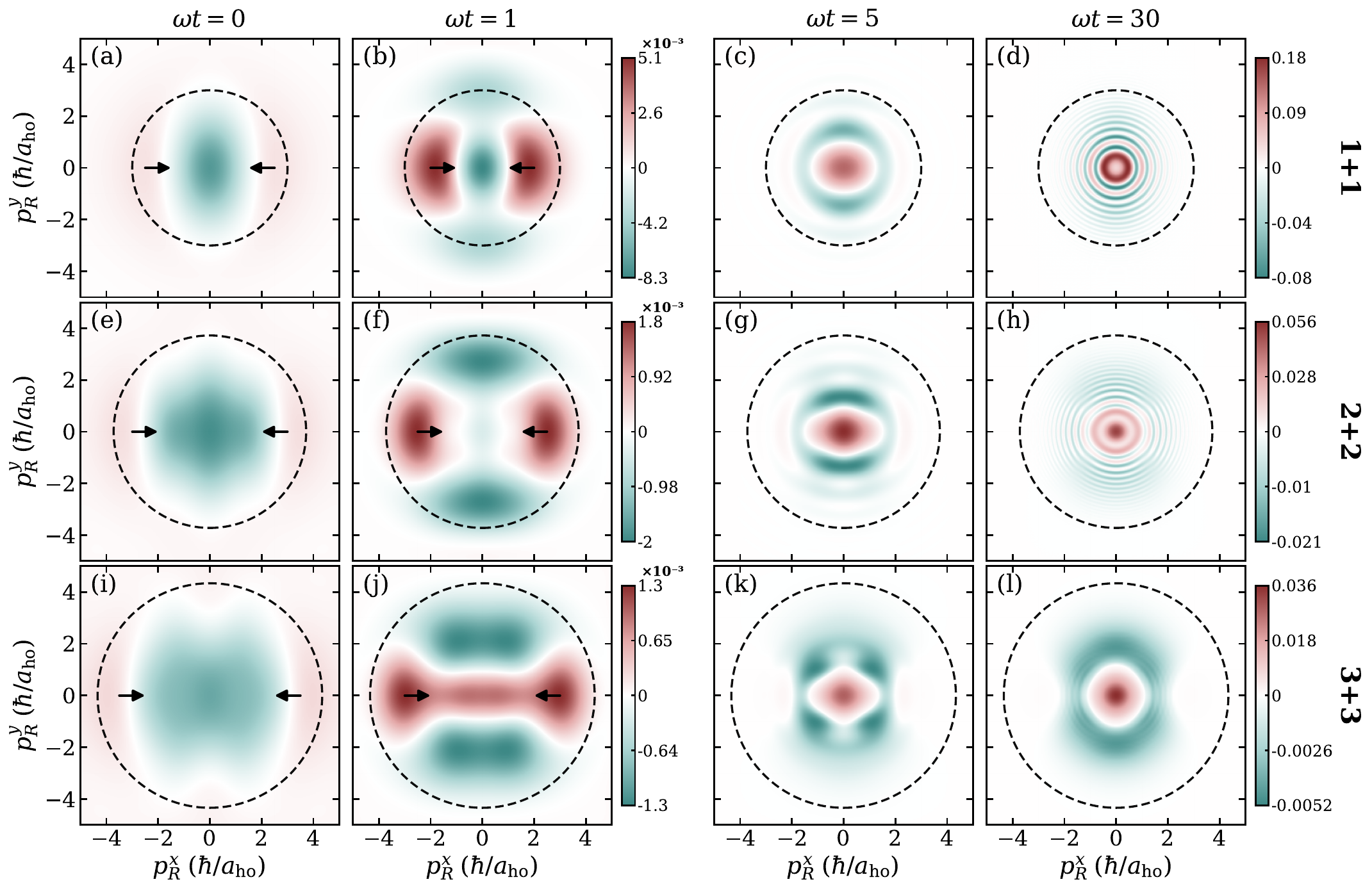}
    \caption{\textbf{Dynamical BCS-BEC crossover visualized through relative momentum correlations.}
Evolution of the relative momentum correlator $C^{(2)}_R(\mathbf{p}_R)$ for the $1{+}1$, $2{+}2$, and $3{+}3$ systems in the anisotropic trap, shown in the top, middle, and bottom rows, respectively. Red (green) regions correspond to enhanced (depleted) opposite-spin coincidence at relative momentum $\mathbf{p}_R$.
From left to right, the four columns correspond to $\omega t=0,1,5,30$.
The dashed circles mark the eye-guide radius $|\mathbf{p}_R|=2\hbar k_F$, and the black arrows in panels (a),(b),(e),(f),(i),(j) indicate the inward migration of the back-to-back positive lobes toward $\mathbf{p}_R=0$.
}
    \label{fig:3}
\end{figure*}

\textit{Results.---}
First, we prepare the initial state as the ground state by imaginary-time propagation.
In real space [Fig.~\ref{fig:overview}(a)], the six fermions spread across the anisotropic ground-state density without tight spin-$\uparrow$/spin-$\downarrow$ pairing, while the momentum distribution [Fig.~\ref{fig:overview}(d)] fills a nearly isotropic disk bounded by the non-interacting Fermi momentum $\hbar k_F = 2.17\,\hbar/a_{\mathrm{ho}}$, despite the trap anisotropy $\omega_y/\omega_x\approx 2.65$, as confirmed by $\delta p_x(0)\simeq\delta p_y(0)$ in Fig.~\ref{fig:overview}(f).
We vary the basis size up to $N_b=48$ to verify convergence: the right inset of Fig.~\ref{fig:overview}(f) shows the ground-state energy stable to two-to-three significant digits by $N_b=48$.
The left inset benchmarks against a stochastic variational method (SVM)~\cite{suzuki1998stochastic} sharing the same matrix elements; even at $N_b=384$ the stochastic basis remains well above the imaginary-time value at $N_b=48$, showing that imaginary-time propagation builds a more efficient correlated basis for the $3{+}3$ problem than stochastic sampling.

Then, we release the trap and the cloud expands.
Panels~(b) and (c) of Fig.~\ref{fig:overview} display the real-space density $n(\mathbf{r})$ at $\omega t=5$ and $30$: the initially anisotropic cloud first expands into a nearly circular profile and then inverts its major and minor axes, a hallmark of collective hydrodynamic behavior.
By $\omega t=5$, the Monte Carlo snapshots reveal tight spin-$\uparrow$/spin-$\downarrow$ pairs, signaling the spontaneous formation of molecules.
Correspondingly, by $\omega t=5$ the single-particle momentum distribution $n(\mathbf{p})$ in panel~(e) has already contracted well inside the initial Fermi momentum disk: as opposite-spin particles bind into dimers, the broad BCS-like occupation extending up to $k_F$ is replaced by a narrower, anisotropic momentum spread.
We monitor the cloud widths $\delta p_{x/y}(t)=\sqrt{\tfrac{1}{N}\sum_{k}\langle(p^{x/y}_{k})^{2}\rangle}$ and $\delta r_{x/y}(t)=\sqrt{\tfrac{1}{N}\sum_{k}\langle(r^{x/y}_{k})^{2}\rangle}$, evaluated analytically from $\mathbf{z}$ thanks to the ECG basis~\cite{companion_pra}.
Figure~\ref{fig:overview}(f) shows $\delta p_{x/y}(t)$ for $N_b=40$, $44$, and $48$; the curves are visually indistinguishable.
The momentum widths are initially nearly isotropic, but they decrease and saturate at different asymptotic values, generating an interaction-induced anisotropy, which is direct evidence for the emergence of elliptic flow~\cite{floerchinger2022qualifying}.
Figure~\ref{fig:overview}(g) shows the corresponding real-space widths, where convergence is even better: results for all three $N_b$ values lie on top of one another.
At long times, $\delta r_{x/y}(t)\propto v_{x/y}t$, consistent with ballistic expansion.
Notably, the free-particle relation $v_{x/y}= \delta p_{x/y}/M$ is no longer obeyed, which is another hint of pairing formation.
We extract the asymptotic aspect ratio $\delta r_x/\delta r_y$ by fitting this linear tail; as illustrated in the inset of Fig.~\ref{fig:overview}(g), the ratio is insensitive to $N_b$ once $N_b\gtrsim 28$.

To identify genuine elliptic flow in few-body systems, we must distinguish interaction-driven anisotropy from that arising purely from Heisenberg uncertainty in anisotropic traps.
Figure~\ref{fig:overview}(h) compares long-time aspect ratios $\delta r_x/\delta r_y$ between interacting and non-interacting systems for $N=2,4,6$ particles.
For $N=2$ and $4$, the interacting and non-interacting aspect ratios nearly coincide, whereas $N=6$ shows a significant deviation, marking the emergence of true elliptic flow~\cite{bayha2020observing}.
Our TDECG results agree well with the experimental data~\cite{brandstetter2025emergent} for the theoretically predicted parameters.
The $N=2$ system behaves similarly to the non-interacting reference because it already forms a weakly bound molecule at $t=0$ [negative relative energy, inset of Fig.~\ref{fig:overview}(h)], effectively reducing the expansion to a single-body problem.
In contrast, the $N=4$ and $6$ systems start with positive relative energies, allowing interactions to drive non-trivial dynamics, although only $N=6$ reaches the level of complexity needed for collective behavior.

To understand the evolution microscopically, we analyze the two-body momentum correlator~\cite{altman2004probing}:
\begin{equation}
    C^{(2)}(\mathbf{p},\mathbf{p}')=n^{(2)}(\mathbf{p},\uparrow;\mathbf{p}',\downarrow)-n(\mathbf{p})n(\mathbf{p}'),
\end{equation}
where $n^{(2)}(\mathbf{p},\uparrow;\mathbf{p}',\downarrow)$ is the joint momentum-space probability density for finding a spin-up particle at momentum $\mathbf{p}$ and a spin-down particle at momentum $\mathbf{p}'$.
Transforming to relative [$\mathbf{p}_R=\mathbf{p}-\mathbf{p}'$] and center-of-mass [$\mathbf{p}_C=(\mathbf{p}+\mathbf{p}')/2$] coordinates and integrating over the conjugate variable yields the relative correlator $C^{(2)}_R(\mathbf{p}_R)$, which directly encodes pairing correlations.

Figure~\ref{fig:3} reveals a striking particle-number dependence in the relative momentum correlator $C^{(2)}_R$.
At $\omega t=0$, all three systems exhibit a pair of positive back-to-back lobes near $|\mathbf{p}_R|\approx 2\hbar k_F$ [arrows in panels (a),(e),(i)], characteristic of Cooper-pair correlations in which opposite-spin fermions occupy nearly opposite momenta close to the Fermi surface~\cite{holten2022observation,hartke2023direct}.
By $\omega t=1$, these lobes have migrated inward toward $\mathbf{p}_R=0$ [arrows in (b),(f),(j)], signaling the onset of molecule formation; the $3{+}3$ system already develops a pronounced positive central feature, whereas the $1{+}1$ and $2{+}2$ systems still show a near-zero central value, indicating a slower crossover to the molecular regime.
At long times, the concentration of $C^{(2)}_R$ near $\mathbf{p}_R=0$ in all three rows confirms that each $N{+}N$ system asymptotes to $N$ tightly bound molecules.
This evolution represents a dynamical BCS-BEC crossover: as the cloud expands, the effective Fermi wavevector $k_F'(t)$ decreases with the cloud density, driving the system from the BCS regime ($a_\mathrm{2D}k_F>1$) toward the BEC limit ($a_\mathrm{2D}k_F'\ll 1$), with the expansion producing stable molecules~\cite{zhang2021transition}.

\begin{figure}[tbp]
    \hspace*{-0.02\textwidth}\includegraphics[width=0.49\textwidth]{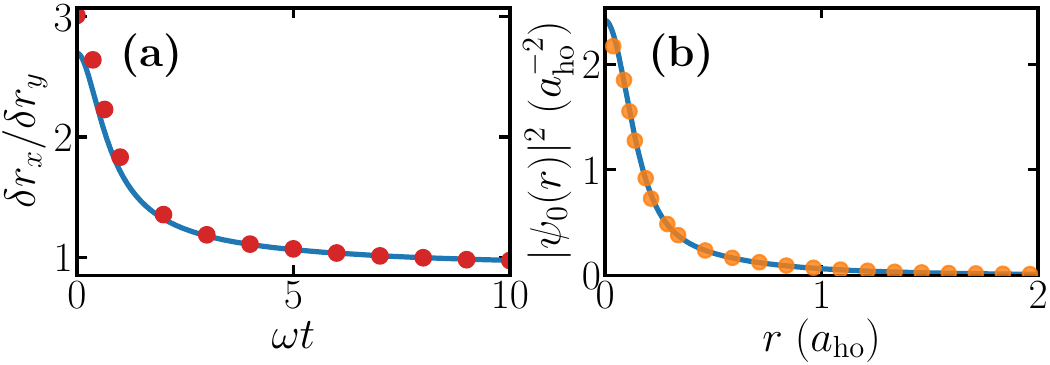}
\caption{\textbf{Quantitative evidence for molecular formation.} (a)~Aspect ratio $\delta r_x/\delta r_y$: blue line (red circles) shows the full ECG many-body calculation (Monte Carlo sampling that treats the system as three distinguishable molecules).
(b)~Radial probability density of the pair-relative coordinate: orange circles show Monte Carlo sampling at $\omega t=30$; blue line shows the s-wave bound-state probability density $|\psi_0(r)|^2$ of the two-body Schr\"odinger equation with potential $V(r)$, Eq.~(\ref{pseudopotential}).
}
\label{fig:4}
\end{figure}

A qualitative distinction among the three particle numbers emerges, however, in the structure of $C^{(2)}_R$ around the central molecular peak at $\omega t=30$ (rightmost column).
For the integrable $1{+}1$ system, $C^{(2)}_R$ exhibits pronounced interference fringes; for $2{+}2$, the fringes remain visible but are blurred, suggesting the possibility of an effective two-molecule description whose dynamics remains close to integrable due to the molecular binding-energy gap; for $3{+}3$, the fringes almost vanish and the long-time correlator becomes smooth, as expected when three emergent molecules provide the minimum complexity for chaotic few-body motion.
This molecular interpretation is checked directly by the late-time diagnostic in Fig.~\ref{fig:4}, where the sampled pair-relative distribution agrees with the two-body bound-state probability density.
Thus, the visibility of fringes in $C^{(2)}_R$ tracks an integrable--nearly-integrable--chaotic hierarchy across $1{+}1$, $2{+}2$, and $3{+}3$.
This hierarchy is consistent with the diagnostics in Figs.~\ref{fig:overview}--\ref{fig:4}: the $N=2$ and $N=4$ aspect ratios stay close to the non-interacting prediction, the $2{+}2$ correlator retains only partially blurred fringes, and only $N=6$ shows both a strong elliptic-flow departure and a smooth long-time correlator.
It is further substantiated by the out-of-time-order correlator calculation~\cite{companion_pra}, where at short times the normalized correlator remains pinned near unity for $1{+}1$, decays only perturbatively for $2{+}2$, and departs much earlier for $3{+}3$, directly indicating the onset of fast operator growth in the six-particle system.

To further substantiate this molecular picture, we perform a Monte Carlo analysis by sampling the many-body wavefunction at each time step: for each sample, we identify the three spin-up--spin-down pairs and extract the molecular center-of-mass coordinates.
Figure~\ref{fig:4}(a) compares the aspect ratio $\delta r_x/\delta r_y$ computed from this molecular decomposition with the full ECG analytical result.
At early times, the two curves deviate, reflecting genuine multi-particle correlations.
For $\omega t\gtrsim 5$, however, they overlap, confirming that the system is quantitatively described as three independent molecules.
Figure~\ref{fig:4}(b) shows the pair-relative probability density at $\omega t=30$, which agrees precisely with $|\psi_0(r)|^2$, the s-wave bound-state probability density of the two-body Schr\"odinger equation with potential $V(r)$, Eq.~(\ref{pseudopotential}).
Together, these results establish that the expansion produces well-defined molecules in their internal ground states.

\textit{Conclusion.---}
We applied the TDECG method to simulate the time-of-flight expansion of strongly interacting two-dimensional few-fermion systems.
By combining imaginary-time relaxation with real-time variational evolution, we quantitatively reproduced the interaction-induced elliptic flow observed experimentally and confirmed that six particles constitute the minimal system size exhibiting genuine collective behavior.
The algorithm converges with only a few dozen basis functions, granting real-time access to the full many-body wave function and its correlation functions.
This capability, combined with Monte Carlo sampling of the wavefunction, allowed us to visualize a dynamical BCS-BEC crossover in which chemical freeze-out manifests as a continuous transformation from Cooper pairing to tightly bound dimers occupying the two-body ground state.
Our results establish TDECG as a powerful \textit{ab initio} tool for nonequilibrium few-body physics, bridging microscopic dynamics and emergent hydrodynamics, and open the way to larger $N$, other dimensionalities, and richer interaction protocols.

\begin{acknowledgments}
    We acknowledge financial support from the National Natural Science Foundation of China under Grant Nos.~124B2074 and 12204395,
    Hong Kong RGC Early Career Scheme (Grant No. 24308323) and Collaborative Research Fund (Grant No.~C4050-23GF), 
    the Space Application System of China Manned Space Program, 
    Guangdong Provincial Quantum Science Strategic Initiative GDZX2404004, 
    and CUHK Direct Grant No. 4053731.
    Q.G.\ acknowledges support from the NSF through Grant No.~PHY-2409600 and from Washington State University through the Claire May \& William Band Distinguished Professorship Award.
\end{acknowledgments}

\section*{DATA AVAILABILITY}
The data that support the findings of this letter are openly available at \cite{data}. The source code and scripts required to reproduce the numerical results are available at \cite{TDECG_code}.

\clearpage
\onecolumngrid
\setcounter{section}{0}
\setcounter{equation}{0}
\setcounter{figure}{0}
\setcounter{table}{0}
\setcounter{footnote}{0}
\renewcommand{\thefigure}{\arabic{figure}}
\renewcommand{\theequation}{\arabic{equation}}
\renewcommand{\thetable}{\arabic{table}}
\renewcommand{\thesection}{\arabic{section}}
\renewcommand{\theHfigure}{II.\arabic{figure}}
\renewcommand{\theHequation}{II.\arabic{equation}}
\renewcommand{\theHtable}{II.\arabic{table}}
\renewcommand{\theHsection}{II.\arabic{section}}
\renewcommand{\theHsubsection}{II.\arabic{section}.\arabic{subsection}}
\vspace*{1em}
\phantomsection\label{arxiv:long:start}
\begin{center}
{\large Companion long paper}\\[0.6em]
{\Large\bfseries Time-Dependent Explicitly Correlated Gaussian Method for Two-Dimensional Few-Fermion Systems}\\[1.2em]
Xin-Yuan Gao$^{1}$, Kai Yuen Lee$^{1}$, Qingze Guan$^{2}$, and Yangqian Yan$^{1,3,4}$\\[0.6em]
{\small
$^{1}$\textit{Department of Physics, The Chinese University of Hong Kong, Shatin, New Territories, Hong Kong, China}\\
$^{2}$\textit{Department of Physics and Astronomy, Washington State University, Pullman, Washington 99164-2814, USA}\\
$^{3}$\textit{State Key Laboratory of Quantum Information Technologies and Materials, The Chinese University of Hong Kong, Hong Kong SAR, China}\\
$^{4}$\textit{The Chinese University of Hong Kong Shenzhen Research Institute, 518057 Shenzhen, China}
}
\end{center}
\vspace{0.8em}
\begin{center}
\begin{minipage}{0.92\textwidth}
\small
\noindent\textbf{Abstract.} We present a detailed methodological account of the time-dependent explicitly correlated Gaussian (ECG) approach for studying the nonequilibrium dynamics of strongly interacting two-dimensional few-fermion systems. Starting from the Lagrangian form of the time-dependent variational principle for complex ECG parameters, we derive the exact equations of motion and provide analytic expressions for the overlap, Hamiltonian, and gradient matrix elements required for imaginary- and real-time propagation. We further document our numerical implementation and post-processing toolkit, which includes convergence benchmarks, the analytical extraction of cloud widths and aspect ratios, the evaluation of reduced density matrices and momentum-space correlators, out-of-time-order correlators as a probe of information scrambling, and Monte Carlo sampling of the many-body wavefunction. This work serves as a comprehensive reference for employing time-dependent ECG method as an \textit{ab initio} tool for nonequilibrium quantum few-body dynamics.
\end{minipage}
\end{center}
\vspace{1em}
\twocolumngrid

\section{Introduction}
\label{pra:sec:introduction}

Strongly interacting many-body quantum systems are inherently challenging to describe, whereas few-body quantum systems offer a controlled, bottom-up framework for understanding the emergence of collective behavior in the many-body limit.
Taking advantage of laser trapping techniques, such as single-atom-resolved optical microtraps and optical tweezers, atomic quantum systems can now be deterministically prepared, manipulated, and measured at the single- and few-particle levels~\cite{serwane2011deterministic,liu2018building,reynolds2020direct,andersen2022optical,florshaim2024spatial,brandstetter2025magnifying,brandstetter2025emergent}, thereby significantly advancing the field.
A striking recent example~\cite{brandstetter2025emergent} is the observation of interaction-induced elliptic flow during the expansion of a strongly interacting two-component few-fermion system released from an anisotropic quasi-two-dimensional optical dipole trap.
In the absence of interactions, ballistic time-of-flight expansion would simply map the in-trap momentum distribution onto real space, and the resulting cloud aspect ratio would remain on the same side of unity as the in-trap one.
Interactions break this picture: collisions during the early stage of expansion redistribute momentum from the strongly confined direction to the weakly confined one, so that the cloud aspect ratio crosses unity and inverts during time-of-flight expansion~\cite{ohara2002observation,cao2011universal,floerchinger2022qualifying,heyen2025quantum,giacalone2025anisotropic}.
At the same time, the strong interactions drive the spontaneous formation of tightly bound opposite-spin molecules~\cite{braun-munzinger2004chemical,andronic2018decoding}.
A rigorous microscopic theory of such dynamics is therefore needed, both to interpret the observed phenomena and to guide future experiments.
However, with increasing particle number, the exponential growth of the Hilbert space,
together with the dynamical buildup of few-body correlations, renders this task computationally challenging.
As demonstrated in this work, the time-dependent explicitly correlated Gaussian (TDECG) approach, which employs a basis that efficiently adapts to strongly interacting relative motion, provides an effective framework for simulating such dynamics.

The explicitly correlated Gaussian (ECG) approach encodes particle correlations
at the level of the Gaussian basis functions,
which enables analytical expressions for all the relevant matrix elements.
The basis set can be generated and optimized using stochastic optimization algorithms.
These analytical and stochastic features make the ECG approach an accurate and efficient numerical tool for studying few-body systems.
For stationary problems, the ECG approach has been widely used to investigate a variety of physical systems, ranging from atomic and molecular to nuclear few-body systems~\cite{suzuki1998stochastic,mitroy2013theory,vonstecher2007spectrum}.

Applying the ECG approach to time-dependent problems is considerably more challenging. Several ECG-based methods for nonequilibrium few-body systems have been proposed in the literature, including interaction-picture formulations~\cite{varga2012solution,sekine2017timedependent}, variational propagation approaches~\cite{varga2019optimization,rowan2020simulation}, and per-step optimization schemes~\cite{schrader2024time,schrader2025multidimensional}. However, these methods are largely restricted to linear-response regimes and to systems containing only one to three particles.
In our companion Letter~\cite{companion_prl} and in this work, we develop the TDECG approach and demonstrate that these limitations can be overcome, achieving quantitative agreement with experimentally observed elliptic flow in Ref.~\cite{brandstetter2025emergent},
as well as the dynamical BCS--BEC crossover in two-component few-fermion systems consisting of $1{+}1$, $2{+}2$, and $3{+}3$ particles.

In this paper,
we provide:
(i) the derivation of the TDECG equations of motion using the Lagrangian variational principle~\cite{kramer1981geometry,kramer2008review},
together with a proof of their continuous-time equivalence to the McLachlan formulation~\cite{mclachlan1964variational,broeckhove1988equivalence} and a numerical demonstration that the Lagrangian form is considerably more robust under finite-step discretization;
(ii) analytical expressions for the Fubini--Study metric~\cite{brody2001geometric,hackl2020geometry} and the Hamiltonian gradient required for imaginary- and real-time propagation under a complex ECG ansatz;
(iii) a discussion of the tunability of a double-Gaussian potential calibrated to reproduce the two-dimensional scattering length and effective range relevant to the experiment;
(iv) regularization strategies for handling the ill-conditioned symplectic matrix arising in the algorithm;
(v) a convergence study for two-component few-fermion systems consisting of $1{+}1$, $2{+}2$, and $3{+}3$ particles;
and
(vi) a post-processing toolbox through which experimentally accessible observables and dynamical diagnostics,
such as cloud widths, aspect ratios, pair correlations, reduced density matrices, and out-of-time-order correlators that probe few-body information scrambling,
are reconstructed either analytically or via Monte Carlo sampling techniques.

The paper is organized as follows.
Section~\ref{pra:sec:formalism} presents the general variational framework.
Section~\ref{pra:sec:model} defines the model Hamiltonian and discusses the interaction-potential parameters.
Section~\ref{pra:sec:ecg_ansatz} introduces the ECG ansatz, the H-tree Jacobi coordinates, and derives the analytical matrix elements.
Section~\ref{pra:sec:regularization} discusses the treatment of the ill-conditioned symplectic matrix.
Section~\ref{pra:sec:numerical} describes the numerical method and presents convergence benchmarks for the $1{+}1$, $2{+}2$, and $3{+}3$ systems.
Section~\ref{pra:sec:postprocessing} presents the post-processing toolkit used to reconstruct observables from the ECG wavefunction---including the analytical cloud widths and aspect ratios, momentum correlators, out-of-time-order correlators as a diagnostic of information scrambling, molecular diagnostics, and reduced-density-matrix analysis---together with their demonstrations.
Section~\ref{pra:sec:conclusion} concludes.
Appendix~\ref{pra:app:variational_principles} provides a proof of the equivalence between the Lagrangian variational principle and the McLachlan variational principle for holomorphic parametrizations.
Appendix~\ref{pra:app:method_comparison} provides a direct numerical comparison of fixed-step Euler propagation under the Lagrangian and McLachlan formulations for the $3{+}3$ system.

\section{General Formalism}
\label{pra:sec:formalism}

We employ the Lagrangian formulation of quantum mechanics, which provides a natural framework for deriving the equation of motion for the Schr\"odinger field using the variational principle~\cite{kramer1981geometry}.
The Lagrangian $L$ for the time-dependent Schr\"odinger field $|\Psi\rangle$ is
\begin{equation}
    L(|\Psi\rangle,\langle\Psi|)=\frac{i\hbar}{2}\frac{\langle\Psi|\dot{\Psi}\rangle-\langle\dot{\Psi}|\Psi\rangle}{\langle\Psi|\Psi\rangle}-\frac{\langle\Psi|\hat{H}|{\Psi}\rangle}{\langle\Psi|\Psi\rangle},
    \label{pra:eq:Lagrangian}
\end{equation}
where
$\hat{H}$ is the Hamiltonian
and the dot denotes the time derivative.
For a state subject to certain constraints and depending parametrically on a set of time-dependent parameters,
the equations of motion for those parameters can be obtained by minimizing the action $S=\int_{t_1}^{t_2} L\, dt$ with fixed
spacetime
boundary conditions~\cite{kramer2008review}.

In this work,
we assume $|\Psi\rangle$ follows a parametrized ansatz
\begin{equation}
|\Psi\rangle=|\Psi(\bar{z}_1,\bar{z}_2,\ldots,\bar{z}_{N_p})\rangle,
\end{equation}
where $z_\alpha$ $(\alpha=1,2,\ldots, N_p)$ are complex parameters and
$\bar{\cdot}$
denotes complex conjugation.
The resulting dynamics of the parameters $z_i$ are governed by a set of classical canonical equations of motion in symplectic form,
\begin{equation}
i\hbar \mathcal{C} \frac{d \bar{\mathbf{z}}}{dt}=\frac{\partial \mathcal{H}}{\partial \mathbf{z}},\quad
-i\hbar \bar{\mathcal{C}} \frac{d \mathbf{z}}{dt}=\frac{\partial \mathcal{H}}{\partial \bar{\mathbf{z}}},
\label{pra:eq:real_time_eqn}
\end{equation}
where the symplectic matrix $\mathcal{C}$ characterizes the geometry of the variational manifold,
\begin{equation}
\mathcal{C}_{\alpha \beta}=\frac{\partial^2}{\partial z_\alpha \partial \bar{z}_\beta}\ln \langle \Psi(\mathbf{z}) | \Psi(\bar{\mathbf{z}}) \rangle,
\label{pra:eq:C_matrix}
\end{equation}
and the energy is defined as 
\begin{eqnarray}
\mathcal{H} = \frac{\langle \Psi(\mathbf{z}) | \hat{H} | \Psi(\bar{\mathbf{z}}) \rangle }{ \langle \Psi(\mathbf{z}) | \Psi(\bar{\mathbf{z}}) \rangle}.
\end{eqnarray}
The matrix $\mathcal{C}$ is Hermitian and positive semidefinite, and its rank equals the effective dimensionality of the variational manifold.

To obtain the ground state,
we perform imaginary-time evolution in the $z_i$ parameter space.
The corresponding equations of motion are obtained by applying the
Wick rotation $t\rightarrow -i\tau$ to Eq.~\eqref{pra:eq:real_time_eqn},
which yields
\begin{equation}
    \hbar\mathcal{C} \frac{d \bar{\mathbf{z}}}{d\tau}=-\frac{\partial \mathcal{H}}{\partial \mathbf{z}},\quad
    \hbar\bar{\mathcal{C}} \frac{d {\mathbf{z}}}{d\tau}=-\frac{\partial \mathcal{H}}{\partial \bar{\mathbf{z}}}.
    \label{pra:eq:imag_time_eqn}
\end{equation}
As a gradient-descent flow on the constrained energy surface,
imaginary-time evolution drives the system toward the variational ground state as $\tau\to\infty$.

The TDECG approach presented here differs from other time-dependent ECG methods in the literature.
The interaction-picture formulations~\cite{varga2012solution,sekine2017timedependent} propagate the expansion coefficients while keeping the basis functions fixed or assigning them prescribed time-dependent widths; these methods are well-suited for perturbative regimes but do not allow the nonlinear parameters to adapt dynamically.
Rothe's method~\cite{schrader2024time,schrader2025multidimensional} recasts time evolution as a discrete optimization problem at each time step; while conceptually elegant, this approach has so far been applied to single-particle problems.
In contrast, the TDECG approach derives continuous equations of motion for \textit{all} ECG parameters directly from the Lagrangian, allowing the basis to adapt smoothly to strongly correlated few-body dynamics without repeated optimization.

In this regard, our continuous parameter-evolution framework is closely related to earlier variational approaches applied to finding the ground state via imaginary-time evolution~\cite{varga2019optimization} and the real-time dynamics of a single hydrogen atom~\cite{rowan2020simulation}, 
which derived the equations of motion using the McLachlan variational principle~\cite{mclachlan1964variational}.
For holomorphic parametrizations where the exact Dirac-Frenkel tangent-space condition is strictly satisfied, 
the Lagrangian formulation and the McLachlan formulation are mathematically equivalent and yield the same continuous equations of motion (a detailed derivation is provided in Appendix~\ref{pra:app:variational_principles}).
However, the Lagrangian formulation
provides a crucial structural advantage for large systems and long-time propagation compared to the McLachlan formulation.
In practice, 
finite time steps, regularization of ill-conditioned matrices, 
and floating-point errors inevitably cause weak violations of the tangent-space condition.
Because the McLachlan equation directly instantiates the Dirac-Frenkel condition, 
its numerical implementation lacks a structural constraint to prevent error accumulation.
Conversely, the Lagrangian equations~(\ref{pra:eq:real_time_eqn}) inherit a Hamiltonian character through the explicit symplectic metric $\mathcal{C}$ [the Fubini--Study metric, see Eq.~(\ref{pra:eq:fubini_study_app})], which makes the conservation laws manifest at the continuous level, better constrains numerical drift when the tangent-space condition is only weakly violated,
confines the effects of numerical errors within a geometrically stable framework, 
and therefore suppresses the secular drift of conserved quantities~\cite{broeckhove1988equivalence,hackl2020geometry}. 
We provide a direct numerical demonstration of this enhanced stability in Appendix~\ref{pra:app:method_comparison}, underscoring why the Lagrangian formulation is advantageous for the long-time evolution of complex few-body systems.

\section{Model System}
\label{pra:sec:model}

\begin{figure*}[t!]
    \centering
\includegraphics[width=\textwidth]{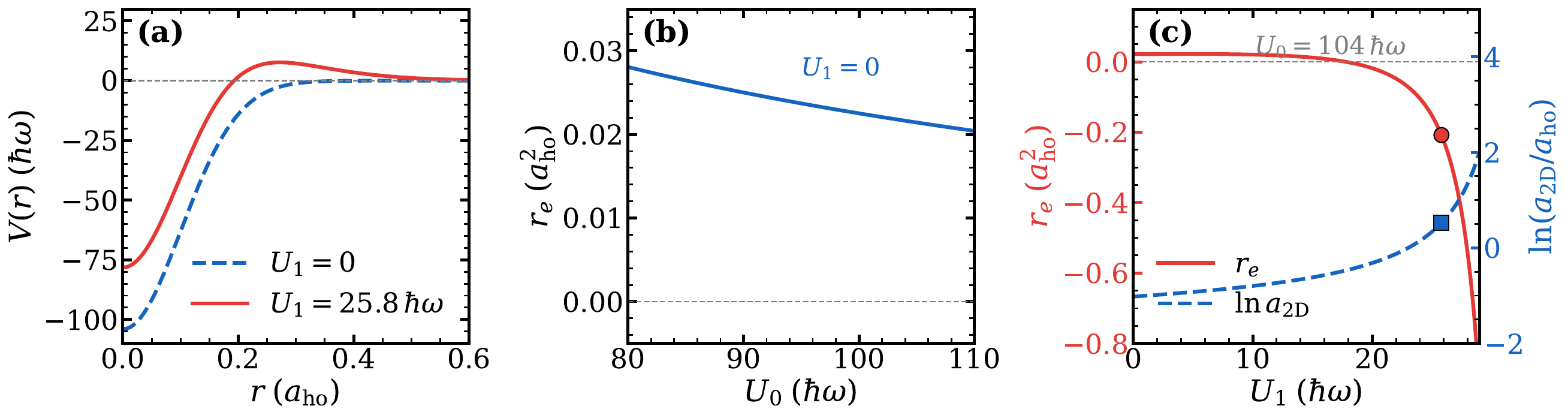}
\caption{Tunability of the 2D scattering length $a_{\text{2D}}$ and the effective-range parameter $r_e$ for the double-Gaussian potential.
(a) Potential shape for a single Gaussian ($U_1=0$, blue dashed) and the double Gaussian with a potential barrier ($U_1=25.8\hbar\omega$, red solid), both with $U_0=104\hbar\omega$ and $\sigma=0.1a_{\mathrm{ho}}$.
(b) Effective-range parameter $r_e$ for a single-Gaussian potential ($U_1=0$) as a function of $U_0$, where $r_e$ remains strictly positive regardless of the value of $U_0$.
(c) Effective-range parameter $r_e$ (left axis, red solid) and $\ln (a_\text{2D}/a_{\text{ho}})$ (right axis, blue dashed) for the double Gaussian as a function of $U_1$ with fixed $U_0=104\hbar\omega$.
As $U_1$ increases, $r_e$ decreases and becomes negative for $U_1\gtrsim 15\hbar\omega$.
The red circle and the blue square mark the values of $r_e$ and $a_{\text{2D}}$ used in this work.}
\label{pra:fig:pseudopotential}
\end{figure*}

We consider $N$ fermions with mass $M$ ($N/2$ spin-up, $N/2$ spin-down) in an anisotropic two-dimensional harmonic trap with frequencies $\omega_x$ and $\omega_y$, interacting via a short-range potential $V$ between opposite spins.
In laboratory coordinates, the total Hamiltonian can be written in the standard single-particle form
\begin{equation}
\begin{aligned}
    \hat{H}={}&
    \sum_{k=1}^{N}\left[
    \frac{\hat{\mathbf{p}}^2_k}{2M}
    +\frac{M}{2}\left(\omega_x^2 (r^x_k)^2+\omega_y^2 (r^y_k)^2\right)
    \right]\\
    &+\sum_{k=1}^{N/2}\sum_{l=N/2+1}^{N}V(|\mathbf{r}_k-\mathbf{r}_l|).
\end{aligned}
\label{pra:eq:Hamiltonian_total}
\end{equation}
The trap term in Eq.~(\ref{pra:eq:Hamiltonian_total}) is the ordinary one-body potential
$U_{\mathrm{trap}}(\mathbf{r})=M(\omega_x^2 x^2+\omega_y^2 y^2)/2$.
Separating center-of-mass and relative motion gives $\hat{H}=\hat{H}_{\mathrm{CM}}+\hat{H}_\text{rel}$, with
\begin{equation}
    \hat{H}_{\mathrm{CM}}=\hat{T}_\mathrm{CM}
    +\frac{NM}{2}\left[\omega_x^2(r^x_\mathrm{CM})^2+\omega_y^2(r^y_\mathrm{CM})^2\right],
\end{equation}
and the relative Hamiltonian reads
\begin{equation}
\begin{aligned}
    \hat{H}_\text{rel}&=\sum_{k=1}^{N}\frac{\hat{\mathbf{p}}^2_k}{2M}-\hat{T}_\text{CM}+\sum_{k=1}^{N/2}\sum_{l=N/2+1}^{N}V(|\mathbf{r}_k-\mathbf{r}_l|)\\
    &+\frac{M}{4N}\sum_{k=1}^N\sum_{l=1}^N\left[\omega_x^2(r^x_k-r^x_l)^2+\omega_y^2(r^y_k-r^y_l)^2\right],
\end{aligned}
\label{pra:eq:Hamiltonian}
\end{equation}
where $\hat{\mathbf{p}}_k=-i\hbar\nabla_{\mathbf{r}_k}$, $\mathbf{r}_\mathrm{CM}=N^{-1}\sum_{k=1}^N\mathbf{r}_k$, the center-of-mass kinetic energy is $\hat{T}_\text{CM}=(\sum_{k=1}^N\hat{\mathbf{p}}_k)^2/2NM$,
and the indices $k=1$ to $N/2$ ($N/2+1$ to $N$) label spin-up (spin-down) particles.
The pairwise coordinate form in the last line of Eq.~(\ref{pra:eq:Hamiltonian}) is therefore simply the trap potential with the center-of-mass contribution removed.

The two-dimensional scattering is characterized by the scattering length $a_{\text{2D}}$ and the effective-range parameter $r_e$~\cite{adhikari1986quantum,galea2017fermions,schonenberg2017effectiverange}:
\begin{equation}
\cot[\delta(k)]=\frac{2}{\pi}\ln\left(k a_{\text{2D}}\right)+\frac{k^2r_e}{2}+ \cdots.
\end{equation}
We define the harmonic-oscillator length $a_{\mathrm{ho}}=\sqrt{\hbar/(M\omega)}$ with the geometric mean trap frequency $\omega=\sqrt{\omega_x\omega_y}$.
Guided by the experimental system realized with fermionic $^6\mathrm{Li}$ atoms, for which $M=m_{^6\mathrm{Li}}\simeq6.015\,\mathrm{u}$ with $\mathrm{u}$ the atomic mass unit, we use 
$\omega/(2\pi)\simeq2.08\,\mathrm{kHz}$ and set $\ln(a_{\text{2D}}k_F)= 1.15$ and $r_e=-0.2a_{\mathrm{ho}}^2$~\cite{petrov2001interatomic,hu2019reduced,yang2024two,brandstetter2025emergent}.
The remaining trap and Fermi scales are specified by $k_F=\sqrt{2ME_F/\hbar^2}$, $\omega_x=0.615\omega$, and $\omega_y=1.63\omega$, placing the system on the strongly interacting BCS side~\cite{ries2015observation,galea2017fermions,murthy2018hightemperature,sobirey2021observation}.
The Fermi energy $E_F$ is defined as the highest single-particle energy occupied in the non-interacting ground state of the anisotropic trap.
Throughout the following discussion, 
lengths are reported in units of $a_{\mathrm{ho}}$ and momenta in units of $\hbar/a_{\mathrm{ho}}$.

We model the pairwise two-body interaction between particles with opposite spins using a double-Gaussian potential (particles with the same spin do not interact) via
\begin{equation}
    V(r)=-U_0\exp\left(-\frac{r^2}{2\sigma^2}\right)+U_1\exp\left[-\frac{r^2}{2(2\sigma)^2}\right],
\label{pra:eq:pseudopotential}
\end{equation}
with $\sigma = 0.1a_{\mathrm{ho}}$
and $r$ denoting the relative distance between the two particles.
We choose $U_0>0$ and $U_1>0$.
The first term provides a potential well of depth $U_0$ and range $\sigma$, while the second term adds a potential barrier of height $U_1$ and a larger range $2\sigma$ [Fig.~\ref{pra:fig:pseudopotential}(a)].

For a given choice of $(U_0,U_1)$, the scattering length $a_{\text{2D}}$ and effective-range parameter $r_e$ are computed directly from the zero-energy two-body problem. 
In particular, 
we numerically integrate the $s$-wave radial Schr\"odinger equation at zero energy,
\begin{equation}
    \psi_{\mathrm{sc}}''(r)+\frac{1}{r}\psi_{\mathrm{sc}}'(r)-\frac{2M_r}{\hbar^2}V(r)\,\psi_{\mathrm{sc}}(r)=0,
\label{pra:eq:zero_energy_radial}
\end{equation}
with reduced mass $M_r=M/2$, starting from a regular initial condition at a small inner radius $r_0\ll\sigma$ and evolving outward to a matching radius $r_{\max}\gg\sigma$ where the potential is negligible.
In the asymptotic region, the solution takes the universal logarithmic form $\psi_{\mathrm{sc}}(r)\simeq A\bigl[\ln(r/a_{\text{2D}})\bigr]$, from which $A=r\,\psi_{\mathrm{sc}}'(r)|_{r_{\max}}$ and $a_{\text{2D}}=r_{\max}\exp[-\psi_{\mathrm{sc}}(r_{\max})/A]$ are read off.
The effective range is then obtained from the integral identity~\cite{yin2020fewbody,adhikari1986quantum,galea2017fermions,schonenberg2017effectiverange}
\begin{equation}
    r_e=2\int_0^{\infty}\!\!\left[\ln^2(r/a_{\text{2D}})-u^2(r)\right] r\,\mathrm{d}r,
\label{pra:eq:reff_integral}
\end{equation}
with $u(r)=\psi_{\mathrm{sc}}(r)/A$, evaluated numerically over $[r_0,r_{\max}]$.
This procedure yields $a_{\text{2D}}$ and $r_e$ to high precision for any $(U_0,U_1)$ and underlies the curves shown in Figs.~\ref{pra:fig:pseudopotential}(b) and \ref{pra:fig:pseudopotential}(c).

The barrier-like structure of the double-Gaussian potential is essential for reproducing the experimentally relevant \textit{negative} effective range.
As shown in Fig.~\ref{pra:fig:pseudopotential}(b), 
a single-Gaussian potential ($U_1=0$) yields only positive $r_e$
when $U_0$ is swept over a wide range.
Therefore, a potential well with a monotonically increasing depth without a barrier generically gives $r_e > 0$ in 2D.
Adding the second repulsive Gaussian potential introduces a barrier that modifies the low-energy scattering phase shift, allowing $r_e$ to be negative.
As demonstrated in Fig.~\ref{pra:fig:pseudopotential}(c), as $U_1$ is increased while $U_0=104\hbar\omega$ is fixed,
$r_e$
decreases monotonically,
crossing zero at $U_1\approx 15\hbar\omega$, while $\ln(a_\text{2D}/a_{\text{ho}})$ increases.
The experimental values of $r_e \simeq -0.2a_{\mathrm{ho}}^2$ and $\ln(a_\text{2D}k_F)\simeq 1.15$ are simultaneously matched at $U_1\simeq 25.8\hbar\omega$ [markers in Fig.~\ref{pra:fig:pseudopotential}(c)].

The double-Gaussian form thus provides two free parameters ($U_0$, $U_1$) with which to match both $a_\text{2D}$ and $r_e$ simultaneously.
Since the Fermi energy $E_F$ (and hence $k_F$) changes with system size, the double-Gaussian potential parameters are adjusted for each particle number:
\begin{itemize}
    \item $N=2$: $U_0=85.9\hbar\omega$, $U_1=19.2\hbar\omega$,
    \item $N=4$: $U_0=95.9\hbar\omega$, $U_1=22.7\hbar\omega$,
    \item $N=6$: $U_0=104\hbar\omega$, $U_1=25.8\hbar\omega$.
\end{itemize}
All parameter sets yield $\ln(a_{\text{2D}}k_F)\simeq 1.15$ and $r_e\simeq -0.2a_{\mathrm{ho}}^2$.

The anisotropic-trap parameter sets listed above are used for every quantitative comparison with experiment in this paper, including the convergence benchmarks of Sec.~\ref{pra:sec:numerical} and the aspect-ratio, OTOC, and reduced-density-matrix observables presented throughout Sec.~\ref{pra:sec:postprocessing}.
For a small subset of the post-processing demonstrations---specifically the coordinate-space density snapshot of Fig.~\ref{pra:fig:density} and the relative-momentum correlator maps and line cuts of Figs.~\ref{pra:fig:relative_correlator} and \ref{pra:fig:relative_correlator_linecuts}---we instead simulate the same $N{+}N$ systems in an \textit{isotropic} two-dimensional harmonic trap with $\omega_x=\omega_y=\omega$.

Because the non-interacting Fermi energy depends on the trap shape, the double-Gaussian parameters $(U_0,U_1)$ that realize the same dimensionless interaction strength $\ln(a_{\text{2D}}k_F)\simeq 1.15$ at fixed $r_e\simeq -0.2a_{\mathrm{ho}}^2$ also shift slightly in the isotropic geometry.
Consequently, the isotropic-trap simulations require only two distinct parameter sets:
\begin{itemize}
    \item $N=2$: $U_0=83.6\hbar\omega$, $U_1=18.4\hbar\omega$,
    \item $N=4$ and $N=6$: $U_0=99.7\hbar\omega$, $U_1=24.1\hbar\omega$.
\end{itemize}
The shifts relative to the anisotropic values reflect the modest change in $k_F$ between the two geometries; all other aspects of the calculation are identical to those in the anisotropic case.

\section{Time-dependent explicitly correlated Gaussian framework}
\label{pra:sec:ecg_ansatz}

\subsection{H-tree Jacobi coordinate transformation}
\label{pra:ssec:htree}

We employ the H-tree Jacobi coordinate transformation~\cite{yin2015thesis} to convert the Cartesian coordinates of $N$ particles into $N-1$ relative coordinates that naturally capture correlations between particles
with different spin states.
The Hamiltonian and permutation sums use the block spin ordering introduced in Sec.~\ref{pra:sec:model}, with particles $1,\ldots,N/2$ and $N/2+1,\ldots,N$ belonging to the two spin sectors.
For the H-tree construction it is convenient to relabel these same particles into an interleaved auxiliary ordering in which odd and even labels denote opposite spins.
Starting with the interleaved coordinates $L^{(0)}_{\mathrm{cart}}\equiv\{\mathbf{r}_1,\mathbf{r}_2,\ldots,\mathbf{r}_N\}$
(odd and even indices denote opposite spins in this auxiliary H-tree ordering, with all particles having the same mass) and
two empty lists $L_{\text{rel}}=\{\}$ and $L_{\text{com}}=\{\}$, 
we construct Jacobi coordinates using the hierarchical clustering algorithm described below to recursively fill the lists $L_{\text{rel}}$ and $L_{\text{com}}$.
At the end of this algorithm, 
the $L_{\text{rel}}$ and $L_{\text{com}}$
contain all the Jacobi vectors and the center-of-mass vector. 

We first define an operation $J[\mathbf{r}_k,\mathbf{r}_l]$ applied to a list $L$ that
(i) appends $\mathbf{r}_k-\mathbf{r}_l$ to the relative coordinate list $L_{\mathrm{rel}}$, and (ii) returns the center-of-mass coordinate $({m_k\mathbf{r}_k + m_l \mathbf{r}_l})/({m_k+m_l})$, where $m_k$ and $m_l$ are the total masses of the respective clusters.
The H-tree transformation proceeds as follows:

\begin{enumerate}
\item Apply $J[\mathbf{r}_{2k-1},\mathbf{r}_{2k}]$ for all valid $k$ in $L^{(0)}_{\mathrm{cart}}$, generating $L^{(1)}_{\mathrm{cart}}$ and populating $L_{\mathrm{rel}}$.
For $N=6$:
\begin{equation}
\begin{aligned}
    &\text{\textbf{Step 1:}}\\
    &L^{(1)}_{\mathrm{cart}}=\left\{\frac{\mathbf{r}_1+\mathbf{r}_2}{2},\frac{\mathbf{r}_3+\mathbf{r}_4}{2},\frac{\mathbf{r}_5+\mathbf{r}_6}{2}\right\},\\
    &L_{\mathrm{rel}}=\{\mathbf{r}_1-\mathbf{r}_2,\mathbf{r}_3-\mathbf{r}_4,\mathbf{r}_5-\mathbf{r}_6\}.
\end{aligned}
\end{equation}

\item Recursively apply $J$ to pairs in $L^{(n)}_{\mathrm{cart}}$ to form $L^{(n+1)}_{\mathrm{cart}}$.
When the list has an odd number of elements, move the last element to the list $L_{\mathrm{com}}$.
Continue until $L^{(n)}_{\mathrm{cart}}$ is empty.
For $N=6$:
\begin{equation}
\begin{aligned}
    &\text{\textbf{Step 2:}}\\
    &L^{(2)}_{\mathrm{cart}}=\left\{\frac{\mathbf{r}_1+\mathbf{r}_2+\mathbf{r}_3+\mathbf{r}_4}{4}\right\},\\
    &L_{\mathrm{rel}}\leftarrow L_{\mathrm{rel}}\cup\left\{\frac{\mathbf{r}_1+\mathbf{r}_2}{2}-\frac{\mathbf{r}_3+\mathbf{r}_4}{2}\right\},\\
    &L_{\mathrm{com}}=\left\{\frac{\mathbf{r}_5+\mathbf{r}_6}{2}\right\}.\\[6pt]
    &\text{\textbf{Step 3:}}\\
    &L^{(3)}_{\mathrm{cart}}=\left\{\right\},\\
    &L_{\mathrm{com}}\leftarrow L_{\mathrm{com}}\cup\left\{\frac{\mathbf{r}_1+\mathbf{r}_2+\mathbf{r}_3+\mathbf{r}_4}{4}\right\}.
\end{aligned}
\end{equation}

\item For $L_{\mathrm{com}}=\{\mathbf{R}_1,\mathbf{R}_2,\ldots,\mathbf{R}_{N_{\mathrm{cl}}}\}$ (ordered by ascending cluster mass), sequentially apply $J$ from tail to head: $J(\mathbf{R}_{N_{\mathrm{cl}}-1},\mathbf{R}_{N_{\mathrm{cl}}})$, then $J(\mathbf{R}_{N_{\mathrm{cl}}-2},J(\mathbf{R}_{N_{\mathrm{cl}}-1},\mathbf{R}_{N_{\mathrm{cl}}}))$, etc.
The last coordinate always represents the total center-of-mass.
For $N=6$:
\begin{equation}
\begin{aligned}
    \text{\textbf{Step 4:}}\quad
    L_{\mathrm{rel}}&\leftarrow L_{\mathrm{rel}} \cup\left\{\frac{\mathbf{r}_5+\mathbf{r}_6}{2}-\frac{\mathbf{r}_1+\mathbf{r}_2+\mathbf{r}_3+\mathbf{r}_4}{4}\right\},\\
    L_{\mathrm{com}}&=\left\{\frac{\mathbf{r}_1+\mathbf{r}_2+\mathbf{r}_3+\mathbf{r}_4+\mathbf{r}_5+\mathbf{r}_6}{6}\right\}.
\end{aligned}
\end{equation}
\end{enumerate}

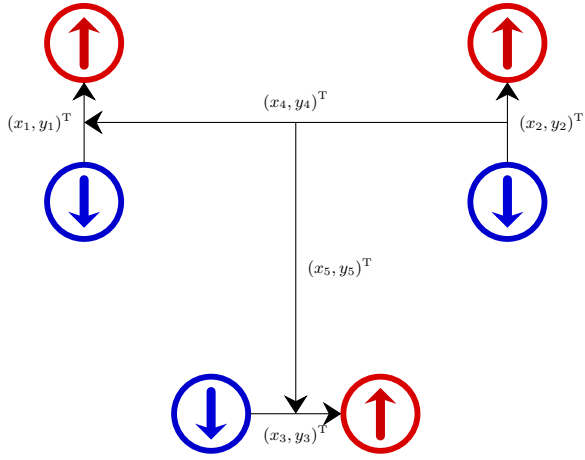
\begin{figure}[tbp]
\centering
\begin{tikzpicture}[scale=0.7, every node/.style={transform shape}, >=Stealth, line cap=round, line join=round]
\tikzset{
site/.style={circle, draw, line width=2.2pt, minimum size=14mm, inner sep=0pt},
arrowin/.style={line width=3.4pt, -{Stealth[length=6pt,width=12pt]}},
spinup/.style={site, draw=red!85!black,
path picture={\draw[arrowin, red!80!black]
([yshift=8pt]path picture bounding box.south)
-- ([yshift=-6pt]path picture bounding box.north);}},
spindown/.style={site, draw=blue!80!black,
path picture={\draw[arrowin, blue!85!black]
([yshift=-8pt]path picture bounding box.north)
-- ([yshift=6pt]path picture bounding box.south);}},
}
\node[spinup]   (uL) at (-4, 4) {};
\node[spindown] (dL) at (-4, 1) {};
\node[spinup]   (uR) at ( 4, 4) {};
\node[spindown] (dR) at ( 4, 1) {};
\node[spindown] (bL) at (-1.6, -3) {};
\node[spinup]   (bR) at ( 1.6, -3) {};
\coordinate (tL) at ($(dL.north)!0.50!(uL.south)$);
\coordinate (tR) at ($(dR.north)!0.50!(uR.south)$);
\coordinate (tC) at ($(tL)!0.5!(tR)$);
\coordinate (bC) at ($(bL.east)!0.5!(bR.west)$);
\draw[-{Stealth[length=7pt,width=9pt]}] (dL.north) -- (uL.south)
    node[midway, left=3pt] {$(x_1,y_1)^\mathrm{T}$};
\draw[-{Stealth[length=7pt,width=9pt]}] (dR.north) -- (uR.south)
    node[midway, right=3pt] {$(x_2,y_2)^\mathrm{T}$};
\draw[-{Stealth[length=7pt,width=9pt]}] (bL.east) -- (bR.west)
    node[midway, below=3pt] {$(x_3,y_3)^\mathrm{T}$};
\draw[-{Stealth[length=7pt,width=9pt]}] (tR) -- (tL)
    node[midway, above=3pt] {$(x_4,y_4)^\mathrm{T}$};
\draw[-{Stealth[length=7pt,width=9pt]}] (tC) -- (bC)
    node[midway, right=3pt] {$(x_5,y_5)^\mathrm{T}$};
\end{tikzpicture}
\caption{H-tree Jacobi coordinate transformation for $N=3+3$ particles.
Red (blue) circles denote spin-up (spin-down) fermions.
Arrows indicate the relative coordinates $(x_m,y_m)^\mathrm{T}$.}
\label{pra:fig:htree}
\end{figure}

The final output $L_{\mathrm{rel}}$ contains all $N-1$ relative Jacobi coordinates and $L_{\mathrm{com}}$ contains the total center-of-mass coordinate.
The transformation defines a full $N\times N$ matrix $\widetilde U$ such that its first $N-1$ rows generate the relative coordinates, $x_m=\sum_k \widetilde U_{mk}r^x_k$ and $y_m=\sum_k \widetilde U_{mk}r^y_k$, and its final row generates the center-of-mass coordinate.
In the formulas below, $U$ denotes the relative-coordinate block formed by the first $N-1$ rows of $\widetilde U$.
For the $N=3+3$ system, the transformation is illustrated in Fig.~\ref{pra:fig:htree}.
Compared with the more widely used K-tree Jacobi coordinates~\cite{suzuki1998stochastic,mitroy2013theory}, which typically 
define the $n$-th Jacobi vector by the relative coordinates between the $(n+1)$-th particle and the center-of-mass position of the other $n$ particles,
the H-tree construction groups particles with opposite spin states into pairs at the lowest level of the hierarchy, with all remaining Jacobi vectors defined as the relative coordinates between the centers of these pairs. 
For interacting two-component fermions, 
this pairing-oriented structure provides a more natural coordinate system for describing dimer formation and pair fluctuations, and therefore tends to represent pairing physics more efficiently within the ECG framework~\cite{yin2015thesis}.

\subsection{Wavefunction ansatz}
\label{pra:ssec:ansatz}

The wavefunction $\Psi_{\text{rel}}$ in the relative coordinates follows a generalized ECG ansatz,
\begin{equation} \Psi_\text{rel}=\sum_{j=1}^{N_b}\hat{\mathcal{A}}\left[\bar{u}_j\exp\left(-\frac{\mathbf{x}^\mathrm{T}\bar{A}_j\mathbf{x}}{2}\right)\exp\left(-\frac{\mathbf{y}^\mathrm{T}\bar{B}_j\mathbf{y}}{2}\right)\right],
    \label{pra:eq:ECG_ansatz}
\end{equation}
where $\mathbf{x}=(x_1,\ldots,x_{N-1})^\mathrm{T}$ and $\mathbf{y}=(y_1,\ldots,y_{N-1})^\mathrm{T}$ are $(N-1)$-dimensional Jacobi coordinates (see Sec.~\ref{pra:ssec:htree}), the operator $\hat{\mathcal{A}}$ fully antisymmetrizes the wavefunction for fermionic statistics, and $N_b$ is the number of basis functions.
The complex prefactors $\bar{u}_j$ and the $(N-1)\times(N-1)$ symmetric matrices $\bar{A}_j$ and $\bar{B}_j$ encode the single-particle distributions and the inter-particle correlations in the $j$-th basis function.
Following the convention of Refs.~\cite{hiyama2003gaussian,bubin2007relativistic}, we allow these parameters to be complex-valued, which significantly improves basis efficiency compared to real-valued ECG, particularly for scattering and time-dependent problems.

The total number $N_p$ of independent variational parameters is
\begin{equation}
    N_p = N_b\left[1 + N(N-1)\right],
\end{equation}
where the ``1'' and the $N(N-1)$ account for the parameter $u_j$ and the parameters in the symmetric matrices $A_j$ and $B_j$, respectively.
These parameters collectively form the variational vector $\bar{\mathbf{z}} = (\bar{u}_1,\ldots,\bar{u}_{N_b}, (\bar{A}_1)_{11},\ldots, (\bar{B}_1)_{11},\ldots)$.
We adopt Wirtinger coordinates, in which the ket is holomorphic in $\bar{\mathbf{z}}$ and the bra in $\mathbf{z}$; $\mathbf{z}$ and $\bar{\mathbf{z}}$ are varied as independent coordinates and are related by complex conjugation on physical trajectories.

\subsection{Permutation matrices and antisymmetrization}
\label{pra:ssec:permutation}

Using Cauchy's two-line notation for a size-$N$ permutation $P$,
\begin{equation}
    P\equiv\begin{pmatrix}
    1&2&3&\cdots&N\\
    p_1&p_2&p_3&\cdots&p_N
\end{pmatrix},
\end{equation}
the antisymmetrizer acts as
\begin{equation}
    \mathcal{A}[f(\mathbf{x},\mathbf{y})]=\sum_{P}(-1)^Pf(T_P\mathbf{x},T_P\mathbf{y}),
\end{equation}
where $P$ runs over permutations within each spin sector, i.e., $P\in S_{N/2}^{(\uparrow)}\times S_{N/2}^{(\downarrow)}$.
The permutation matrix $T_{P}$ in Jacobi coordinates is
\begin{equation}
(T_P)_{mn}=\sum_{k=1}^N U_{mk}(\widetilde U^{-1})_{p_kn},\qquad m,n=1,\ldots,N-1.
\end{equation}
As described in Sec.~\ref{pra:ssec:htree},
the matrix $U$ 
denotes the relative-coordinate block of the full H-tree transformation $\widetilde U$.

\subsection{Convention of index labels}
\label{pra:ssec:indices}

Different levels of indexing appear in the analytical matrix-element formulas below.
To avoid confusion there, we adopt the conventions in Table~\ref{pra:tab:indices},
where 
the flattened index $\alpha$ maps to a quadruplet $(\alpha_0, \alpha_1, \alpha_2, \alpha_3)$ via
\begin{equation}
z_\alpha\equiv z_{\alpha_0, \alpha_1, \alpha_2, \alpha_3}\equiv \begin{cases}
u_{\alpha_1} & \text{if }\alpha_0=0\\
(A_{\alpha_1})_{\alpha_2,\alpha_3}& \text{if }\alpha_0=1\\
(B_{\alpha_1})_{\alpha_2,\alpha_3}& \text{if }\alpha_0=2
\end{cases},
\end{equation}
where $\alpha_2$ and $\alpha_3$ are undefined when $\alpha_0=0$.
For matrix parameters only the independent symmetric entries are included in $\mathbf{z}$, so $\alpha_2\leq\alpha_3$ and $\beta_2\leq\beta_3$.

\begin{table}[h]
\renewcommand{\arraystretch}{1.5}
\begin{tabular}{|l|l|l|}
\hline
Index label    & Range        & Explanation                                           \\ \hline
$\alpha,\beta$ & $1$ to $N_p$ & Flattened index of $\mathbf{z}$, $\mathcal{C}$, and $\frac{\partial \mathcal{H}}{\partial \mathbf{z}}$      \\ \hline
$i,j,\alpha_1,\beta_1$          & $1$ to $N_b$ & ECG basis index \\ \hline
$k,l$          & $1$ to $N$   & Particle index                          \\ \hline
$m,n,\alpha_2,\alpha_3,\beta_2,\beta_3$          & $1$ to $N-1$ & Relative coordinate index                         \\ \hline
\end{tabular}
\caption{Convention of index labels used in the analytical matrix-element formulas.}
\label{pra:tab:indices}
\end{table}

\subsection{Matrix elements of $\mathcal{C}$}
\label{pra:ssec:C_matrix}

To represent the expressions compactly, we define the following shorthand notation,
\begin{align}
    &D^O_{P,i j}\equiv\det(O_{i}+T_P^\mathrm{T}\bar{O}_{j}T_P)^{1/2},\label{pra:eq:DO}\\
    &M^{O,1}_{P,i j}=(O_{i}+T_P^\mathrm{T}\bar{O}_jT_P)^{-1},\label{pra:eq:MO1}\\
    &M^{O,2}_{P,i j}=T_P(O_{i}+T_P^\mathrm{T}\bar{O}_jT_P)^{-1}T_P^\mathrm{T},\label{pra:eq:MO2}\\
    &M^{O,3}_{P,i j}=T_P(O_{i}+T_P^\mathrm{T}\bar{O}_jT_P)^{-1},\label{pra:eq:MO3}
\end{align}
where $O$ stands for either $A$ or $B$.
The common prefactor is
\begin{equation}
\mathcal{N}_{P,ij} \equiv \frac{(\pm)^{P} u_i \bar{u}_j\, (2\pi)^{N-1}}{D^A_{P,ij}\,D^B_{P,ij}},
    \label{pra:eq:calN}
\end{equation}
where $(\pm)^P = (-1)^P$.
The symmetric-matrix derivative factor is
\begin{equation}
\chi_\alpha\equiv-\frac{1}{1+\delta_{\alpha_2,\alpha_3}},
\end{equation}
where $\chi_\alpha=-1$ for $\alpha_2\neq\alpha_3$ and $\chi_\alpha=-1/2$ for $\alpha_2=\alpha_3$.

The matrix elements of the matrix $\mathcal{C}$ are calculated via
\begin{equation}
\mathcal{C}_{\alpha \beta}=\bar{\mathcal{C}}_{\beta \alpha}=\frac{1}{\langle\Psi|\Psi\rangle}\frac{\partial^2 \langle\Psi|\Psi\rangle}{\partial z_\alpha \partial \bar{z}_\beta}-\frac{1}{\langle\Psi|\Psi\rangle^2}\frac{\partial \langle\Psi|\Psi\rangle}{\partial z_\alpha}\frac{\partial \langle\Psi|\Psi\rangle}{\partial \bar{z}_\beta},
\end{equation}
where the norm $\langle\Psi|\Psi\rangle$ together with its first and second derivatives with respect to the parameters $z_\alpha$ and $\bar{z}_{\beta}$ can be expressed using the quantities defined in Eqs.~\eqref{pra:eq:DO}--\eqref{pra:eq:calN}.
The norm of the wavefunction is
\begin{equation}
    \langle\Psi|\Psi\rangle=\sum_{P,i,j}\mathcal{N}_{P,ij}.\label{pra:eq:norm}
\end{equation}
The first-order derivatives are
\begin{equation}
\frac{\partial\langle\Psi|\Psi\rangle}{\partial z_\alpha}=\begin{cases}
\sum_{P,j}\mathcal{N}_{P,\alpha_1 j}/u_{\alpha_1} & \alpha_0=0 \\[8pt]
\chi_\alpha\sum_{P,j}\mathcal{N}_{P,\alpha_1 j}\,
  (M^{A,1}_{P,\alpha_1 j})_{\alpha_3\alpha_2} & \alpha_0=1\\[8pt]
\chi_\alpha\sum_{P,j}\mathcal{N}_{P,\alpha_1 j}\,
  (M^{B,1}_{P,\alpha_1 j})_{\alpha_3\alpha_2} & \alpha_0=2
\end{cases},
\label{pra:eq:dnormdz1}
\end{equation}
and
\begin{equation}
\frac{\partial\langle\Psi|\Psi\rangle}{\partial \bar{z}_\beta}=\begin{cases}
\sum_{P,i}\mathcal{N}_{P,i\beta_1}/\bar{u}_{\beta_1} & \beta_0=0 \\[8pt]
\chi_\beta\sum_{P,i}\mathcal{N}_{P,i\beta_1}\,
  (M^{A,2}_{P,i\beta_1})_{\beta_3\beta_2} & \beta_0=1\\[8pt]
\chi_\beta\sum_{P,i}\mathcal{N}_{P,i\beta_1}\,
  (M^{B,2}_{P,i\beta_1})_{\beta_3\beta_2} & \beta_0=2
\end{cases}.
\label{pra:eq:dnormdz2}
\end{equation}
The second-order derivatives take several different forms depending on the values of $\alpha_0$ and $\beta_0$.
To make the expressions more compact, in the formulas below we write $(M^{O,s})_P\equiv M^{O,s}_{P,\alpha_1\beta_1}$ whenever the two basis indices coincide with the fixed values $\alpha_1$ and $\beta_1$.
For $\beta_0=0$, $\alpha_0=0$:
\begin{equation}
    \frac{\partial^2 \langle\Psi|\Psi\rangle}{\partial z_\alpha \partial \bar{z}_\beta}=\sum_P\frac{\mathcal{N}_{P,\alpha_1\beta_1}}{u_{\alpha_1}\bar{u}_{\beta_1}}.
\end{equation}
For $\beta_0=1$, $\alpha_0=0$:
\begin{equation}
    \frac{\partial^2 \langle\Psi|\Psi\rangle}{\partial z_\alpha \partial \bar{z}_\beta}= \chi_\beta\sum_P\frac{\mathcal{N}_{P,\alpha_1\beta_1}}{u_{\alpha_1}}\,(M^{A,2}_{P})_{\beta_3\beta_2}.
\end{equation}
For $\beta_0=2$, $\alpha_0=0$: the same with $A\to B$.
For $\beta_0=2$, $\alpha_0=1$:
\begin{equation}
    \frac{\partial^2 \langle\Psi|\Psi\rangle}{\partial z_\alpha \partial \bar{z}_\beta}= \chi_\alpha \chi_\beta\sum_P\mathcal{N}_{P,\alpha_1\beta_1}\,(M^{A,1}_{P})_{\alpha_3\alpha_2}\,(M^{B,2}_{P})_{\beta_3\beta_2}.
\end{equation}
For $\beta_0=\alpha_0=1$ (both acting on $A$):
\begin{widetext}
\begin{equation}
\begin{aligned}
    \frac{\partial^2 \langle\Psi|\Psi\rangle}{\partial z_\alpha \partial \bar{z}_\beta}\bigg|_{\alpha_0=\beta_0=1}\!\!=& \sum_P\mathcal{N}_{P,\alpha_1\beta_1} \bigg[\chi_\alpha \chi_\beta\,(M^{A,1}_{P})_{\alpha_3\alpha_2}\,(M^{A,2}_{P})_{\beta_3\beta_2}
    -\chi_\alpha\Big[(M^{A,3}_{P})_{\beta_2\alpha_2}(M^{A,3}_{P})_{\beta_3\alpha_3}\\
    &+(M^{A,3}_{P})_{\beta_3\alpha_2}(M^{A,3}_{P})_{\beta_2\alpha_3}-\delta_{\beta_2\beta_3}(M^{A,3}_{P})_{\beta_2\alpha_2}(M^{A,3}_{P})_{\beta_2\alpha_3}\Big]\bigg].
\end{aligned}
\end{equation}
\end{widetext}
The case $\beta_0=\alpha_0=2$ is obtained from the equation above by replacing every $A$ with $B$ in the matrix labels (i.e., $M^{A,1}_P\to M^{B,1}_P$, $M^{A,2}_P\to M^{B,2}_P$, and $M^{A,3}_P\to M^{B,3}_P$), since the $\mathbf{x}$ and $\mathbf{y}$ sectors of the Gaussian basis factorize and play symmetric roles in $\langle\Psi|\Psi\rangle$.
All remaining sector combinations (e.g., $\alpha_0{=}1$, $\beta_0{=}0$ or $\alpha_0{=}2$, $\beta_0{=}1$) are determined by the Hermiticity relation $\mathcal{C}_{\alpha\beta}=\bar{\mathcal{C}}_{\beta\alpha}$: the formula for sector labels $(\alpha_0,\beta_0)$ is obtained from the formula for $(\beta_0,\alpha_0)$ by taking the complex conjugate and exchanging $\alpha\leftrightarrow\beta$.

\subsection{Hamiltonian gradient $\partial \mathcal{H}/\partial \mathbf{z}$}
\label{pra:ssec:dHdz}

The elements of the gradient vector are
\begin{equation}
\frac{\partial\mathcal{H}}{\partial {z}_\alpha}=\frac{1}{\langle\Psi|\Psi\rangle}\frac{\partial \langle\Psi|\hat{H}|\Psi\rangle}{\partial {z}_\alpha}-\frac{\langle\Psi|\hat{H}|\Psi\rangle}{\langle\Psi|\Psi\rangle^2}\frac{\partial \langle\Psi|\Psi\rangle}{\partial {z}_\alpha}.
\end{equation}
Since $\langle\Psi|\Psi\rangle$ and its derivatives are derived in Eqs.~(\ref{pra:eq:norm})--(\ref{pra:eq:dnormdz2}), we focus on the Hamiltonian expectation value $\langle\Psi|\hat{H}|\Psi\rangle$ and its derivative with respect to $z_{\alpha}$.
The Hamiltonian [Eq.~(\ref{pra:eq:Hamiltonian})] is  $\hat{H}=\hat{H}_\text{kin}+\hat{H}_\text{trap}+\hat{H}_\text{int}$, where the kinetic energy term,
the harmonic trap potential, 
and the two-body interaction term are 
\begin{equation}
\begin{aligned}
\hat{H}_\text{kin}&=\sum_{k=1}^{N}\frac{\hat{\mathbf{p}}^2_k}{2M}-\hat{T}_\text{CM},\;\hat{H}_\text{int}=\sum_{k=1}^{N/2}\sum_{l=N/2+1}^{N}V(|\mathbf{r}_k-\mathbf{r}_l|).\\
\hat{H}_\text{trap}&=\frac{M}{4N}\sum_{k=1}^N\sum_{l=1}^N\left[\omega_x^2(r^x_k-r^x_l)^2+\omega_y^2(r^y_k-r^y_l)^2\right].
\end{aligned}
\label{pra:eq:H_decomposition}
\end{equation}
Next, we present the analytical matrix elements for each of these terms, expressed using the quantities defined in Eqs.~\eqref{pra:eq:DO}--\eqref{pra:eq:calN}.

\subsubsection{Kinetic energy term}

The kinetic energy expectation value is
\begin{equation}
    \langle\Psi|\hat{H}_{\text{kin}}|\Psi\rangle=\frac{\hbar^2}{2M}\sum_{P,i,j} \mathcal{N}_{P,ij}\,(R^A_{P,ij}+R^B_{P,ij})
\end{equation}
with 
\begin{equation}
\label{pra:eq:ROP_ij}
R^O_{P,ij}=\mathrm{Tr}(O_i(O_i+T_P^\mathrm{T}\bar{O}_jT_P)^{-1}T_P^\mathrm{T}\bar{O}_jT_P\Lambda),
\end{equation}
where we define
\begin{equation}
\label{pra:eq:Lambda}
\Lambda_{mn} = \sum_{k=1}^N U_{mk}U_{nk}.
\end{equation}
The derivatives of $\langle\Psi|\hat{H}_\text{kin}|\Psi\rangle$ with respect to $z_{\alpha}$, depending on the value of $\alpha_0$, are
\begin{widetext}
\begin{equation}
\frac{\partial\langle\Psi|\hat{H}_\text{kin}|\Psi\rangle}{\partial {z}_\alpha}=\frac{\hbar^2}{2M}\begin{cases}
    \sum_{P,j} \dfrac{\mathcal{N}_{P,\alpha_1 j}}{u_{\alpha_1}}\,R^\pm_{P,\alpha_1 j}& \alpha_0 = 0\\[8pt]
\chi_\alpha\sum_{P,j}\mathcal{N}_{P,\alpha_1 j}
\left[R^\pm_{P,\alpha_1 j}\,M^{A,1}_{P,\alpha_1 j}
-2\,\mathcal{K}^A_{P,\alpha_1 j}\right]_{\alpha_3 \alpha_2} & \alpha_0 = 1\\[8pt]
\chi_\alpha\sum_{P,j}\mathcal{N}_{P,\alpha_1 j}
\left[R^\pm_{P,\alpha_1 j}\,M^{B,1}_{P,\alpha_1 j}
-2\,\mathcal{K}^B_{P,\alpha_1 j}\right]_{\alpha_3 \alpha_2} & \alpha_0 = 2
\end{cases},
\end{equation}
\end{widetext}
where we use the shorthand notation
\begin{align}
R^\pm_{P,ij}\equiv R^A_{P,ij}+R^B_{P,ij}\end{align}
and the kinetic-energy kernel is
\begin{equation}
\mathcal{K}^O_{P,ij}\equiv M^{O,1}_{P,ij}\,T_P^\mathrm{T}\bar{O}_jT_P\,\Lambda\, T_P^\mathrm{T}\bar{O}_jT_P\, M^{O,1}_{P,ij}.
\end{equation}
The expressions for $R^A_{P,ij}$, $R^B_{P,ij}$, and $\Lambda$ are given in Eqs.~\eqref{pra:eq:ROP_ij}--\eqref{pra:eq:Lambda}.

\subsubsection{Harmonic trap potential}

The expectation value of the harmonic trap term is
\begin{equation}
\langle\Psi|\hat{H}_\text{trap}|\Psi\rangle=\frac{M}{2}\sum_{P,i,j,k}\mathcal{N}_{P,ij}\left[\omega_x^2 U^A_{P,ijk}+\omega_y^2 U^B_{P,ijk}\right]
\end{equation}
with 
\begin{equation}
U^O_{P,ijk}=\operatorname{Tr}[(O_i+T_P^\mathrm{T}\bar{O}_jT_P)^{-1}\vec{\omega}^{k}[\vec{\omega}^{k}]^\mathrm{T}],
\end{equation}
where we define
the vector
\begin{equation}
[\vec{\omega}^k]^\mathrm{T} =  ((\widetilde U^{-1})_{k1},(\widetilde U^{-1})_{k2},\cdots,(\widetilde U^{-1})_{k(N-1)}).
\end{equation}
The derivatives of $\langle\Psi|\hat{H}_{\text{trap}}|\Psi\rangle$ with respect to $z_{\alpha}$, depending on the value of $\alpha_0$,
are
\begin{widetext}
\begin{equation}
\frac{\partial\langle\Psi|\hat{H}_\text{trap}|\Psi\rangle}{\partial {z}_\alpha}=\frac{M}{2}\begin{cases}
\sum_{P,j,k} \dfrac{\mathcal{N}_{P,\alpha_1 j}}{u_{\alpha_1}}\,\mathcal{U}^\pm_{P,\alpha_1jk}& \alpha_0 = 0\\[8pt]
\chi_\alpha\sum_{P,j,k}\mathcal{N}_{P,\alpha_1 j}
\left[\mathcal{U}^\pm_{P,\alpha_1jk}\,M^{A,1}_{P,\alpha_1 j}
+2\omega_x^2\,\mathcal{P}^{A}_{P,\alpha_1 jk}\right]_{\alpha_3 \alpha_2} & \alpha_0 = 1\\[8pt]
\chi_\alpha\sum_{P,j,k}\mathcal{N}_{P,\alpha_1 j}
\left[\mathcal{U}^\pm_{P,\alpha_1jk}\,M^{B,1}_{P,\alpha_1 j}
+2\omega_y^2\,\mathcal{P}^{B}_{P,\alpha_1 jk}\right]_{\alpha_3 \alpha_2} & \alpha_0 = 2
\end{cases},
\end{equation}
\end{widetext}
where we use the shorthand notation  
\begin{align}
\mathcal{U}^\pm_{P,ijk}\equiv\omega_x^2 U^A_{P,ijk}+\omega_y^2 U^B_{P,ijk}
\end{align}
and the potential-energy kernel is
\begin{align}
\mathcal{P}^{O}_{P,ijk}\equiv M^{O,1}_{P,ij}\,\vec{\omega}^{k}[\vec{\omega}^{k}]^\mathrm{T}\,M^{O,1}_{P,ij}.
\end{align}

\subsubsection{Two-body interaction term}

The expression for the interaction matrix element requires additional quantities beyond those in Eqs.~\eqref{pra:eq:DO}--\eqref{pra:eq:calN}.
We define the vector
\begin{equation}
[\vec{\omega}^{kl}]^\mathrm{T} \!=\!
\left((\widetilde U^{-1})_{k1}\!-\!(\widetilde U^{-1})_{l1},\,\ldots,\,(\widetilde U^{-1})_{k(N-1)}\!-\!(\widetilde U^{-1})_{l(N-1)}\right),
\end{equation}
and the interaction-modified quantities, similar to those in Eqs.~\eqref{pra:eq:DO}--\eqref{pra:eq:calN},  
\begin{align}
    \Sigma^O_{P,ijkl}(\sigma) &\equiv O_i+T_P^\mathrm{T}\bar{O}_jT_P+\vec{\omega}^{kl}[\vec{\omega}^{kl}]^\mathrm{T}\!/\sigma^2,\\
D^{O,\text{I}}_{P,ijkl}(\sigma) &\equiv [\det\Sigma^O_{P,ijkl}(\sigma)]^{1/2},\label{pra:eq:DI}\\
M^{O,\text{I}}_{P,ijkl}(\sigma) &\equiv [\Sigma^O_{P,ijkl}(\sigma)]^{-1}.\label{pra:eq:MI}
\end{align}
We further define the interaction prefactor
\begin{equation}
\label{pra:eq:I_Pijkl}
\mathcal{I}_{P,ijkl} \equiv \frac{-U_0}{D^{A,\text{I}}_{P,ijkl}(\sigma)\,D^{B,\text{I}}_{P,ijkl}(\sigma)}+\frac{U_1}{D^{A,\text{I}}_{P,ijkl}(2\sigma)\,D^{B,\text{I}}_{P,ijkl}(2\sigma)}.
\end{equation}
With these quantities, the expectation value of the two-body interaction energy reads
\begin{equation}
\langle\Psi|\hat{H}_\mathrm{int}|\Psi\rangle=\sum_{\substack{P,i,j\\k\leq N/2<l}}(\pm)^{P}u_i \bar{u}_j(2\pi)^{N-1}\,\mathcal{I}_{P,ijkl},
\end{equation}
where the index restriction $k\leq N/2<l$ enforces that $k$ runs over spin-up particles and $l$ over spin-down particles, consistent with the convention used in the Hamiltonian [Eq.~(\ref{pra:eq:Hamiltonian})].
The derivatives of $\langle\Psi|\hat{H}_{\text{int}}|\Psi\rangle$ with respect to $z_{\alpha}$, depending on the value of $\alpha_0$, are
\begin{equation}
\frac{\partial\langle\Psi|\hat{H}_\mathrm{int}|\Psi\rangle}{\partial {z}_\alpha}=\begin{cases}
\sum_{\substack{P,j\\k\leq N/2<l}}\dfrac{\mathcal{N}'_{P,\alpha_1 j}}{u_{\alpha_1}}\;\mathcal{I}_{P,\alpha_1 jkl}& \alpha_0 = 0\\[8pt]
\chi_\alpha\sum_{\substack{P,j\\k\leq N/2<l}}\mathcal{N}'_{P,\alpha_1 j}\;\mathcal{I}^{A}_{P,\alpha_1 jkl;\alpha} & \alpha_0 = 1\\[8pt]
\chi_\alpha\sum_{\substack{P,j\\k\leq N/2<l}}\mathcal{N}'_{P,\alpha_1 j}\;\mathcal{I}^{B}_{P,\alpha_1 jkl;\alpha} & \alpha_0 = 2
    \end{cases},
\end{equation}
where 
we define 
\begin{align}
\mathcal{N}'_{P,ij}\equiv (\pm)^P u_i\bar{u}_j(2\pi)^{N-1},
\end{align}
i.e., the prefactor without $D$-denominators
compared to Eq.~\eqref{pra:eq:calN}.
The corresponding matrix-parameter interaction terms are
\begin{equation}
\begin{aligned}
\mathcal{I}^{O}_{P,ijkl;\alpha} \equiv&
\frac{-U_0\,(M^{O,\text{I}}_{P,ijkl}(\sigma))_{\alpha_3\alpha_2}}
{D^{A,\text{I}}_{P,ijkl}(\sigma)\,D^{B,\text{I}}_{P,ijkl}(\sigma)}
\\
&+\frac{U_1\,(M^{O,\text{I}}_{P,ijkl}(2\sigma))_{\alpha_3\alpha_2}}
{D^{A,\text{I}}_{P,ijkl}(2\sigma)\,D^{B,\text{I}}_{P,ijkl}(2\sigma)},
\qquad O=A,B .
\end{aligned}
\end{equation}

\section{Regularization of the Symplectic Matrix}
\label{pra:sec:regularization}

The symplectic matrix $\mathcal{C}$ [Eq.~(\ref{pra:eq:C_matrix})] is generically ill-conditioned in the ECG framework due to two distinct mechanisms:
\begin{enumerate}
    \item \textit{Complex-scaling invariance.} The Lagrangian~(\ref{pra:eq:Lagrangian}) is invariant under simultaneous complex scaling of all basis prefactors $u_j \to \lambda u_j$, which generates a null direction in $\mathcal{C}$.
This gauge freedom is fixed by setting $\bar{u}_1 = 1$ throughout the propagation.
    \item \textit{Over-parametrization.} Approximate linear dependencies among the ECG basis functions produce near-zero eigenvalues in $\mathcal{C}$ that grow as the basis is refined.
\end{enumerate}

We have implemented two complementary regularization strategies to handle the second issue.

\subsection{Tikhonov regularization}
We introduce a Tikhonov-regularized inverse by adding a small positive constant $\varepsilon$ to the diagonal matrix elements before inversion,
\begin{equation}
\mathcal{C}^+_{\text{reg}}(\varepsilon) = \left(\mathcal{C}+\varepsilon \mathbbm{1}\right)^{-1},
\end{equation}
where the eigenvalues of $\mathcal{C}$ are shifted by $\varepsilon$ before inversion, so the regularized inverse has eigenvalues $1/(\lambda_\alpha+\varepsilon)$.
The regularization parameter $\varepsilon$ must be chosen small enough not to perturb the dynamics; typical values range from $10^{-7}$ to $10^{-10}$ depending on the system size and basis.

\subsection{Moore-Penrose pseudo-inverse}
The pseudo-inverse is constructed from a singular-value decomposition $\mathcal{C}=U_\mathrm{svd}\Sigma V_\mathrm{svd}^\dagger$ and inverts only singular values above a relative threshold:
\begin{equation}
    \Sigma_{\alpha\alpha}^+(\varepsilon) = \begin{cases}
    1/\sigma_\alpha  & \sigma_\alpha > \varepsilon\sigma_{\max}, \\
    0  & \sigma_\alpha \leq \varepsilon\sigma_{\max},
    \end{cases}
\end{equation}
yielding the regularized version $\mathcal{C}_{\text{pseudo}}^+(\varepsilon)=V_\mathrm{svd}\Sigma^+(\varepsilon)U_\mathrm{svd}^\dagger$.
This approach projects out the null space entirely rather than regularizing it, and is particularly useful when the number of near-zero eigenvalues is known (e.g., from the gauge fixing).

In practice, both methods yield equivalent physical results when $\varepsilon$ is appropriately chosen.
Results are considered converged when physical observables are insensitive to $\varepsilon$ over at least two orders of magnitude.

\section{Numerical Implementation}
\label{pra:sec:numerical}

\subsection{Real and imaginary-time evolution}

The ground state is obtained by propagating Eq.~(\ref{pra:eq:imag_time_eqn}) from a random initial guess for $\mathbf{z}$.
The time step for imaginary-time evolution is adapted: larger steps ($\Delta\tau \sim 10^{-2}\omega^{-1}$) are used initially when the energy is far from convergence and are progressively reduced ($\Delta\tau \sim 10^{-4}\omega^{-1}$) as the energy stabilizes.
The basis size is incrementally increased during imaginary-time propagation: starting from a small $N_b$ (e.g., 8--16) allows rapid relaxation, after which additional basis functions are added with random parameters and the propagation continues.
Convergence is monitored by tracking the energy $\mathcal{H}(\tau)$ and verifying that it plateaus to a stable value.
The final ground-state energy and wavefunction serve as initial conditions for the subsequent real-time evolution.

At $t=0$,
the harmonic trap is instantaneously quenched off
and the system undergoes real-time evolution governed by
Eq.~(\ref{pra:eq:real_time_eqn}).
The integration is performed using an embedded Runge-Kutta-Fehlberg 4(5) [RKF45] scheme with adaptive time stepping.
During real-time evolution, the total energy $\mathcal{H}$ is expected to be strictly conserved by the exact equations of motion.
Therefore,
monitoring the conservation of the
energy during the time evolution serves as a diagnostic for the accuracy of both the time integration and the regularization.
In our calculations, the relative error $|\Delta \mathcal{H}/\mathcal{H}|$ of the total energy is maintained below $10^{-4}$ up to $\omega t = 30$.

The dominant computational cost at each time step is the evaluation of $\mathcal{C}$ and $\partial\mathcal{H}/\partial\mathbf{z}$, which requires summing over all permutations $P$, all basis pairs $(i,j)$, and---for the potential and interaction terms---all particle indices.
The total cost scales as $\mathcal{O}(N_{\mathrm{perm}} \cdot N_b^2 \cdot N^2)$ when the opposite-spin interaction-pair sum is included, where $N_{\mathrm{perm}} = [(N/2)!]^2$ is the number of permutations for a system with $N/2$ particles per spin species.
For $N=6$ with $N_b = 48$, each time step involves $N_{\mathrm{perm}} = 36$ permutations and $N_b^2 = 2304$ basis pairs, making the computation manageable on a single workstation.

\subsection{Convergence benchmarks}
\label{pra:ssec:convergence}

\begin{figure*}[t!]
    \centering
\includegraphics[width=0.99\textwidth]{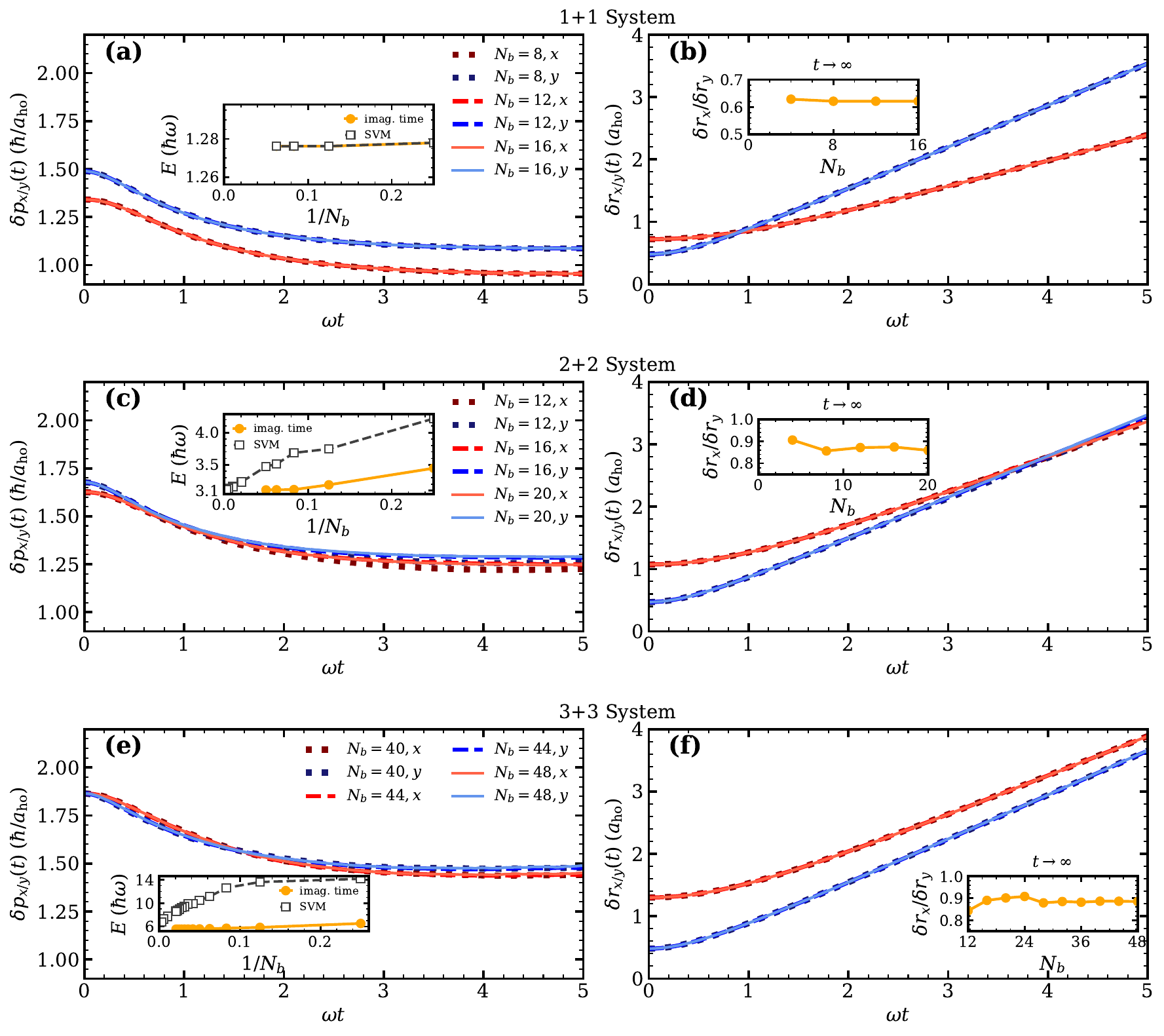}
\caption{
Convergence benchmarks for the TDECG dynamics in the $1{+}1$, $2{+}2$, and $3{+}3$ systems.
Panels (a,b), (c,d), and (e,f) correspond to the $1{+}1$, $2{+}2$, and $3{+}3$ systems, respectively.
The left column shows the momentum-space widths $\delta p_{x/y}(t)$ after trap release, with insets showing the total ground-state energy $E=\langle\Psi|\hat{H}_{\mathrm{rel}}|\Psi\rangle/\langle\Psi|\Psi\rangle+E_{\mathrm{CM}}$ as a function of $1/N_b$ and comparing imaginary-time TDECG with standalone stochastic variational method (SVM) ECG benchmarks, where $E_{\mathrm{CM}}=\hbar(\omega_x+\omega_y)/2$ is the center-of-mass ground-state contribution.
The right column shows the real-space widths $\delta r_{x/y}(t)$, with insets showing the asymptotic aspect ratio $\delta r_x/\delta r_y$ as a function $N_b$.
The TDECG dynamical curves use $N_b=8,12,16$ for $1{+}1$, $N_b=12,16,20$ for $2{+}2$, and $N_b=40,44,48$ for $3{+}3$; the SVM energy benchmarks extend to larger $N_b$ where shown.
}
    \label{pra:fig:convergence_benchmarks}
\end{figure*}

With these analytical tools in hand, we benchmark the basis-set convergence for the three systems treated in this work, namely, the $1{+}1$, $2{+}2$, and $3{+}3$ systems.
For each of these systems, 
we monitor the ground-state energy obtained from imaginary-time evolution.
For the dynamics,
we benchmark the convergence by monitoring
the momentum-space widths $\delta p_{x/y}(t)$
and the real-space widths $\delta r_{x/y}(t)$
after trap release, 
together with 
the asymptotic aspect ratio $\delta r_x/\delta r_y$ in real space that is extracted from the long-time ballistic tail.
Note that the analytical expressions for 
$\delta p_{x/y}(t)$
and $\delta r_{x/y}(t)$
from the ECG wavefunction are derived in Sec.~\ref{pra:ssec:aspect_ratios}.

Figure~\ref{pra:fig:convergence_benchmarks} shows that the smaller systems converge rapidly.
For the $1{+}1$ system, 
the curves for $N_b=8,12,16$ are already nearly indistinguishable in both momentum and real space, and the insets confirm that the ground-state energy and the asymptotic aspect ratio have stabilized within the displayed precision.
The $2{+}2$ system is slightly more demanding, but the results for $N_b=16$ and $20$ already lie essentially on top of each other.

The $3{+}3$ system remains the most demanding case and therefore sets the computational scale for the remainder of the paper. The widths obtained with $N_b=40,44,48$ overlap almost perfectly, the ground-state energy has converged to a few significant digits, and the long-time aspect ratio changes only at the sub-percent level over the largest basis sizes shown, indicating convergence for $N_b\gtrsim 40$.
To quantify the benefit of the time-dependent optimization, we also implement a standalone stochastic variational ECG solver (SVM) using the same analytical overlap and Hamiltonian matrix elements as the TDECG code.
Each trial basis function is generated with real positive-definite width matrices.
Two complementary strategies established in the SVM literature~\cite{suzuki1998stochastic,mitroy2013theory} are used.
The first is a Cholesky sampler in which the symmetric positive-definite blocks are written as $\mathrm{Re}\,A=L_A L_A^\mathrm{T}$ and $\mathrm{Re}\,B=L_B L_B^\mathrm{T}$, automatically enforcing positive-definiteness without rejection.
The diagonal entries of $L_{A/B}$ are drawn from a log-uniform distribution because the relevant Gaussian widths span several orders of magnitude---from the short-range pair-binding scale set by $\sigma$ up to the trap length $a_{\mathrm{ho}}$---so sampling uniformly in the logarithm of the width gives even coverage of all physically relevant length scales~\cite{suzuki1998stochastic}.
The off-diagonal entries are bounded to suppress nearly degenerate width matrices that would lead to ill-conditioned overlap matrices.
The second strategy is a pair-distance sampler~\cite{suzuki1998stochastic,mitroy2013theory}, in which $A$ and $B$ are constructed by drawing target pair distances $d_{ij}$ directly from a log-uniform grid; this provides an independent cross-check that the comparison is not biased by a particular parametrization.
Importantly, the SVM benchmark is initialized from scratch using these stochastic samplers and is \textit{not} warm-started from any TDECG-optimized basis, so that the comparison reflects the intrinsic efficiency of stochastic basis construction relative to imaginary-time evolution rather than a hybrid scheme that inherits TDECG-optimized parameters.
For each candidate, the generalized eigenvalue problem in the enlarged basis is solved, the energy-lowering candidate is accepted, and the resulting basis is further improved by replacement/refinement sweeps and by progressively adding more basis functions, with each new enlargement initialized from the previously optimized smaller basis rather than restarted from scratch.
The $1{+}1$ benchmark provides a useful sanity check where the SVM energy agrees with the imaginary-time result within $10^{-4}\hbar\omega$ already for $N_b=4$ and within the displayed precision by $N_b=12$--$16$.
For $2{+}2$, however, the ground-state energy obtained using the same stochastic construction is still substantially above that obtained using imaginary-time TDECG at the basis sizes used for dynamics.
For example, at $N_b=20$ the relative-energy gap is about $0.36\,\hbar\omega$.
Only after the basis is gradually enlarged in this incremental fashion up to $N_b=384$ does the SVM energy approach the imaginary-time value within about $8\times10^{-3}\hbar\omega$, showing that the stochastic method can recover the intermediate system but requires an order of magnitude more basis functions.
The contrast is strongest for $3{+}3$: despite the same aggressive stochastic search, the best SVM energy remains about $3.06\,\hbar\omega$ above imaginary-time TDECG at $N_b=48$ and still more than $1.2\,\hbar\omega$ high even at $N_b=384$.
These comparisons isolate the central practical advantage of imaginary-time propagation: it optimizes all ECG parameters coherently on the variational manifold and reaches efficient ground-state representations that direct stochastic basis construction reaches only inefficiently as the many-body correlations grow.
These benchmarks justify the basis sizes used in the real-time calculations discussed below.

\section{Post-Processing Methods and Demonstrations}
\label{pra:sec:postprocessing}

Given the parameter vector $\mathbf{z}$ after the imaginary- or real-time evolution, 
a separate post-processing stage extracts physical observables from the many-body wavefunction.
In the companion Letter~\cite{companion_prl}, these observables were used mainly to motivate and illustrate the physical interpretation of the dynamical process.
Here, we explain those post-processing methods in more detail 
and provide demonstrations of their application.

\subsection{Analytical evaluation of aspect ratios}
\label{pra:ssec:aspect_ratios}

Using the analytic Gaussian integrals available for ECG basis functions,
we extract the per-particle real-space width $\delta r_{x/y}=\sqrt{N^{-1}\sum_k\langle(r^{x/y}_{k})^{2}\rangle}$ and momentum-space width $\delta p_{x/y}=\sqrt{N^{-1}\sum_k\langle(\hat p^{x/y}_{k})^{2}\rangle}$ without computing the full many-body wavefunction.
The key observation is that
these two widths
are closely related to the potential and kinetic energy terms [Eq.~(\ref{pra:eq:H_decomposition})] via
\begin{align}
\sum_k\langle (r^{x/y}_k)^2\rangle&=N \langle (r^{x/y}_\mathrm{CM})^2\rangle+\frac{1}{\langle\Psi|\Psi\rangle}\sum_{P,i,j,k}\mathcal{N}_{P,ij}\,U^{A/B}_{P,ijk},\label{pra:eq:rxy2}\\
\sum_k\langle(\hat p^{x/y}_k)^2\rangle&= \frac{\langle (\hat p^{x/y}_\mathrm{CM})^2\rangle}{N}+\frac{\hbar^2}{\langle\Psi|\Psi\rangle}\sum_{P,i,j}\mathcal{N}_{P,ij}\,R^{A/B}_{P,ij},
\label{pra:eq:pxy2}
\end{align}
where the center-of-mass coordinate is defined as the average $r^{x/y}_{\text{CM}}=\frac{1}{N}\sum_{k=1}^N r_k^{x/y}$, while the conjugate center-of-mass momentum is the total $\hat p^{x/y}_{\text{CM}} = \sum_{k=1}^N \hat p_k^{x/y}$.
This asymmetric convention is what produces the prefactors $N$ and $1/N$ on the center-of-mass terms in Eqs.~\eqref{pra:eq:rxy2}--\eqref{pra:eq:pxy2}.

The relative parts, i.e., the second terms on the right-hand side of Eqs.~\eqref{pra:eq:rxy2}--\eqref{pra:eq:pxy2}, are obtained directly from the analytical expressions for $\mathcal{N}_{P,ij}$, $U^{A/B}_{P,ijk}$, and $R^{A/B}_{P,ij}$ derived in Secs.~\ref{pra:ssec:C_matrix} and \ref{pra:ssec:dHdz}, which are functions of the
parameter vector $\mathbf{z}$.
The center-of-mass contributions are computed analytically from the non-interacting Schr\"odinger equation for the center-of-mass degree of freedom:
\begin{equation}
    \left[-\frac{\hbar^2\nabla^2_{r_\mathrm{CM}}}{2NM}+V_\mathrm{CM}(r_\mathrm{CM})\right]\Psi_\mathrm{CM}=i\hbar\frac{\partial}{\partial t}\Psi_\mathrm{CM},
\end{equation}
where $V_\mathrm{CM}=\frac{1}{2}NM\omega_x^2 (r^x_\mathrm{CM})^2+\frac{1}{2}NM\omega_y^2 (r^y_\mathrm{CM})^2$.
Starting from the ground state, the center-of-mass wavefunction during time-of-flight expansion evolves as a Gaussian with time-dependent widths,
\begin{equation}
\begin{aligned}
    \Psi_\mathrm{CM}(\mathbf{r}_\mathrm{CM},t)
    &=\sqrt{\frac{NM}{\pi\hbar}}\,A(t) \\
    &\times e^{-\frac{NM}{2\hbar}\left[B_x\omega_x (r^x_\mathrm{CM})^2+B_y\omega_y (r^y_\mathrm{CM})^2\right]},
\end{aligned}
\end{equation}
where
\begin{equation}
    A(t)=\frac{(-1)^{3/2}}{\sqrt{\omega_xt-i}\sqrt{\omega_yt-i}},\qquad B_{x/y}(t)=\frac{i}{i-\omega_{x/y}t}.
\end{equation}
The center-of-mass expectation values, i.e., the first term on the right-hand side of Eqs.~\eqref{pra:eq:rxy2}--\eqref{pra:eq:pxy2}, are
\begin{align}
&\langle (r^{x/y}_\mathrm{CM})^2\rangle=\frac{\hbar}{2NM\omega_{x/y}}[1+(\omega_{x/y} t)^2],\\
&\langle (\hat p^{x/y}_\mathrm{CM})^2\rangle=\frac{NM\hbar\omega_{x/y}}{2}.
\end{align}
Using the compact notation $\mathcal{N}_{P,ij}$ [Eq.~(\ref{pra:eq:calN})], the cloud widths are finally
\begin{align}
    \delta r_{x/y}&=\sqrt{\frac{\hbar[1+(\omega_{x/y} t)^2]}{2NM\omega_{x/y}}+\frac{1}{N\langle\Psi|\Psi\rangle}\sum_{P,i,j,k}\mathcal{N}_{P,ij}\,U^{A/B}_{P,ijk}},\label{pra:eq:dr}\\
    \delta p_{x/y}&=\sqrt{\frac{M\hbar\omega_{x/y}}{2N}+\frac{\hbar^2}{N\langle\Psi|\Psi\rangle}\sum_{P,i,j}\mathcal{N}_{P,ij}\,R^{A/B}_{P,ij}}.\label{pra:eq:dp}
\end{align}

\subsection{Wavefunction reconstruction in laboratory real and momentum coordinates}
\label{pra:ssec:wf_reconstruction}

The ECG ansatz [Eq.~(\ref{pra:eq:ECG_ansatz})] is written in Jacobi coordinates.
To compute observables involving individual particle coordinates, we transform back to the laboratory frame.
Let $\mathbf{X}=(x_1,\ldots,x_{N-1},r^x_\mathrm{CM})^\mathrm{T}$ and $\mathbf{Y}=(y_1,\ldots,y_{N-1},r^y_\mathrm{CM})^\mathrm{T}$ denote the relative-plus-center-of-mass coordinates, and $\mathbf{r}_x=(r^x_1,\ldots,r^x_N)^\mathrm{T}$ and $\mathbf{r}_y=(r^y_1,\ldots,r^y_N)^\mathrm{T}$ the laboratory-coordinate vectors.
We denote by $\widetilde{U}$ the full $N\times N$ transformation defined by $\mathbf{X}=\widetilde{U}\mathbf{r}_x$ and $\mathbf{Y}=\widetilde{U}\mathbf{r}_y$, i.e., the H-tree Jacobi transformation augmented by the center-of-mass row.
The width matrices in laboratory coordinates are
\begin{equation}
    \bar{A}_j^{(\mathrm{lab})} = \widetilde{U}^\mathrm{T} \bar{A}_j^{(\mathrm{COM})} \widetilde{U}, \quad \bar{B}_j^{(\mathrm{lab})} = \widetilde{U}^\mathrm{T} \bar{B}_j^{(\mathrm{COM})} \widetilde{U},
\end{equation}
where
\begin{equation}
\begin{aligned}
    \bar{A}_j^{(\mathrm{COM})} &= \begin{pmatrix} \bar{A}_j & 0 \\ 0 & NM\omega_x(t)/\hbar \end{pmatrix}, \\
    \bar{B}_j^{(\mathrm{COM})} &= \begin{pmatrix} \bar{B}_j & 0 \\ 0 & NM\omega_y(t)/\hbar \end{pmatrix},
\end{aligned}
\end{equation}
with time-dependent center-of-mass frequencies $\omega_{x/y}(t) = \omega_{x/y}^{(0)}/(1+i\omega_{x/y}^{(0)} t)$ during expansion.
To permute particle labels in the laboratory frame, we use the $N\times N$ permutation matrix $\Pi_P$ with elements $(\Pi_P)_{kl}=\delta_{k,p_l}$, which is distinct from the Jacobi-space matrix $T_P$ introduced in Sec.~\ref{pra:ssec:permutation}.

Defining $\tilde{A}_{P,j}\equiv \Pi_P^\mathrm{T}\bar{A}_j^{(\mathrm{lab})}\Pi_P$ and $\tilde{B}_{P,j}\equiv \Pi_P^\mathrm{T}\bar{B}_j^{(\mathrm{lab})}\Pi_P$, the full wavefunction in laboratory coordinates reads
\begin{equation}
    \Psi(\mathbf{r}_1,\ldots,\mathbf{r}_N) = \sum_{P,j} (\pm)^P \bar{u}_j\, e^{-\frac{1}{2}\mathbf{r}_x^\mathrm{T}\tilde{A}_{P,j}\,\mathbf{r}_x}\, e^{-\frac{1}{2}\mathbf{r}_y^\mathrm{T}\tilde{B}_{P,j}\,\mathbf{r}_y},
\end{equation}
where $\mathbf{r}_x = (r^x_1, \ldots, r^x_N)^\mathrm{T}$ and $\mathbf{r}_y = (r^y_1, \ldots, r^y_N)^\mathrm{T}$.
This laboratory-frame representation is the starting point for both direct Monte Carlo sampling and analytical spectator integration.

The momentum-space wavefunction is obtained by analytically Fourier-transforming each Gaussian basis function.
Throughout this section we use the non-unitary Fourier convention: the forward transform carries no prefactor, while the inverse transform and all momentum-space marginalizations carry the phase-space measure $d^2\mathbf{p}/(2\pi\hbar)^2$ per particle.
Defining the exponent of a single term as $-\frac{1}{2}\mathbf{r}_x^\mathrm{T} \tilde{A}\, \mathbf{r}_x$, the Fourier transform with respect to the physical momentum $\mathbf{p}_x$ yields
\begin{equation}
\begin{aligned}
    &\int d^N\mathbf{r}_x\, e^{-\frac{1}{2}\mathbf{r}_x^\mathrm{T} \tilde{A}\, \mathbf{r}_x - i\mathbf{p}_x^\mathrm{T}\mathbf{r}_x/\hbar} \\
    &\qquad = \frac{(2\pi)^{N/2}}{\sqrt{\det \tilde{A}}} \exp\left[-\tfrac{1}{2\hbar^2}\mathbf{p}_x^\mathrm{T} \tilde{A}^{-1} \mathbf{p}_x\right],
\end{aligned}
\end{equation}
and analogously for the $y$-direction.
This gives the momentum-space wavefunction as a sum of Gaussians in $(\mathbf{p}_x, \mathbf{p}_y)$ with width matrices $\tilde{A}^{-1}/\hbar^2$ and $\tilde{B}^{-1}/\hbar^2$.
All momentum-space observables below inherit this Gaussian structure.

\subsection{Monte Carlo sampling and coordinate-space diagnostics}
\label{pra:ssec:mcmc}

Analytic Gaussian integration is complemented by direct sampling of $|\Psi(\mathbf{r}_1,\ldots,\mathbf{r}_N)|^2$ in laboratory coordinates.
We use Metropolis--Hastings Markov-chain Monte Carlo (MCMC) with dimer-aware moves, including five types of stochastic proposals.
For all moves involving pair selection, available opposite-spin partners are assigned stochastically with weights $\eta_{ij}=e^{-2r_{ij}/a}$:
here $r_{ij}=|\mathbf{r}_i-\mathbf{r}_j|$, $a$ is a tunable pairing length used in the proposal weights, and $\Delta_S$, $\Delta_R$, and $\Delta$ are proposal step sizes.
\begin{enumerate}
    \item \textit{Single-particle move}: displace a randomly chosen particle by $\boldsymbol{\delta} \in [-\Delta_S, \Delta_S]^2$.
    \item \textit{Pair-relative move}: select an opposite-spin pair $(i\!\uparrow, j\!\downarrow)$ with probability proportional to $\eta_{ij}$ and displace symmetrically ($\mathbf{r}_i \to \mathbf{r}_i + \boldsymbol{\delta}$, $\mathbf{r}_j \to \mathbf{r}_j - \boldsymbol{\delta}$), preserving the pair center-of-mass.
    \item \textit{Pair center-of-mass move}: same selection, but displace both particles by $+\boldsymbol{\Delta}_R$.
    \item \textit{Two-dimer recoil}: select two opposite-spin pairs sequentially with the same stochastic weighted rule and displaced by $\pm\boldsymbol{\Delta}$, conserving total momentum.
    \item \textit{Three-dimer recoil} ($N/2=3$): select all three opposite-spin pairs sequentially with the same stochastic weighted rule and displaced to conserve total momentum.
\end{enumerate}

The Metropolis--Hastings acceptance probability is $\min(1,\, |\Psi'|^2/|\Psi|^2 \cdot q_{\mathrm{back}}/q_{\mathrm{fwd}})$. For the single-particle move, the proposal is symmetric ($q_{\mathrm{back}}/q_{\mathrm{fwd}} = 1$). For moves involving non-uniform pair selection (moves~2--5), each pair $(i_k, j_k)$ selected at step $k$ of the sequential selection contributes a factor $(\eta'_{i_k j_k}/S'_{i_k})/(\eta_{i_k j_k}/S_{i_k})$ to the Hastings ratio. Here, primes denote evaluation at the proposed configuration, and $S_{i_k} = \sum_{j \in \mathcal{D}_k} \eta_{i_k j}$ sums over the set $\mathcal{D}_k$ of spin-down indices still available at step $k$. The total correction is
\begin{equation}
    \frac{q_{\mathrm{back}}}{q_{\mathrm{fwd}}} = \prod_{k=1}^{N_{\mathrm{sel}}} \frac{\eta'_{i_k j_k}/S'_{i_k}}{\eta_{i_k j_k}/S_{i_k}},
\end{equation}
with $N_{\mathrm{sel}} = 1$ for the pair-relative and pair center-of-mass moves, $N_{\mathrm{sel}} = 2$ for the two-dimer recoil, and $N_{\mathrm{sel}} = 3$ for the three-dimer recoil.

For a representative particle $i_s$ (arbitrary choice, due to the symmetry of the wavefunction) in spin sector $s$, the real-space one-body density used below is
\begin{equation}
    n_s^{(1)}(\mathbf{r}) =
    \frac{1}{\langle\Psi|\Psi\rangle}
    \int \prod_{k\neq i_s} d^2\mathbf{r}_k\,
    \left|\Psi(\mathbf{r}_{i_s}=\mathbf{r},\{\mathbf{r}_k\}_{k\neq i_s})\right|^2 .
    \label{pra:eq:real_space_density}
\end{equation}
This is a per-particle probability density, normalized as $\int d^2\mathbf{r}\, n_s^{(1)}(\mathbf{r})=1$; the corresponding spin-sector density would be $(N/2)n_s^{(1)}(\mathbf{r})$.
Equivalently, $n_s^{(1)}(\mathbf{r})=\rho_s^{(1)}(\mathbf{r},\mathbf{r})$ is the diagonal of the normalized coordinate-space OBRDM defined in Sec.~\ref{pra:ssec:obrdm}.

\begin{figure}
    \centering
\includegraphics[width=0.99\linewidth]{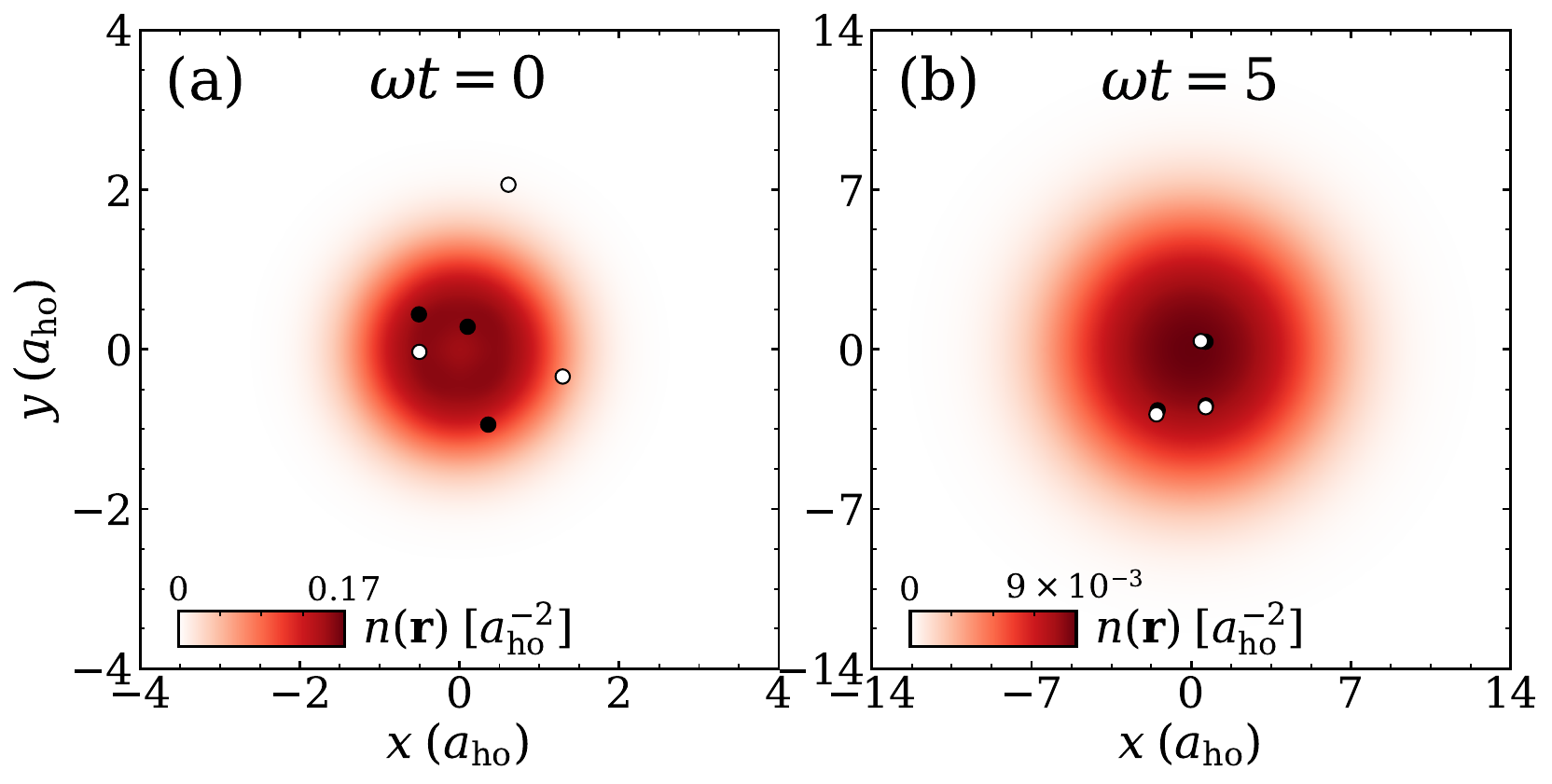}
\caption{
Coordinate-space diagnostic for the $N=3+3$ system obtained from laboratory-frame reconstruction in real space and the Monte Carlo sampling technique.
The background is the per-particle real-space one-body density $n_s^{(1)}(\mathbf{r})$ defined in Eq.~\eqref{pra:eq:real_space_density}, and the markers are sampled particle positions.
Panel (a) and (b) are for $\omega t=0$ and $\omega t=5$, respectively.
Black (white) markers denote spin-up (spin-down) particles.
At later times the samples resolve three nearby opposite-spin pairs, while the smooth density retains the coarse-grained cloud profile.
Each panel uses its own color scale.}
\label{pra:fig:density}
\end{figure}

Figure~\ref{pra:fig:density} illustrates a typical real-space output.
The smooth background is obtained by integrating out spectator coordinates from the laboratory-frame wavefunction,
whereas the markers come from direct Monte Carlo samples of the same state.
At $\omega t=0$ the sampled positions remain strongly interpenetrated; by $\omega t=5$ the configurations already isolate nearby opposite-spin pairs.
For the present methodological discussion, the key point is that analytic marginalization and direct sampling give mutually consistent and complementary coordinate-space diagnostics.

For the molecular diagnostic, opposite-spin partners are assigned in each sampled configuration with the same stochastic weights $\eta_{ij}=e^{-2r_{ij}/a}$ and expressed through the pair-relative coordinate $\mathbf{r}_R=\mathbf{r}_\uparrow-\mathbf{r}_\downarrow$.
The extracted radial amplitude is compared with the corresponding $s$-wave two-body bound-state wavefunction $\phi_b(r)$, using the $k_F$-matched double-Gaussian potential parameters appropriate to the $1{+}1$, $2{+}2$, or $3{+}3$ system.

\begin{figure}[tbp]
    \centering
\includegraphics[width=0.96\linewidth]{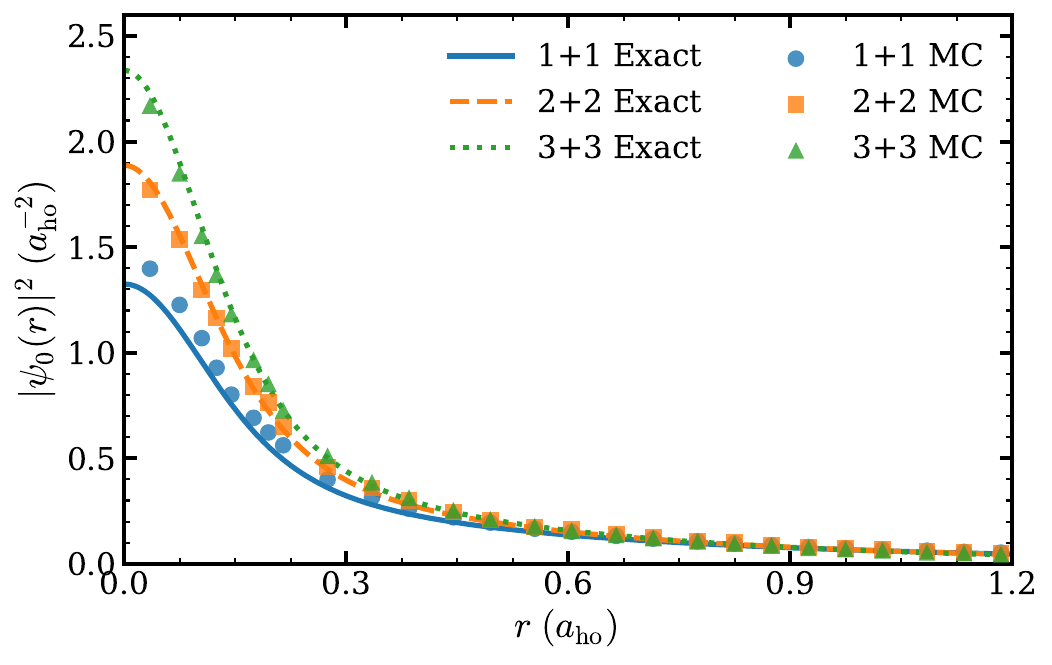}
\caption{Late-time pair-relative amplitude $|\psi_0(r)|^2$ at $\omega t=30$ reconstructed from sampled opposite-spin pairs (markers), compared with the ground-state two-body bound-state wavefunction $\phi_b(r)$ of the corresponding case-matched double-Gaussian potential (lines), for the $1{+}1$ (blue), $2{+}2$ (orange), and $3{+}3$ (green) systems.}
\label{pra:fig:molecular_bound_state}
\end{figure}

Figure~\ref{pra:fig:molecular_bound_state} shows the late-time ($\omega t=30$) pair-relative amplitude $|\psi_0(r)|^2$ reconstructed from the sampled opposite-spin pairs with the ground-state two-body bound-state wavefunction $\phi_b(r)$ of the case-matched double-Gaussian potential, for the $1{+}1$, $2{+}2$, and $3{+}3$ systems.
The close tracking between the sampled pair-relative amplitude and the case-matched two-body bound-state solution $\phi_b(r)$ benchmarks the pair-resolved Monte Carlo post-processing as a faithful tool for extracting the internal molecular wavefunction from the full many-body state.
The system-to-system comparison itself carries information: the $2{+}2$ and $3{+}3$ pairs have converged onto the exact two-body ground eigenstate, while the $1{+}1$ pair has not.
The fact that the emergent molecules take precisely the two-body ground-state form is therefore not automatic: it is a nontrivial consequence of cross-molecule interactions, which appear to assist each pair in relaxing onto the ground bound state.

\subsection{Momentum-space densities and pair correlators}
\label{pra:ssec:reduced_density}

The key analytic step for momentum-space observables is to integrate out spectator particles from the momentum-space many-body density $|\tilde\Psi|^2$, where
\begin{equation}
    \tilde\Psi(\mathbf{p}_1,\ldots,\mathbf{p}_N)=\int\!\prod_{k=1}^N d^2\mathbf{r}_k\,e^{-i\sum_k \mathbf{p}_k\cdot\mathbf{r}_k/\hbar}\,\Psi(\mathbf{r}_1,\ldots,\mathbf{r}_N)
\end{equation}
denotes the Fourier transform of the real-space ECG wavefunction.
With this convention, Parseval's relation reads
\begin{equation}
    \int\!\prod_{k=1}^N d^2\mathbf{r}_k\,|\Psi|^2
    =
    \int\!\prod_{k=1}^N\frac{d^2\mathbf{p}_k}{(2\pi\hbar)^2}\,|\tilde{\Psi}|^2 .
\end{equation}
Since the Fourier transform of a Gaussian is again a Gaussian, each term in $|\tilde\Psi|^2$ remains a product of Gaussians,
and the integration proceeds sequentially using the standard Gaussian integral formula
\begin{equation}
    \int_{-\infty}^{\infty} dx\, e^{\alpha x^2 + bx + c} = \sqrt{\frac{\pi}{-\alpha}}\, e^{c - b^2/(4\alpha)}, \quad \mathrm{Re}(\alpha) < 0.
    \label{pra:eq:gauss_int}
\end{equation}

\subsubsection{One-body and two-body densities in momentum space}

For compactness, we write $\mathbf{Q}_{\bar{i}}=\{\mathbf{q}_k\}_{k\neq i}$ and $\mathbf{Q}_{\overline{ij}}=\{\mathbf{q}_k\}_{k\neq i,j}$, with integration measures
\begin{equation}
    d\Gamma_{\bar{i}}=\prod_{k\neq i}\frac{d^2\mathbf{q}_k}{(2\pi\hbar)^2},
    \qquad
    d\Gamma_{\overline{ij}}=\prod_{k\neq i,j}\frac{d^2\mathbf{q}_k}{(2\pi\hbar)^2}.
    \label{pra:eq:momentum_spectator_measures}
\end{equation}

The normalized single-particle density in momentum space is obtained by integrating $|\tilde\Psi|^2$ over the momenta of all particles other than a representative particle $i_s$ in spin sector $s$,
\begin{equation}
    \begin{aligned}
    n_s^{(1)}(\mathbf{p}) =
    \frac{1}{\langle\Psi|\Psi\rangle}
    \int d\Gamma_{\bar{i}_s}\,
    \big|\tilde\Psi(\mathbf{p}_{i_s}=\mathbf{p},
    \mathbf{Q}_{\bar{i}_s})\big|^2 .
    \end{aligned}
\end{equation}
After integration, the density reduces to a sum of single-variable Gaussians,
\begin{equation}
    \begin{aligned}
    n_s^{(1)}(\mathbf{p})
    &= \frac{1}{\langle\Psi|\Psi\rangle} \sum_\mu w_{\mu,s}\,
    e^{a^x_{\mu,s} p_x^2 + b^x_{\mu,s} p_x + c^x_{\mu,s}} \\
    &\quad \times
    e^{a^y_{\mu,s} p_y^2 + b^y_{\mu,s} p_y + c^y_{\mu,s}},
    \end{aligned}
    \label{pra:eq:n1}
\end{equation}
where $\{w_{\mu,s}, a_{\mu,s}, b_{\mu,s}, c_{\mu,s}\}$ are complex coefficients determined by the ECG parameters and the spin sector.

The normalized opposite-spin two-particle density is defined analogously by integrating $|\tilde\Psi|^2$ over all momenta except those of one spin-up particle $i_\uparrow$ and one spin-down particle $j_\downarrow$,
\begin{equation}
    \begin{aligned}
    n_{\uparrow\downarrow}^{(2)}(\mathbf{p},\mathbf{p}') =
    \frac{1}{\langle\Psi|\Psi\rangle}
    \int d\Gamma_{\overline{i_\uparrow j_\downarrow}}\,
    \big|\tilde\Psi(
    \mathbf{p}_{i_\uparrow}=\mathbf{p},
    \mathbf{p}_{j_\downarrow}=\mathbf{p}',
    \mathbf{Q}_{\overline{i_\uparrow j_\downarrow}})
    \big|^2 .
    \end{aligned}
\end{equation}
The two-body momentum correlator is
\begin{equation}
    C_{\uparrow\downarrow}^{(2)}(\mathbf{p}, \mathbf{p}') =
    n_{\uparrow\downarrow}^{(2)}(\mathbf{p}, \mathbf{p}')
    - n_\uparrow^{(1)}(\mathbf{p})\, n_\downarrow^{(1)}(\mathbf{p}').
    \label{pra:eq:C2}
\end{equation}

\subsubsection{Relative-momentum projection}
\label{pra:ssec:c2r_c2c}

To extract the pairing diagnostic emphasized in the companion Letter~\cite{companion_prl}, we transform to relative and center-of-mass coordinates, $\mathbf{p}_R = \mathbf{p} - \mathbf{p}'$ and $\mathbf{p}_C = (\mathbf{p} + \mathbf{p}')/2$, and integrate the opposite-spin correlator over $\mathbf{p}_C$ by
\begin{equation}
    \begin{aligned}
    C_{R,\uparrow\downarrow}^{(2)}(\mathbf{p}_R)
    &=
    \int \frac{d^2\mathbf{p}_C}{(2\pi\hbar)^2}\,
    C_{\uparrow\downarrow}^{(2)}
    (\mathbf{p}_C + \mathbf{p}_R/2,\,
    \mathbf{p}_C - \mathbf{p}_R/2).
    \end{aligned}
\end{equation}
We denote the projected two-body momentum contribution by
\begin{equation}
    n_{R,\uparrow\downarrow}^{(2)}(\mathbf{p}_R)
    =
    \int \frac{d^2\mathbf{p}_C}{(2\pi\hbar)^2}\,
    n_{\uparrow\downarrow}^{(2)}
    (\mathbf{p}_C + \mathbf{p}_R/2,\,
    \mathbf{p}_C - \mathbf{p}_R/2).
    \label{pra:eq:relative_pair_density}
\end{equation}
The contribution from $n_{\uparrow\downarrow}^{(2)}(\mathbf{p},\mathbf{p}')$ in this integral is computed analytically through the change of variables followed by Gaussian integration.
The disconnected relative background is
\begin{equation}
    n_{R,\mathrm{disc}}^{(2)}(\mathbf{p}_R)
    =
    \int \frac{d^2\mathbf{p}_C}{(2\pi\hbar)^2}\,
    n_\uparrow^{(1)}(\mathbf{p}_C+\mathbf{p}_R/2)\,
    n_\downarrow^{(1)}(\mathbf{p}_C-\mathbf{p}_R/2).
    \label{pra:eq:relative_disconnected}
\end{equation}
Numerically, this is evaluated on a uniform grid by a discrete FFT cross-correlation, equivalently a convolution with the reflected down-spin density,
\begin{equation}
    n_{R,\mathrm{disc}}^{(2)}(\mathbf{p}_R)
    =
    [n_\uparrow^{(1)} \ast n_{\downarrow,\mathrm{ref}}^{(1)}](\mathbf{p}_R),
    \qquad
    n_{\downarrow,\mathrm{ref}}^{(1)}(\mathbf{p})=n_\downarrow^{(1)}(-\mathbf{p}),
\end{equation}
with
\begin{equation}
    [n_\uparrow^{(1)} \ast n_{\downarrow,\mathrm{ref}}^{(1)}](\mathbf{p}) =
    \mathcal{F}^{-1}\!\big[
    \mathcal{F}[n_\uparrow^{(1)}]\cdot
    \mathcal{F}[n_{\downarrow,\mathrm{ref}}^{(1)}]
    \big](\mathbf{p}),
\end{equation}
where $\mathcal{F}$ denotes the discrete Fourier transform, $\ast$ denotes convolution, and the one-body density is sampled on a uniform grid and zero-padded to avoid circular-convolution artifacts.
The final relative correlator is
\begin{equation}
    C_{R,\uparrow\downarrow}^{(2)}(\mathbf{p}_R)
    = n_{R,\uparrow\downarrow}^{(2)}(\mathbf{p}_R)
    - n_{R,\mathrm{disc}}^{(2)}(\mathbf{p}_R).
\end{equation}

\begin{figure*}[t!]
    \centering
    \includegraphics[width=0.99\textwidth]{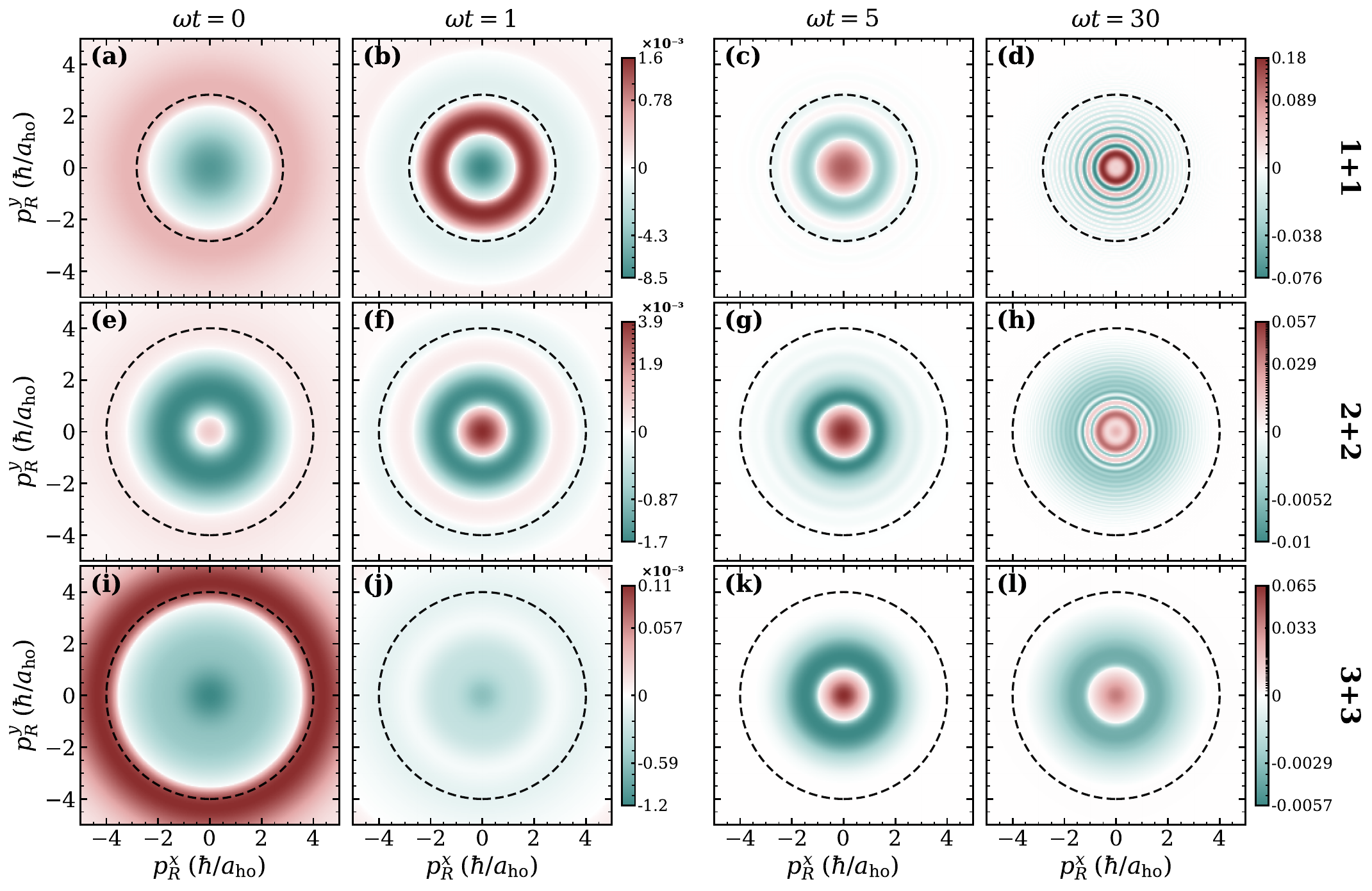}
    \caption{Time evolution of the relative-momentum correlator $C_{R,\uparrow\downarrow}^{(2)}(\mathbf{p}_R)$ in the isotropic trap.
Rows from top to bottom correspond to the $1{+}1$, $2{+}2$, and $3{+}3$ systems, while columns from left to right show snapshots at $\omega t=0,\,1,\,5,\,30$.
Dashed circles indicate the reference radius $|\mathbf{p}_R|=2\hbar k_F$ as a guide to the eye.}
    \label{pra:fig:relative_correlator}
\end{figure*}

\begin{figure*}[t!]
    \centering
    \includegraphics[width=0.99\textwidth]{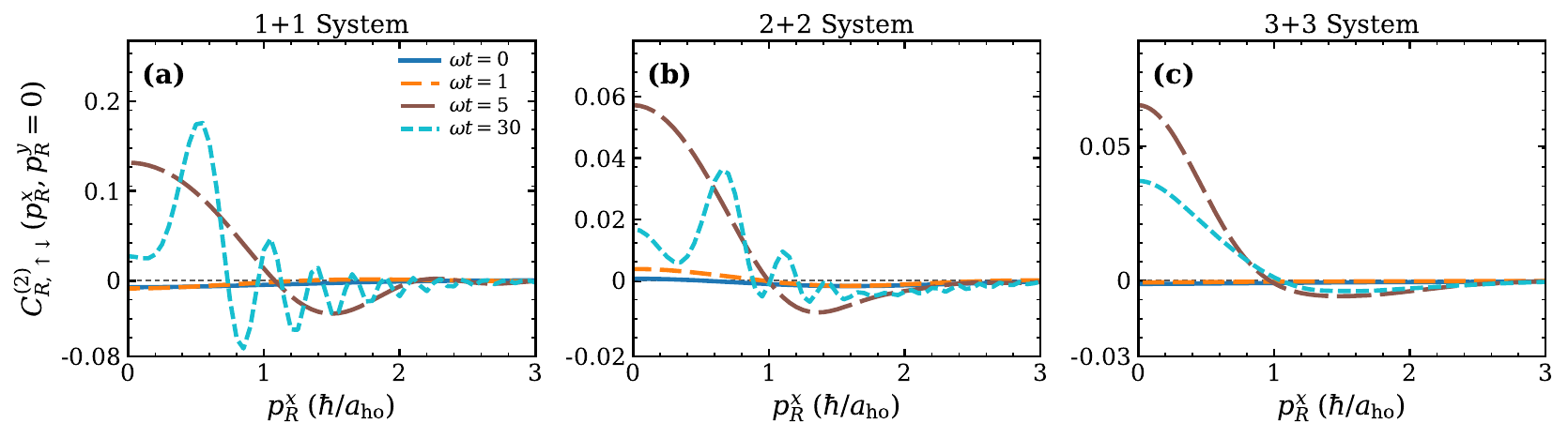}
    \caption{Relative-momentum correlator $C_{R,\uparrow\downarrow}^{(2)}(\mathbf{p}_R)$ as a function of $p_R^x$ with $p_R^y$ fixed at $0$, for the $1{+}1$, $2{+}2$, and $3{+}3$ isotropic systems.}
    \label{pra:fig:relative_correlator_linecuts}
\end{figure*}

Figure~\ref{pra:fig:relative_correlator} shows the relative-correlator construction of isotropic systems as an example.
Methodologically, it verifies that the analytic spectator integration and the FFT-based disconnected subtraction remain numerically stable throughout the time evolution.
The resulting correlator also retains the expected finite-momentum ring and later low-momentum buildup while remaining radially symmetric in the isotropic trap.
The red positive region in $C_{R,\uparrow\downarrow}^{(2)}(\mathbf{p}_R)$ identifies relative momenta where opposite-spin particles are correlated beyond the disconnected one-body background.
At $\omega t=0$, this excess correlation lies near $|\mathbf{p}_R|\simeq 2\hbar k_F$, consistent with pairing across the Fermi surface.
A subtlety, however, emerges when comparing the three rows at $\omega t=0$: for the $1{+}1$ and $3{+}3$ systems the positive lobe is confined to the $2\hbar k_F$ ring, whereas the $2{+}2$ correlator carries an additional positive peak right at $\mathbf{p}_R=0$.
This particle-number selectivity is a direct fingerprint of the shell structure of the isotropic 2D harmonic oscillator.
The single-particle level that fixes $E_F$ in the $2{+}2$ and $3{+}3$ systems is the $n_x+n_y=1$ shell, which has a two-fold orbital degeneracy.
For $3{+}3$ this shell is fully occupied, so the ground state is non-degenerate and supports only the standard back-to-back pairing at $|\mathbf{p}_R|\simeq 2\hbar k_F$.
For $2{+}2$, by contrast, the same shell is only half-filled per spin: each spin places a single particle into the two-dimensional degenerate $n=1$ subspace, and the interacting ground state mixes the two orbital choices coherently between opposite spins, producing pair configurations in which both partners share the same single-particle state and contribute weight at $\mathbf{p}_R=0$ in addition to the Fermi-surface lobes.
The $1{+}1$ system has no such ambiguity because only the non-degenerate $n=0$ shell is involved.
This degeneracy-induced central peak is therefore specific to the isotropic geometry: in the anisotropic trap used in the companion Letter~\cite{companion_prl}, the asymmetry $\omega_x\neq\omega_y$ splits the $n=1$ shell into the non-degenerate $(n_x,n_y)=(1,0)$ and $(0,1)$ levels, removing the orbital choice and eliminating the $\mathbf{p}_R=0$ feature from the $2{+}2$ ground-state correlator at $\omega t=0$.
During expansion the correlated weight moves inward, and by $\omega t=30$, the dominant positive region is concentrated near $\mathbf{p}_R=0$, indicating molecule-like low-relative-momentum correlations.

To sharpen the $N$-dependence visible in the two-dimensional maps, Fig.~\ref{pra:fig:relative_correlator_linecuts} extracts one-dimensional line cuts of $C_{R,\uparrow\downarrow}^{(2)}(\mathbf{p}_R)$ along $p_R^y=0$ for the same three systems and four time slices.
The cuts reveal a nontrivial trend as $N$ increases from $1{+}1$ to $3{+}3$.
At long times ($\omega t=30$), the $1{+}1$ and $2{+}2$ curves both develop pronounced oscillations flanking the central molecular peak, which is the hallmark of coherent interference between molecular center-of-mass plane waves left over from the trap-release dynamics, whereas the $3{+}3$ curve is essentially smooth and the fringes are washed out.
This $N$-dependent loss of long-time interference parallels the anisotropic-trap result reported in the companion Letter~\cite{companion_prl}, where the corresponding two-dimensional correlator displays the same qualitative hierarchy: clear interference fringes for $1{+}1$, partially blurred fringes for $2{+}2$, and a smooth long-time correlator for $3{+}3$.
The agreement between the isotropic line cuts here and the anisotropic two-dimensional maps there indicates that the disappearance of the long-time fringes with growing $N$ is a robust feature of the dynamics, set by particle number rather than by trap geometry.

\subsection{Out-of-time-order correlator and information scrambling}
\label{pra:ssec:otoc}

We present a complementary diagnostic that directly probes differences between systems of different sizes through the rate at which local quantum information spreads under the Hamiltonian dynamics using the 
out-of-time-order correlator (OTOC)~\cite{larkin1969quasiclassical,maldacena2016bound,swingle2018unscrambling},
\begin{equation}
    F(t) = \langle\Psi_0|\,W_x(t)\,W_y(0)\,W_x(t)\,W_y(0)\,|\Psi_0\rangle,
    \label{pra:eq:otoc_def}
\end{equation}
where
the $W_{x/y}$ are 
the two bounded Hermitian and permutation-symmetric Gaussian observables that are
supported on the inter-particle separations of the cloud, 
the operator $W_x(t)$
is evaluated in the Heisenberg picture via $W_x(t)=e^{iHt/\hbar}\,W_x\,e^{-iHt/\hbar}$,
and $|\Psi_0\rangle$ denotes the ground state of the trapped system. 
For Hermitian $W_x$ and $W_y$, 
$F(t)$ coincides with the standard two-operator OTOC form~\cite{roberts2015diagnosing,hashimoto2017otoc} and
is also related to the 
squared commutator OTOC, 
$C(t)=-\langle[W_x(t),W_y(0)]^{2}\rangle$, 
via 
$C(t)=2(\langle W_x^{2}(t)W_y^{2}(0)\rangle-\operatorname{Re}F(t))$.
Introducing the two doubly-shifted ECG states
\begin{equation}
    |\Psi_R(t)\rangle \equiv W_x(t)\,W_y\,|\Psi_0\rangle,
    \qquad
    |\Psi_L(t)\rangle \equiv W_y\,W_x(t)\,|\Psi_0\rangle,
    \label{pra:eq:PsiLR}
\end{equation}
the OTOC of Eq.~\eqref{pra:eq:otoc_def} is simply $F(t)=\langle\Psi_L(t)|\Psi_R(t)\rangle$, and we define its gauge-invariant Cauchy--Schwarz-normalized form as
\begin{equation}
    F_{\mathrm{norm}}(t)
    = \frac{\langle\Psi_L(t)|\Psi_R(t)\rangle}
           {\sqrt{\langle\Psi_L(t)|\Psi_L(t)\rangle\,\langle\Psi_R(t)|\Psi_R(t)\rangle}}.
    \label{pra:eq:Fnorm}
\end{equation}
The Cauchy--Schwarz inequality then guarantees $|F_{\mathrm{norm}}(t)|\le 1$ for all $t$, with equality if and only if the two doubly-shifted states are parallel.
The modulus $|F_{\mathrm{norm}}(t)|$ measures how nontrivially $W_x(t)$ and $W_y(0)$ overlap on the same many-body state;
a rapid decay of $|F_{\mathrm{norm}}(t)|$ from its initial value diagnoses fast information spreading, often called scrambling.

Within the ECG framework, we choose $W_x$ and $W_y$ as permutation-symmetric Gaussians of the inter-particle separations restricted to a single Cartesian direction,
\begin{equation}
    \begin{aligned}
    W_x &= \exp\!\Big[-\frac{1}{2\sigma_{\mathrm{meas}}^{2}}\sum_{k<l}^{N}(x_k-x_l)^{2}\Big], \\
    W_y &= \exp\!\Big[-\frac{1}{2\sigma_{\mathrm{meas}}^{2}}\sum_{k<l}^{N}(y_k-y_l)^{2}\Big],
    \end{aligned}
    \label{pra:eq:otoc_Wxy}
\end{equation}
with width $\sigma_{\mathrm{meas}}$ controlling how strongly each operator perturbs the ground state, and the sums running over all distinct particle pairs.
For convenience, we set $\sigma_{\mathrm{meas}}=1\,a_{\mathrm{ho}}$ throughout the calculations reported below.
Because $W_x$ depends only on $\{x_k\}$ and $W_y$ only on $\{y_k\}$, the canonical commutators $[x_k,y_l]=0$ imply that $[W_x,W_y]=0$ holds identically as an operator equation, independent of the state, the Hamiltonian, and any factorization assumption.
At $t=0$, this gives $|\Psi_R(0)\rangle = W_x W_y|\Psi_0\rangle = W_y W_x|\Psi_0\rangle = |\Psi_L(0)\rangle$, so the Cauchy--Schwarz inequality~\eqref{pra:eq:Fnorm} is saturated and $|F_{\mathrm{norm}}(0)|=1$ exactly for any choice of $|\Psi_0\rangle$.

The form~\eqref{pra:eq:otoc_Wxy} is also compatible with the ECG ansatz: applying $W_x$ to the state $\Psi_0$
leaves the $y$-width matrices $B_j$ unchanged and shifts each $x$-width matrix $A_j$ by $\sum_{k<l}\Omega^{kl}/\sigma_{\mathrm{meas}}^{2}$, where $\Omega^{kl}$ is the Jacobi-frame projector onto the $(k,l)$ pair-relative coordinate;
$W_y$ acts symmetrically by
shifting $B_j$ alone in the same manner.
The action is therefore exact within the variational manifold and does not require any basis expansion at the operator step.
A further useful property follows from this directional split: in the non-interacting limit the Hamiltonian is $H = H_x + H_y$, so $W_x(t)$ stays purely a function of $\{x_k\}$ at all times, $[W_x(t),W_y(0)]=0$ for every $t$, and $|F_{\mathrm{norm}}(t)|\equiv 1$.
Any departure from unity in the interacting calculation at $t>0$ is therefore a direct dynamical fingerprint of the inter-particle interaction coupling the two Cartesian directions through $|\mathbf{r}_i-\mathbf{r}_j|$.

We calculated $F(t)$ defined in Eq.~\eqref{pra:eq:otoc_def} numerically by three real-time TDECG propagations: 
a forward evolution of $W_y|\Psi_0\rangle$ to time $t$ to get $e^{-iHt/\hbar}W_y|\Psi_0\rangle$ followed by a backward evolution of $W_x\,e^{-iHt/\hbar}W_y|\Psi_0\rangle$ from $t$ to $0$, and an independent backward evolution of $W_x|\Psi_0\rangle$ from $t$ to $0$.
The OTOC is then the overlap of the two resulting ECG states, evaluated by the same Gaussian-determinant kernel that gives the static norm $\langle\Psi|\Psi\rangle$ in Sec.~\ref{pra:ssec:dHdz}.
$F_{\mathrm{norm}}(t)$ then follows directly from Eq.~\eqref{pra:eq:Fnorm} together with two additional static overlaps of the doubly-shifted states with themselves.

\begin{figure}[tbp]
    \centering
    \includegraphics[width=0.99\linewidth]{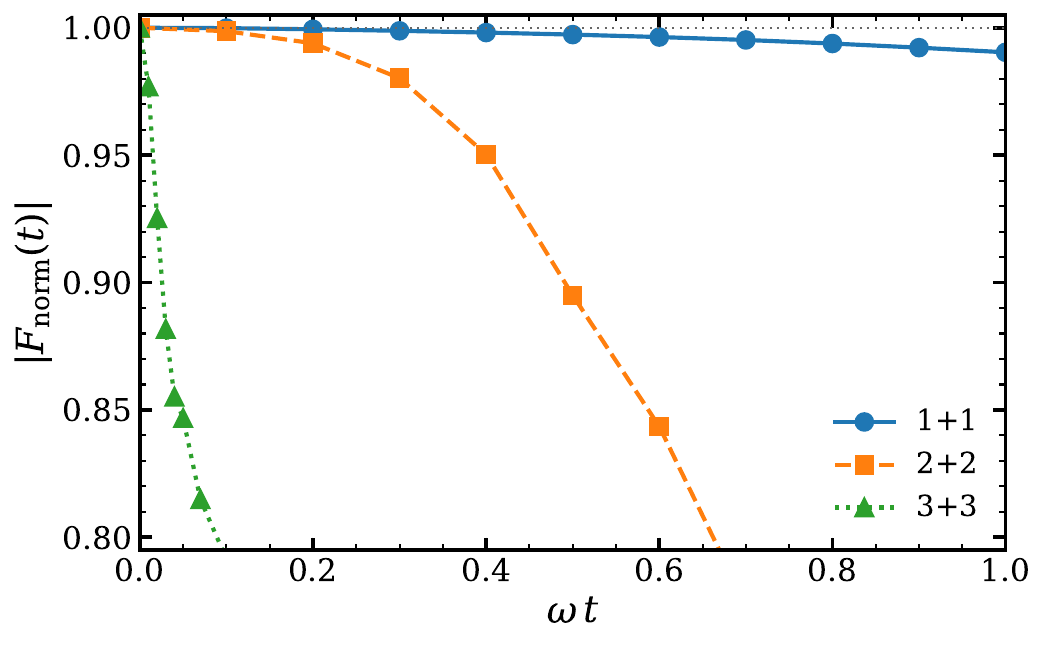}
    \caption{Modulus of the normalized OTOC $|F_{\mathrm{norm}}(t)|$ from Eq.~\eqref{pra:eq:otoc_def} for the $1{+}1$, $2{+}2$, and $3{+}3$ systems, with the directional permutation-symmetric Gaussian operators of Eq.~\eqref{pra:eq:otoc_Wxy} ($W_x$ in $x$, $W_y$ in $y$) of width $\sigma_{\mathrm{meas}}=1\,a_{\mathrm{ho}}$ applied to the anisotropic-trap ground state.
    All curves are anchored at the exact value $|F_{\mathrm{norm}}(0)|=1$, which follows from $[W_x,W_y]=0$ and is satisfied to machine precision in the actual interacting calculation (see text).
    The horizontal dotted line marks the Cauchy--Schwarz upper bound.
    }
    \label{pra:fig:otoc_three_systems}
\end{figure}

Figure~\ref{pra:fig:otoc_three_systems} compares $|F_{\mathrm{norm}}(t)|$ for the three systems of different sizes.
For the $1{+}1$ system, $|F_{\mathrm{norm}}(t)|$ remains within $1\%$ of unity over the entire window $\omega t\le 1$, so $W_x(t)$ and $W_y(0)$ effectively commute at all times accessible here and there is no measurable information spreading.
For $2{+}2$, the OTOC decays smoothly from unity to roughly $0.9$ by $\omega t=0.5$, with a clear leading $t^{2}$ growth of $1-|F_{\mathrm{norm}}(t)|$, as expected for unitary dynamics.
The $3{+}3$ system departs much faster: a comparable departure from unity is already reached at $\omega t\simeq 0.05$, almost an order of magnitude earlier than in the $2{+}2$ system.

We interpret this hierarchy as a few-body realization of information scrambling.
Because $W_x$ and $W_y$ probe orthogonal Cartesian directions, the only mechanism by which $W_x(t)$ can develop overlap with $W_y(0)$ is the inter-particle interaction, which couples $\{x_k\}$ and $\{y_k\}$ through $|\mathbf{r}_i-\mathbf{r}_j|$ on each pair.
The number of such pair channels grows as $N(N-1)/2$, so $W_x(t)$ acquires support along progressively more directions in operator space as the system size increases, accelerating the growth of $[W_x(t),W_y(0)]$ away from zero.
The $1{+}1$ result, with $|F_{\mathrm{norm}}(t)|$ pinned to unity over the entire window, sets a clear baseline: a single pair separation transfers operator weight only trivially between $x$ and $y$, and there is essentially no scrambling on the observable timescale.
The smooth quadratic onset in $2{+}2$ corresponds to the perturbative regime of operator growth, in which $1-|F_{\mathrm{norm}}(t)|$ tracks the squared-commutator $C(t)/(2\langle W_x^2\rangle\langle W_y^2\rangle)$ and grows as $t^{2}$ from $0$.
The substantially earlier departure for $3{+}3$ comes from the additional pair channels, which open a markedly faster route for an initially $x$-supported operator to spread into the $y$ sector.
The feature is the operational signature of scrambling in few-body systems analogous to the operator-growth picture established for large many-body systems~\cite{maldacena2016bound,swingle2018unscrambling}.

\subsection{One-body reduced density matrix and velocity field}
\label{pra:ssec:obrdm}

The momentum-space one-body reduced density matrix (OBRDM) $\rho_s^{(1)}(\mathbf{p};\mathbf{p}')$ retains the bra and ket momenta of one particle in spin sector $s$ while integrating out all spectator momenta:
\begin{equation}
    \begin{aligned}
    \rho_s^{(1)}(\mathbf{p};\mathbf{p}') =
    \frac{1}{\langle\Psi|\Psi\rangle}
    \int d\Gamma_{\bar{i_s}}\,
    \tilde\Psi(\mathbf{p},\mathbf{Q}_{\bar{i_s}})
    \tilde\Psi^*(\mathbf{p}',\mathbf{Q}_{\bar{i_s}}).
    \end{aligned}
    \label{pra:eq:obrdm_def}
\end{equation}
Here, $\mathbf{p}=(p_x,p_y)$ and $\mathbf{p}'=(p_x',p_y')$ denote the bra and ket single-particle momenta.
Because the ECG kernel factorizes into $x$ and $y$ directions after spectator integration, the full two-dimensional OBRDM can be written as
\begin{equation}
    \begin{aligned}
    \rho_s^{(1)}(p_x,p_y;p_x',p_y')
    &=
    \frac{1}{\langle\Psi|\Psi\rangle}
    \sum_{\mu=1}^{K} w_\mu\,
    \rho_{x,\mu}^{(1)}(p_x,p_x') \\
    &\qquad\times
    \rho_{y,\mu}^{(1)}(p_y,p_y') .
    \end{aligned}
    \label{pra:eq:obrdm}
\end{equation}
Here, $K=(N_bN_{\mathrm{perm}})^2$ with $N_{\mathrm{perm}}=[(N/2)!]^2$ is the number of Gaussian product terms generated by the bra-ket ECG expansion after antisymmetrization and spectator integration, before any algebraic merging of identical kernels.
Equivalently, $N_bN_{\mathrm{perm}}$ is the number of Gaussian product terms in the antisymmetrized wavefunction.
The coefficient $w_\mu$ is the corresponding scalar weight, including ECG coefficients, permutation signs, and Gaussian prefactors but excluding the explicit global normalization shown in Eq.~\eqref{pra:eq:obrdm}.
The one-dimensional Gaussian kernels are
\begin{equation}
    \begin{aligned}
    \rho_{\nu,\mu}^{(1)}(p,p') &=
    \exp\!\Big[
        a_{11}^{\nu,\mu}p^2
        + a_{22}^{\nu,\mu}p'^2 \\
        &\qquad
        + b_{12}^{\nu,\mu}p\,p'
        + c^{\nu,\mu}
    \Big],\qquad \nu=x,y .
    \end{aligned}
    \label{pra:eq:obrdm_1d_kernel}
\end{equation}
The OBRDM serves two distinct diagnostic purposes in our analysis: extracting natural-orbital occupations via diagonalization and encoding the local hydrodynamic velocity field through its phase gradients.

For the diagonalization of $\rho_s^{(1)}$, we represent it in a discrete Gauss-Hermite basis $\{\phi_n\}_{n=0}^{N_{\max}}$ with tunable length scale $\ell$:
\begin{equation}
    D^{(\nu,\mu)}_{nm} =
    \int dp\, dp'\, \phi_n(p/\ell)\,
    \rho_{\nu,\mu}^{(1)}(p,p')\,
    \phi_m(p'/\ell).
    \label{pra:eq:rho1_1d_matrix}
\end{equation}
For a fixed $(\nu,\mu)$, we suppress these labels on the local quantities in the recurrence below.
The integral in Eq.~\eqref{pra:eq:rho1_1d_matrix} is evaluated analytically as
\begin{equation}
   D_{nm}\equiv
   D^{(\nu,\mu)}_{nm}
    = \ell\sqrt{\frac{\pi}{\Delta}}\, F_{nm}(\alpha, \beta, \gamma),
    \label{pra:eq:D_from_F}
\end{equation}
where $\alpha = a_{11}^{\nu,\mu}\ell^2 - 1/2$, $\beta = b_{12}^{\nu,\mu}\ell^2$, $\gamma = a_{22}^{\nu,\mu}\ell^2 - 1/2$, and $\Delta = \alpha\gamma - \beta^2/4$.
Defining $\kappa_1 = 1 + \gamma/\Delta$, $\kappa_2 = 1 + \alpha/\Delta$, and $\kappa_{12} = \beta/\Delta$, 
the $F_{nm}$ satisfies
\begin{multline}
    (n{+}m)\,F_{nm} = -\kappa_1\sqrt{n(n{-}1)}\,F_{n-2,m}\\
    - \kappa_2\sqrt{m(m{-}1)}\,F_{n,m-2} + \kappa_{12}\sqrt{nm}\,F_{n-1,m-1},
    \label{pra:eq:Fnm_recurrence}
\end{multline}
with the initial condition of $F_{00} = 1$ and the convention of $F_{nm} = 0$ for negative indices and odd $n+m$.

The resulting one-dimensional matrices $D^{(\nu,\mu)}_{nm}$ give the two-dimensional representation as a sum of Kronecker products,
\begin{equation}
    \rho^{(1)}_{s;(n_x,n_y),(m_x,m_y)}
    =
    \sum_{\mu=1}^{K} w_\mu\,
    D^{(x,\mu)}_{n_x m_x}\,
    D^{(y,\mu)}_{n_y m_y}.
    \label{pra:eq:rho1_kron}
\end{equation}
For the coordinate-space velocity field, we use the OBRDM with bra coordinate $\mathbf{r}'$ and ket coordinate $\mathbf{r}$,
\begin{equation}
    \begin{aligned}
    \rho_s^{(1)}(\mathbf{r}',\mathbf{r})
    &=
    \frac{1}{\langle\Psi|\Psi\rangle}
    \int \prod_{k\neq i_s} d^2\mathbf{r}_k\\
    &\qquad\times
    \Psi^*(\mathbf{r}',\{\mathbf{r}_k\}_{k\neq i_s})
    \Psi(\mathbf{r},\{\mathbf{r}_k\}_{k\neq i_s}) .
    \end{aligned}
    \label{pra:eq:obrdm_coordinate_def}
\end{equation}

For the velocity field reconstruction, the OBRDM encodes the local velocity field via the probability current,
\begin{equation}
    \mathbf{j}_s(\mathbf{r}) = \frac{\hbar}{M}\,\mathrm{Im}\!\left[\nabla_{\mathbf{r}}\,\rho_s^{(1)}(\mathbf{r}',\mathbf{r})\Big|_{\mathbf{r}'=\mathbf{r}}\right],
\end{equation}
and $\mathbf{v}_s(\mathbf{r}) = \mathbf{j}_s(\mathbf{r})/n_s^{(1)}(\mathbf{r})$.
In the central high-density region of the freely expanding cloud, the reconstructed flow is well described by the leading self-similar, linear component
\begin{equation}
    v_x(x,y) \approx b_x\, x, \qquad v_y(x,y) \approx b_y\, y,
\end{equation}
with $b_x \neq b_y$ quantifying the elliptic component of anisotropic expansion.

\begin{figure}[tbp]
    \centering
\includegraphics[width=\linewidth]{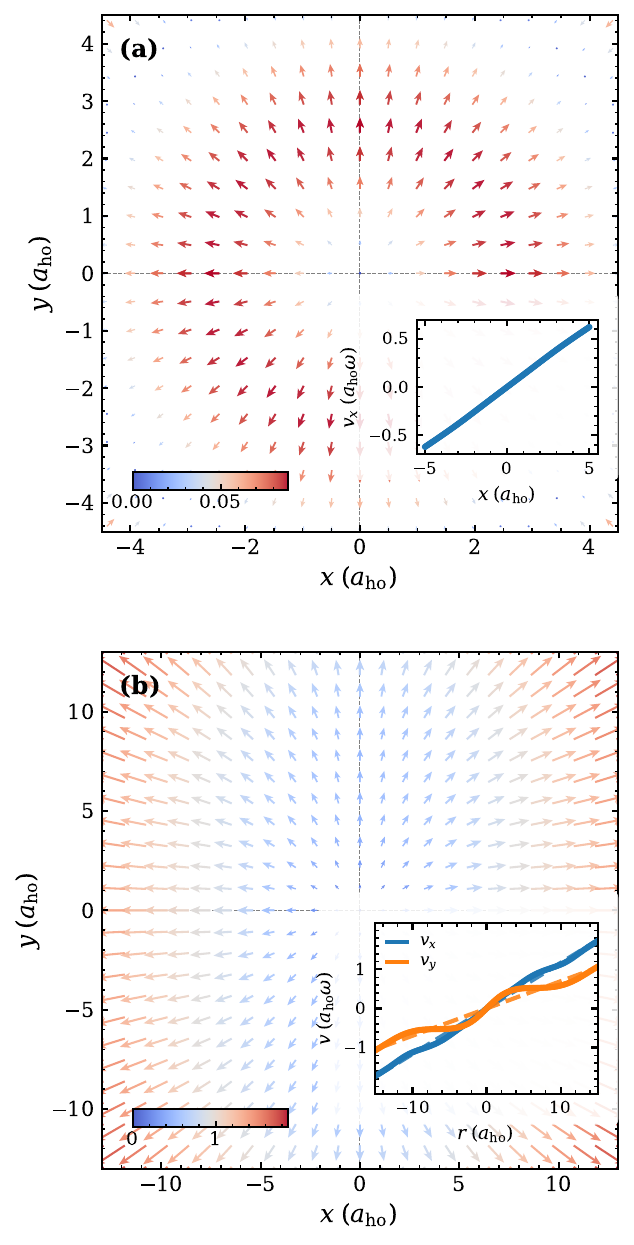}
\caption{Velocity fields reconstructed from the OBRDM for the $N=3+3$ system at $\omega t=5$.
The arrows show $\mathbf{v}_s(\mathbf{r})=\mathbf{j}_s(\mathbf{r})/n_s^{(1)}(\mathbf{r})$ in the high-density region, while the color scale indicates the local speed.
Insets: one-dimensional cuts (solid markers) with linear fits (dashed lines).
For the isotropic case in panel (a), $b_x=b_y\simeq 0.126\,\omega$.
For the anisotropic case in panel (b), $b_x\simeq 0.117\,\omega$ and $b_y\simeq 0.070\,\omega$.}
\label{pra:fig:vectorfield}
\end{figure}

Figure~\ref{pra:fig:vectorfield} shows outputs of the current-reconstruction procedure for the $N=3+3$ system at $\omega t=5$.
The isotropic example isolates the OBRDM current reconstruction from trap-induced anisotropy and exposes the self-similar radial expansion.
The anisotropic example shows the corresponding elliptic expansion, with structure beyond the linear component along the initially more confined y-direction.
Overall, the arrows are well approximated by a flow field proportional to position.
The insets quantify this through linear fits to the one-dimensional velocity cuts, demonstrating that the OBRDM-based current is sufficiently smooth for extracting expansion rates and anisotropies.

\subsection{Two-body reduced density matrix}
\label{pra:ssec:tbrdm}

The momentum-space two-body reduced density matrix (TBRDM) $\rho^{(2)}$ retains the bra and ket momenta of two particles while integrating out all spectators.
For the opposite-spin pair channel used here, the momentum-space object is
\begin{equation}
    \begin{aligned}
    \rho^{(2)}(\mathbf{p}_i,\mathbf{p}_j;\mathbf{p}_i',\mathbf{p}_j')
    &=
    \frac{1}{\langle\Psi|\Psi\rangle}
    \int d\Gamma_{\overline{ij}}\,
    \tilde\Psi(\mathbf{p}_i,\mathbf{p}_j,\mathbf{Q}_{\overline{ij}}) \\
    &\qquad\times
    \tilde\Psi^*(\mathbf{p}_i',\mathbf{p}_j',\mathbf{Q}_{\overline{ij}}).
    \end{aligned}
    \label{pra:eq:tbrdm_def}
\end{equation}
After spectator integration, the TBRDM has the same Cartesian factorization as Eq.~\eqref{pra:eq:obrdm}. For one Cartesian direction $\nu=x,y$, we write the retained momentum components as $(p_{i\nu},p_{i\nu}',p_{j\nu},p_{j\nu}')$ and use the one-dimensional kernel
\begin{equation}
    \begin{aligned}
    \rho_{\nu,\mu}^{(2)}(p_{i\nu},p_{i\nu}';p_{j\nu},p_{j\nu}')
    &=
    \exp\!\left[
    \sum_{a,b=1}^{4} R_{ab}^{\nu,\mu}\,\xi_a \xi_b
    + c_{cc}^{\nu,\mu}
    \right],\\
    (\xi_1,\xi_2,\xi_3,\xi_4)
    &=(p_{i\nu},p_{i\nu}',p_{j\nu},p_{j\nu}').
    \end{aligned}
\end{equation}
Thus, $(p_{i x},p_{i x}',p_{j x},p_{j x}')$ represents only the $x$-direction momentum components; the $y$-direction has the same structure with its own coefficients.

The pair matrix for this one-dimensional kernel has row index $(n_i,n_j)$ and column index $(m_i,m_j)$ and is defined, in direct analogy to Eq.~\eqref{pra:eq:rho1_1d_matrix}, by
\begin{equation}
    \begin{aligned}
    \mathsf{M}_{(n_i,n_j),(m_i,m_j)}^{(\nu,\mu)}
    &=
    \int dp_{i\nu}\,dp_{i\nu}'\,dp_{j\nu}\,dp_{j\nu}'\\
    &\quad\times
    \rho_{\nu,\mu}^{(2)}(p_{i\nu},p_{i\nu}';p_{j\nu},p_{j\nu}')\,
    \phi_{n_i}(p_{i\nu}/\ell)\\
    &\quad\times
    \phi_{m_i}(p_{i\nu}'/\ell)\,
    \phi_{n_j}(p_{j\nu}/\ell)\,
    \phi_{m_j}(p_{j\nu}'/\ell).
    \end{aligned}
\end{equation}
The four-index recurrence below evaluates the Gaussian part of this integral with coefficient index $\mathbf{n}=(n_1,n_2,n_3,n_4)=(n_i,m_i,n_j,m_j)$.

For a fixed $(\nu,\mu)$, 
we define the local recurrence matrices $\mathsf{A} = \alpha_\ell \mathsf{I}_4 - 2R^{\nu,\mu}$ and $\mathsf{B} = 2\alpha_\ell\mathsf{A}^{-1} - \mathsf{I}_4$ with $\alpha_\ell = 1/\ell^2$.
As in the OBRDM recurrence, the fixed $(\nu,\mu)$ labels are suppressed on local quantities.
The normalized coefficients $C_{\mathbf{n}}\equiv C_{\mathbf{n}}^{\nu,\mu}$ satisfy
\begin{multline}
    |\mathbf{n}|\,C_{\mathbf{n}} = \sum_{r=1}^{4} \mathsf{B}_{rr}\sqrt{n_r(n_r{-}1)}\,C_{\mathbf{n}-2\mathbf{e}_r}\\
    + 2\!\sum_{r<s} \mathsf{B}_{rs}\sqrt{n_r n_s}\,C_{\mathbf{n}-\mathbf{e}_r-\mathbf{e}_s},
    \label{pra:eq:C4_recurrence}
\end{multline}
where $|\mathbf{n}| = n_1+n_2+n_3+n_4$, $\mathbf{e}_r$ is the unit vector in the $r$th direction, and $C_{\mathbf{0}} = 1$.
The convention is that $C_{\mathbf{m}}=0$ if any component of $\mathbf{m}$ is negative. For example, Eq.~\eqref{pra:eq:C4_recurrence} gives
\begin{equation}
    C_{2\mathbf{e}_r}=\frac{\mathsf{B}_{rr}}{\sqrt{2}},
    \qquad
    C_{\mathbf{e}_r+\mathbf{e}_s}=\mathsf{B}_{rs}\quad (r<s),
\end{equation}
starting from $C_{\mathbf{0}}=1$; higher-order coefficients then follow by increasing $|\mathbf{n}|$.
The actual matrix element is the normalized coefficient multiplied by the zeroth-order Gaussian integral,
\begin{equation}
    \mathsf{M}_{(n_i,n_j),(m_i,m_j)}^{(\nu,\mu)}
    =
    \frac{4\pi\alpha_\ell\,e^{c_{cc}^{\nu,\mu}}}
    {\sqrt{\det\mathsf{A}}}\,
    C_{(n_i,m_i,n_j,m_j)}^{\nu,\mu}.
    \label{pra:eq:M_from_C4}
\end{equation}
Thus, for each Gaussian term $\mu$, these coefficients define one-dimensional pair matrices $\mathsf{M}^{(x,\mu)}$ and $\mathsf{M}^{(y,\mu)}$, whose composite row and column indices are the two retained particles' oscillator indices, e.g. $(n_i,n_j)$ and $(m_i,m_j)$.
The full TBRDM is then assembled as a sum of Kronecker products,
\begin{equation}
    \rho^{(2)} = \sum_{\mu=1}^{K} \tilde{w}_\mu \left(\mathsf{M}^{(x,\mu)} \otimes \mathsf{M}^{(y,\mu)}\right),
    \label{pra:eq:rho2_kron}
\end{equation}
Here, $\tilde{w}_\mu$ is the scalar weight of the $\mu$th two-body Gaussian kernel after spectator integration; it absorbs the bra-ket ECG coefficients, Gaussian prefactors, the global normalization $1/\langle\Psi|\Psi\rangle$, and any permutation or spin-pair factors associated with the selected opposite-spin channel.
This construction is analogous to the OBRDM decomposition in Eq.~\eqref{pra:eq:rho1_kron}.
We diagonalize this matrix to obtain the pair-natural-orbital spectrum.

\begin{figure}[tbp]
    \centering
    \includegraphics[width=\linewidth]{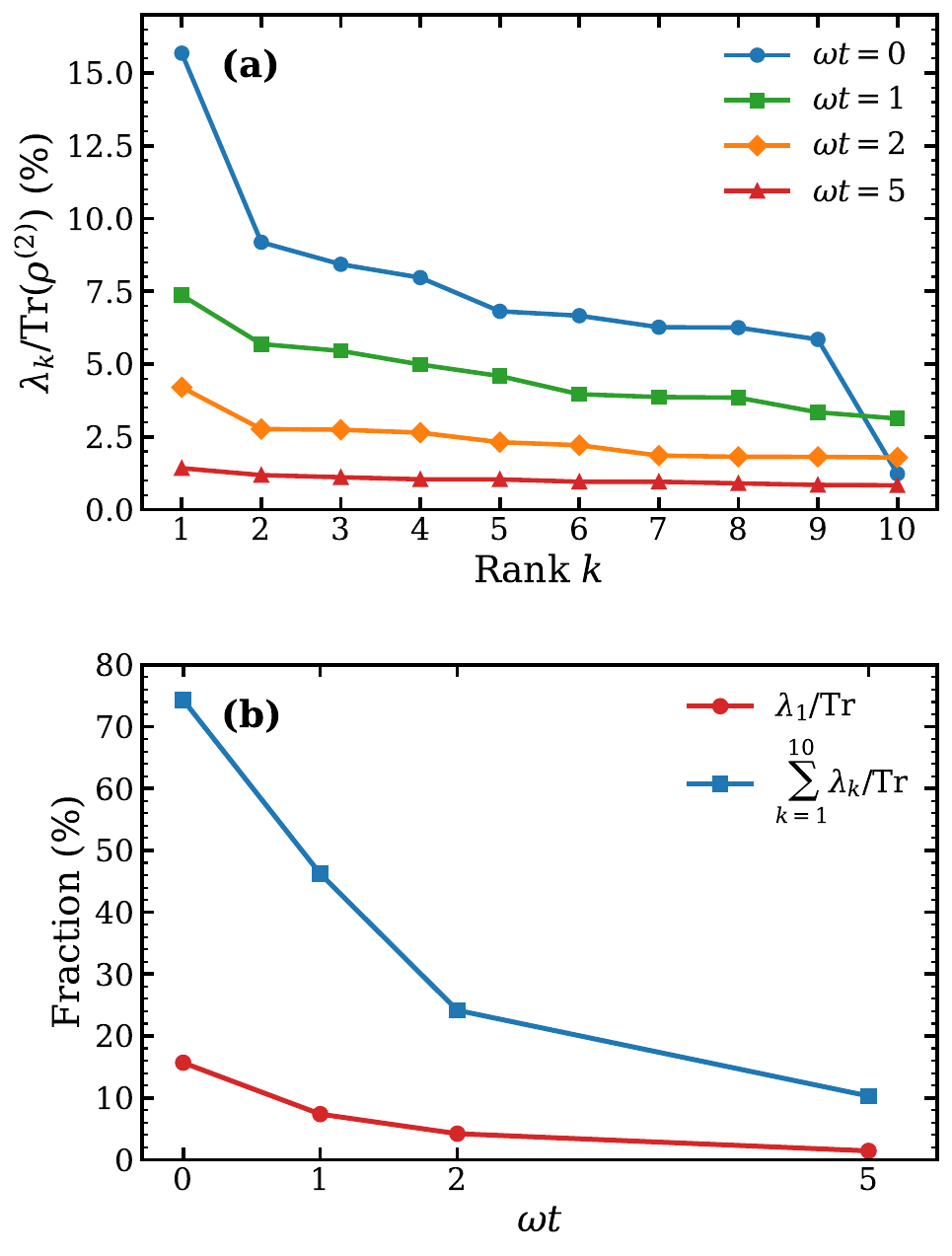}
    \caption{TBRDM diagonalization for the $3{+}3$ system during expansion.
    (a)~The ten largest eigenvalues, normalized by $\mathrm{Tr}(\rho^{(2)})$ and shown in descending rank $k$, at $\omega t=0,1,2,5$.
    (b)~Leading eigenvalue fraction $\lambda_1/\mathrm{Tr}(\rho^{(2)})$ and cumulative top-ten fraction $\sum_{k=1}^{10}\lambda_k/\mathrm{Tr}(\rho^{(2)})$ as a function of time,
    where $\lambda_k$ corresponds to the $k$th largest eigenvalue of $\rho^{(2)}$.
The spectrum broadens continuously during expansion, showing that the two-body sector remains distributed over many pair modes rather than collapsing onto a single dominant orbital.
}    \label{pra:fig:tbrdm_eigenvalues}
\end{figure}

Figure~\ref{pra:fig:tbrdm_eigenvalues} illustrates the output of this diagonalization for the $3{+}3$ system.
As a demonstration of the method, panel~(a) shows that even the initial state already has a broad pair spectrum: the leading eigenvalue carries only about $15.7\%$ of $\mathrm{Tr}(\rho^{(2)})$, while the ten largest eigenvalues together account for about $74\%$.
During expansion, the spectrum flattens further rather than sharpening; by $\omega t=5$, the leading fraction is reduced to about $1.4\%$ and the cumulative top-ten weight to about $10\%$, as summarized in panel~(b).
Methodologically, this means that the TBRDM cannot be represented accurately by only one or a few pair orbitals, so any truncated post-processing based on $\rho^{(2)}$ must retain a comparatively large subspace.
At the same time, the figure shows that the expanding state is not approaching a trivial product of independent dimers: substantial weight remains spread across many two-body eigenmodes, signaling persistent inter-pair correlations and a strongly nontrivial two-body sector.

\section{Conclusion and Outlook}
\label{pra:sec:conclusion}

We have presented a comprehensive account of the time-dependent explicitly correlated Gaussian (TDECG) method, from its theoretical foundations to its application to the nonequilibrium dynamics of strongly interacting two-dimensional few-fermion systems.
The method combines the Lagrangian variational principle with a complex ECG ansatz, yielding classical canonical equations of motion on a finite-dimensional symplectic manifold.
Complete analytical expressions for the symplectic matrix and Hamiltonian gradient---including kinetic, potential, and interaction contributions---have been derived and documented in full.

From a numerical standpoint, the method combines imaginary-time state preparation, adaptive RKF45 propagation, and a post-processing toolkit that reconstructs observables directly from the evolving ECG state.
These tools include analytical extraction of widths and aspect ratios, laboratory real and momentum-space wavefunction reconstruction, Monte Carlo sampling, reduced-density-matrix analysis, correlator evaluation through analytic Gaussian integration supplemented by FFT-based subtraction, and out-of-time-order correlators that quantify few-body information scrambling.

Several directions for future work are evident. First, to access larger particle numbers, the method could be extended to $N=8$--$10$ by incorporating improved permutation sampling or symmetry-adapted bases, which would help mitigate the factorial scaling with $N$ inherent in the permutation sum. Second, the formalism generalizes straightforwardly to systems in other dimensions. Additionally, more complex interaction protocols, such as time-dependent trapping potentials, interaction quenches, and periodic driving, can be incorporated by modifying the Hamiltonian during real-time evolution. Finally, the few-body results presented here provide exact benchmarks for approximate many-body methods, including time-dependent BCS/Bogoliubov--de Gennes theory~\cite{giorgini2008theory,tonini2006formation}, hydrodynamic descriptions of strongly interacting Fermi gases~\cite{ohara2002observation,cao2011universal,giorgini2008theory}, and diagrammatic Monte Carlo techniques such as the bold scheme applied to the unitary Fermi gas~\cite{vanhoucke2012feynman}.

\begin{acknowledgments}
    We acknowledge financial support from the National Natural Science Foundation of China under Grant Nos.~124B2074 and 12204395,
    Hong Kong RGC Early Career Scheme (Grant No.~24308323) and Collaborative Research Fund (Grant No.~C4050-23GF),
    the Space Application System of China Manned Space Program,
    Guangdong Provincial Quantum Science Strategic Initiative GDZX2404004,
    and CUHK Direct Grant No.~4053731.
    Q.G.\ acknowledges support from the NSF through Grant No.~PHY-2409600 and from Washington State University through the Claire May \& William Band Distinguished Professorship Award.
\end{acknowledgments}

\section*{DATA AVAILABILITY}
The data that support the findings of this article are openly available at \cite{data}. The source code and scripts required to reproduce the numerical results are available at \cite{TDECG_code}.

\appendix

\section{Equivalence of Variational Principles for Holomorphic Parametrizations}
\label{pra:app:variational_principles}

In this appendix, we derive the equations of motion from both the Lagrangian and McLachlan variational principles and show their equivalence for holomorphic parametrizations when the Dirac-Frenkel condition is rigorously satisfied.

\subsection{The Dirac-Frenkel condition}

The Dirac-Frenkel variational principle~\cite{dirac1930note,frenkel1934wave} requires that the residual of the Schr\"odinger equation be orthogonal to all allowed variations:
\begin{equation}
    \langle\delta\Psi|\left(i\hbar\frac{\partial}{\partial t} - \hat{H}\right)|\Psi\rangle = 0,
    \label{pra:eq:dirac_frenkel}
\end{equation}
for all $|\delta\Psi\rangle$ in the tangent space of the variational manifold.
For a parametrized ansatz $|\Psi(\bar{\mathbf{z}})\rangle$, the allowed variations are $|\delta\Psi\rangle = \sum_\alpha \delta\bar{z}_\alpha |\partial_\alpha\Psi\rangle$, where $|\partial_\alpha\Psi\rangle \equiv \partial|\Psi\rangle/\partial\bar{z}_\alpha$.
The Dirac-Frenkel condition thus becomes
\begin{equation}
\langle\partial_\alpha\Psi|\left(i\hbar\sum_\beta\dot{\bar{z}}_\beta|\partial_\beta\Psi\rangle - \hat{H}|\Psi\rangle\right) = 0, \quad \forall\,\alpha.
\label{pra:eq:df_parametrized}
\end{equation}

\subsection{McLachlan variational principle}

The McLachlan variational principle~\cite{mclachlan1964variational} minimizes the squared norm of the Schr\"odinger residual,
\begin{equation}
\min_{\dot{\bar{\mathbf{z}}}} \left\|i\hbar|\dot\Psi\rangle - \hat{H}|\Psi\rangle\right\|^2.
\label{pra:eq:mclachlan}
\end{equation}
Taking the variation with respect to $\dot{\bar{z}}_\alpha$
\begin{equation}
    \frac{\partial}{\partial\dot{\bar{z}}_\alpha}\left\|i\hbar\sum_\beta\dot{\bar{z}}_\beta|\partial_\beta\Psi\rangle - \hat{H}|\Psi\rangle\right\|^2 = 0
\end{equation}
yields precisely Eq.~(\ref{pra:eq:df_parametrized}), 
which shows that McLachlan and Dirac-Frenkel are equivalent for complex holomorphic parameters.

By defining the Gram matrix 
\begin{eqnarray}
\mathcal{G}_{\alpha\beta} = \langle\partial_\alpha\Psi|\partial_\beta\Psi\rangle
\end{eqnarray}
and the energy gradient vector 
\begin{eqnarray}
v_\alpha = \langle\partial_\alpha\Psi|\hat{H}|\Psi\rangle,
\end{eqnarray}
we obtain the McLachlan equation of motion in compact form,
\begin{equation}
    i\hbar \mathcal{G} \dot{\bar{\mathbf{z}}} = \mathbf{v}.
    \label{pra:eq:mclachlan_compact}
\end{equation}
This is the formulation adopted by previous time-dependent ECG works~\cite{varga2019optimization,rowan2020simulation}, where the equations of motion are derived from the Dirac-Frenkel/McLachlan principle.

\subsection{Lagrangian variational principle}

The Lagrangian formulation starts from the action,
\begin{equation}
    S = \int_{t_1}^{t_2} L\,dt, \quad L = \frac{i\hbar}{2}\frac{\langle\Psi|\dot\Psi\rangle - \langle\dot\Psi|\Psi\rangle}{\langle\Psi|\Psi\rangle} - \mathcal{H},
\end{equation}
where $\mathcal{H} = \langle\Psi|\hat{H}|\Psi\rangle/\langle\Psi|\Psi\rangle$ is the normalized energy.
Writing the normalization $\mathcal{N} = \langle\Psi|\Psi\rangle$ and K\"ahler potential $\Phi = \ln\mathcal{N}$, we obtain the Lagrangian
\begin{equation}
    L = \frac{i\hbar}{2}\sum_\beta \frac{\partial \Phi}{\partial\bar{z}_\beta}\dot{\bar{z}}_\beta - \frac{i\hbar}{2}\sum_\alpha \frac{\partial \Phi}{\partial z_\alpha}\dot{z}_\alpha - \mathcal{H}.
\end{equation}
The Euler-Lagrange equations 
\begin{eqnarray}
\frac{d}{dt}\frac{\partial L}{\partial\dot{\bar{z}}_\alpha} = \frac{\partial L}{\partial\bar{z}_\alpha}
\end{eqnarray}
yield the canonical equations
\begin{equation}
i\hbar\,\mathcal{C}_{\alpha\beta}\,\dot{\bar{z}}_\beta = \frac{\partial\mathcal{H}}{\partial z_\alpha},
\label{pra:eq:lagrangian_eom}
\end{equation}
where the Fubini--Study metric is
\begin{equation}
\mathcal{C}_{\alpha\beta} = \frac{\partial^2 \Phi}{\partial z_\alpha\partial\bar{z}_\beta} = \frac{\mathcal{G}_{\alpha\beta}}{\mathcal{N}} - \frac{s_\alpha\bar{s}_\beta}{\mathcal{N}^2}
\label{pra:eq:fubini_study_app}
\end{equation}
with $s_\alpha = \langle\partial_\alpha\Psi|\Psi\rangle$.

\subsection{Derivation of equivalence}

Here, we show that Eq.~(\ref{pra:eq:lagrangian_eom}) reduces to Eq.~(\ref{pra:eq:mclachlan_compact}) when the Dirac-Frenkel condition is strictly satisfied.
Using $\mathcal{H} = \langle\Psi|\hat{H}|\Psi\rangle/\mathcal{N}$,
we obtain
\begin{equation}
\frac{\partial\mathcal{H}}{\partial z_\alpha} = \frac{v_\alpha}{\mathcal{N}} - \frac{\mathcal{H}\,s_\alpha}{\mathcal{N}}.
\label{pra:eq:energy_gradient}
\end{equation}
Inserting Eqs.~(\ref{pra:eq:fubini_study_app}) and (\ref{pra:eq:energy_gradient}) into Eq.~(\ref{pra:eq:lagrangian_eom}) and rearranging terms,
we obtain
\begin{equation}
    i\hbar \mathcal{G}_{\alpha\beta}\dot{\bar{z}}_\beta - \frac{i\hbar\,s_\alpha}{\mathcal{N}}\left(\bar{s}_\beta\dot{\bar{z}}_\beta\right) = v_\alpha - \mathcal{H}\,s_\alpha.
\label{pra:eq:intermediate}
\end{equation}

To proceed, we need to evaluate the contraction $\bar{s}_\beta\dot{\bar{z}}_\beta$ that appears on the left-hand side of Eq.~(\ref{pra:eq:intermediate}).
Using $\bar{s}_\beta=\langle\Psi|\partial_\beta\Psi\rangle$ and the chain rule $|\dot\Psi\rangle=\sum_\beta\dot{\bar{z}}_\beta|\partial_\beta\Psi\rangle$, we obtain
\begin{equation}
\bar{s}_\beta\dot{\bar{z}}_\beta=\langle\Psi|\dot\Psi\rangle.
\end{equation}
Its value is fixed by the Dirac-Frenkel condition only if $|\Psi\rangle$ itself happens to lie in the tangent space spanned by $\{|\partial_\alpha\Psi\rangle\}$, i.e.\ if there exist coefficients $c_\alpha$ such that $|\Psi\rangle=\sum_\alpha c_\alpha|\partial_\alpha\Psi\rangle$.
Since Eq.~(\ref{pra:eq:df_parametrized}) imposes orthogonality of the residual to every $|\partial_\alpha\Psi\rangle$, taking the linear combination $\sum_\alpha\bar{c}_\alpha\langle\partial_\alpha\Psi|\,\cdots=\langle\Psi|\,\cdots$ then yields a Dirac-Frenkel-like orthogonality with the test vector $|\Psi\rangle$ itself,
\begin{equation}
\langle\Psi|\left(i\hbar|\dot\Psi\rangle - \hat{H}|\Psi\rangle\right) = 0 \quad\Rightarrow\quad \langle\Psi|\dot\Psi\rangle = -i\langle\Psi|\hat{H}|\Psi\rangle/\hbar.
\label{pra:eq:tangent_condition}
\end{equation}
We refer to the assumption $|\Psi\rangle\in\mathrm{span}\{|\partial_\alpha\Psi\rangle\}$ as the tangent-space condition.

Three logically distinct situations should be kept separate.
First, whenever the linear coefficients $\bar{u}_j$ enter the ansatz as independent variational parameters, the tangent-space condition should be satisfied: writing $|g_k\rangle$ for the $k$th ECG basis function,
\begin{equation}
|\Psi\rangle = \sum_k \bar{u}_k|g_k\rangle = \sum_k \bar{u}_k\,\partial|\Psi\rangle/\partial\bar{u}_k
\end{equation}
expresses $|\Psi\rangle$ as a linear combination of tangent vectors by construction.
Adding the nonlinear ECG parameters $\bar{A}_j,\bar{B}_j$ to the variational set only enlarges the tangent space and therefore cannot invalidate this identity.
Hence, for the unfixed holomorphic linear-plus-nonlinear TDECG ansatz used in this work, the Lagrangian and McLachlan equations of motion are exactly equivalent in continuous time.
Second, projective gauge fixing such as $\bar{u}_1 = 1$ would remove one tangent direction along $|\Psi\rangle$ itself and break the tangent-space condition.
Third, even when the continuous-time identity holds, finite time steps and the regularization of the ill-conditioned Fubini--Study metric (Sec.~\ref{pra:sec:regularization}) break the cancellation between the two subtraction terms in Eq.~(\ref{pra:eq:intermediate}) numerically; the resulting discrepancy is the source of the practical differences between the Lagrangian and McLachlan formulations studied in Appendix~\ref{pra:app:method_comparison}.

By substituting $\bar{s}_\beta\dot{\bar{z}}_\beta = -i\langle\Psi|\hat{H}|\Psi\rangle/\hbar$ into Eq.~(\ref{pra:eq:intermediate}), we obtain
\begin{equation}
    i\hbar \mathcal{G}_{\alpha\beta}\dot{\bar{z}}_\beta + \frac{i\hbar\,s_\alpha}{\mathcal{N}}\left(\frac{i\langle\Psi|\hat{H}|\Psi\rangle}{\hbar}\right) = v_\alpha - \frac{\langle\Psi|\hat{H}|\Psi\rangle}{\mathcal{N}}\,s_\alpha.
\end{equation}
The $s_\alpha$ terms cancel exactly:
\begin{equation}
    i\hbar \mathcal{G}_{\alpha\beta}\dot{\bar{z}}_\beta - \frac{\langle\Psi|\hat{H}|\Psi\rangle}{\mathcal{N}}\,s_\alpha = v_\alpha - \frac{\langle\Psi|\hat{H}|\Psi\rangle}{\mathcal{N}}\,s_\alpha,
\end{equation}
yielding precisely the McLachlan equation of motion~(\ref{pra:eq:mclachlan_compact}).

\begin{figure}[t]
    \centering
\includegraphics[width=\linewidth]{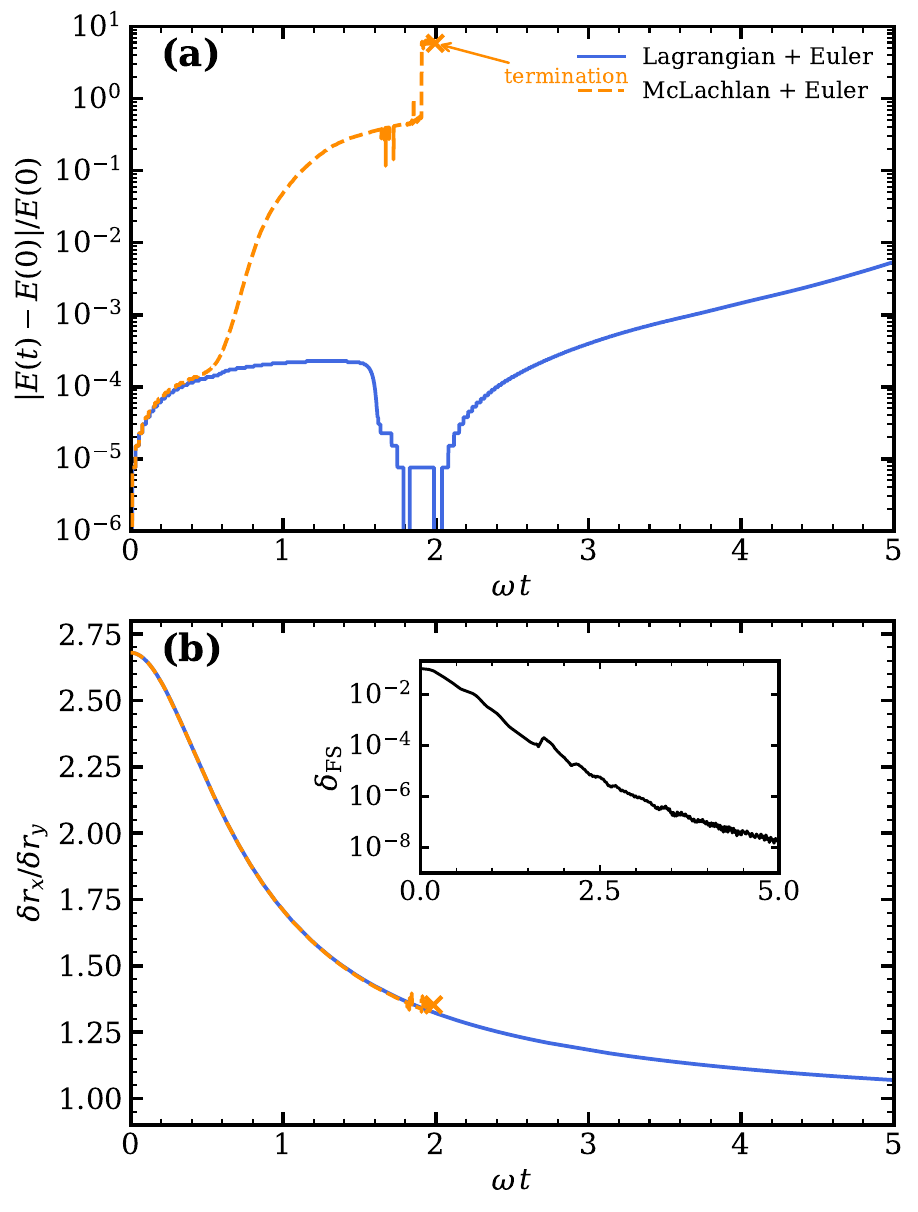}
\caption{
Comparison of fixed-step Euler propagation for the $3{+}3$ fermion time-of-flight expansion ($N_b = 16$, $\Delta t = 10^{-4}\,\omega^{-1}$).
Both runs start from the same initial ECG state and evolve all parameters simultaneously; only the variational matrix differs.
(a)~Relative energy drift $|E(t)-E(0)|/E(0)$ on a logarithmic scale.
The relative error for the Lagrangian run remains below $5.4\times10^{-3}$ through $\omega t = 5$, whereas the error for the McLachlan run grows to order unity and terminates near $\omega t \simeq 2$ after the coefficient matrices lose positive definiteness.
(b)~Real-space aspect ratio $\delta r_x/\delta r_y$ as a function of $\omega t$.
The Lagrangian trajectory evolves smoothly to $\omega t = 5$, while the McLachlan trajectory deviates before the run terminates.
The inset shows the norm-rescaled scalar mismatch $\delta_{\mathrm{FS}}(t)$ defined in Eq.~(\ref{pra:eq:delta_FS_def}) (evaluated in the numerics with $\hbar=1$), which compares the prefactors of the two Fubini--Study subtraction terms in Eq.~(\ref{pra:eq:intermediate}) on a per-norm basis.
The fixed-step Euler method is used here only as a diagnostic of error growth; the large-scale production calculations reported in the main text ($N_b \ge 28$ and $\omega t$ up to $30$) instead use the adaptive RKF45 integrator described in Sec.~\ref{pra:sec:numerical}.}
\label{pra:fig:method_comparison}
\end{figure}

\section{Numerical Comparison of Euler Propagation}
\label{pra:app:method_comparison}

The equivalence derived in Appendix~\ref{pra:app:variational_principles} shows that the Lagrangian and McLachlan variational principles yield identical equations of motion under the exact Dirac-Frenkel condition.
The equivalence, however, does not imply identical behavior after time discretization: finite time steps and regularization of the ill-conditioned symplectic matrix break the cancellation between the two Fubini--Study subtraction terms in Eq.~(\ref{pra:eq:intermediate}) numerically, even though the continuous-time identity holds exactly for the unfixed holomorphic ansatz.
In this section, we numerically show that Lagrangian-type propagation is much more robust than McLachlan-type propagation for the same fixed step size under naive Euler discretization.
The enhanced stability stems from the Fubini--Study subtraction terms, which enforce the symplectic structure more explicitly and thereby stabilize the dynamics.

Both calculations use the same $3{+}3$ fermion system, the same initial ECG ground state, the same basis size $N_b=16$, and the same fixed step size $\Delta t = 10^{-4}\,\omega^{-1}$. 
Figure~\ref{pra:fig:method_comparison} summarizes the result up to $\omega t=5$.
The Lagrangian Euler run remains well behaved throughout this interval.
Its energy drifts only from $E_0 = 1.32546\,\hbar\omega$ to $E(\omega t=5)=1.33260\,\hbar\omega$, corresponding to a relative change of $5.4\times10^{-3}$, while the aspect ratio $\delta r_x/\delta r_y$ evolves smoothly from $2.67982$ at $\omega t=0$ to $1.06913$ at $\omega t=5$.
To diagnose the origin of this improved stability, the inset of Fig.~\ref{pra:fig:method_comparison}(b) shows the norm-rescaled scalar mismatch
\begin{equation}
    \delta_{\mathrm{FS}}(t)\;\equiv\;
    \frac{1}{\mathcal{N}(t)}\left| \frac{i\hbar}{\mathcal{N}(t)}\, \bar{s}_\beta(t)\, \dot{\bar{z}}_\beta(t) - \mathcal{H}(t)\right|,
    \label{pra:eq:delta_FS_def}
\end{equation}
which compares the prefactors of the two subtraction terms proportional to $s_\alpha$ in Eq.~(\ref{pra:eq:intermediate}) on a per-norm basis, with $\mathcal{N}=\langle\Psi|\Psi\rangle$.
If the tangent-space identity $\bar{s}_\beta \dot{\bar{z}}_\beta = -i\langle\Psi|\hat{H}|\Psi\rangle/\hbar$ held exactly, $\delta_{\mathrm{FS}}$ would vanish and the two subtraction terms would cancel identically.
In the Euler run, this cancellation is only approximate: $\delta_{\mathrm{FS}}$ decreases from about $1.02\times10^{-1}$ at $\omega t = 0$ to about $1.55\times10^{-8}$ at $\omega t = 5$, showing that the residual mismatch becomes numerically tiny on the stable part of the Lagrangian trajectory.
The key point is that the Lagrangian formulation retains this subtraction structure explicitly, and the resulting evolution remains numerically stable even when the discrete trajectory does not satisfy the cancellation exactly at every step.

The McLachlan Euler run behaves very differently.
Its energy grows rapidly and reaches $E \simeq 9.09\,\hbar\omega$ by $\omega t \simeq 1.996$, corresponding to a relative drift of $5.86$.
The aspect ratio has already deviated substantially by this point, reaching $\delta r_x/\delta r_y \simeq 1.35$ at $\omega t \simeq 1.98$, after which the run terminates when the coefficient matrices are no longer numerically positive definite.

For the large-scale simulations reported in the main text, where the basis size is increased to $N_b \ge 28$ and the propagation extends to $\omega t \sim 30$, a fixed-step Euler method is not appropriate.
Those production calculations instead use the adaptive RKF45 integrator described in Sec.~\ref{pra:sec:numerical}.
The Euler runs shown in this appendix are included only to expose how rapidly discretization errors accumulate in the two formulations under the same first-order time stepping.

\bibliographystyle{apstest}
\bibliography{reference}

\end{document}